\documentclass[11pt]{article}
\pdfoutput=1

\usepackage{jheppub}

\usepackage{amsmath, amssymb} 
\usepackage{mathtools}
\usepackage{esint}
\usepackage{bm}
\usepackage{dsfont}
\usepackage{braket}
\usepackage{eqparbox}
\usepackage{enumitem}

\numberwithin{equation}{section}

\usepackage[dvipsnames,table]{xcolor}
\usepackage{soul} 
\usepackage{calc}
\usepackage{subcaption}
\usepackage{tikz}
\usepackage{tkz-euclide} 
\usetikzlibrary{arrows.meta}
\tikzset{hidden/.style = {thick, dashed}}
\usepackage{scalefnt}
\usepackage[font=small,labelfont=bf]{caption}
\usepackage{graphicx}

\usepackage[numbers, sort&compress]{natbib}
\usepackage{comment} 
\usepackage[colorinlistoftodos]{todonotes}

\usepackage{hyperref}
\hypersetup{colorlinks=true, linktoc=page, linkcolor=purple, citecolor=blue}

\newcommand{\mcC}{\mathcal{C}}

\newcommand{\mcN}{\mathcal{N}}
\newcommand{\mcO}{\mathcal{O}}

\newcommand{\bbP}{\mathbb{P}}

\newcommand{\bbT}{\mathbb{T}}

\renewcommand{\a}{\alpha}

\newcommand{\g}{\gamma}

\renewcommand{\[}{\left[}

\newcommand{\<}{\left\langle}
\renewcommand{\>}{\right\rangle}

\newcommand{\p}{\partial}
\renewcommand{\d}{\mathrm{d}}

\newcommand{\VS}{\text{VS}}

\renewcommand{\Im}{\operatorname{Im}}

\newcommand{\floor}[1]{\cramped{\left\lfloor #1 \right\rfloor}}

\newcommand{\half}{\tfrac{1}{2}}
    
\newcommand{\no}{\nonumber}

\newcommand{\be}{\begin{equation}}
\newcommand{\ee}{\end{equation}}
\newcommand{\reef}[1]{(\ref{#1})}
\newcommand{\gap}{\text{gap}}

\begin{document}

\title{Bootstrapping Extremal Scalar Amplitudes With and Without Supersymmetry}

\date{\today}

\author[a]{Justin Berman}
\author[a,b]{Henriette Elvang}
\author[a,c]{Nicholas Geiser}
\author[a]{Loki L.~Lin}

\affiliation[a]{
    Leinweber Center for Theoretical Physics, Randall Laboratory of Physics\\
    University of Michigan, Ann Arbor\\
    450 Church St, Ann Arbor, MI 48109-1040, USA}
\affiliation[b]{Niels Bohr International Academy, Niels Bohr Institute \\
University of Copenhagen \\
Blegdamsvej 17, DK-2100 Copenhagen Ø, Denmark}
\affiliation[c]{
    Michigan Center for Applied and Interdisciplinary Mathematics\\
    University of Michigan, Ann Arbor\\
    530 Church St, Ann Arbor, MI 48109-1040, USA}

\emailAdd{jdhb@umich.edu}
\emailAdd{elvang@umich.edu}
\emailAdd{ngeiser@umich.edu}
\emailAdd{lokilin@umich.edu}

\preprint{LCTP-24-23}

\abstract{We re-examine positivity bounds on the $2\to2$ scattering of identical massless real scalars with a novel perspective on how these bounds can be used to constrain the spectrum of UV theories.  We propose that the entire space of consistent weakly-coupled (and generically non-supersymmetric) UV amplitudes is determined as a convex hull of the massive scalar amplitude and a one-parameter family of scalarless ``extremal amplitudes'' parameterized by the ratio of the masses of the two lightest massive states. Further, we propose that the extremal amplitudes can be constructed from a similar one-parameter set of maximally supersymmetric amplitudes, leading to the surprising possibility that the S-matrix bootstrap with maximal supersymmetry may be sufficient to determine the entire allowed space of four-point amplitudes! Finally, we show that minimal spectrum input reduces the allowed space of Wilson coefficients to small islands around the open string Dirac-Born-Infeld tree amplitude and the closed string Virasoro-Shapiro amplitude.}


\maketitle

\section{Introduction}
\label{sec:intro}

Relativistic dispersion relations explicitly display how the spectrum of a theory contributes to its scattering processes, and their importance for understanding the effects of high-energy physics in and beyond quantum field theory has long been appreciated~\cite{Dyson:1956vea}. 
Fundamental physical principles such as unitarity, locality, and analyticity
are straightforwardly implemented on the dispersion relations, and the
dispersive approach (originally developed to describe
the interactions of mesons \cite{Chew:1957tf,Mandelstam:1958xc}) therefore plays an essential role in
the modern S-matrix bootstrap program \cite{Paulos:2016but,Paulos:2017fhb}.

An extremely useful way to organize dispersion relations is in terms of the Wilson coefficients of low-energy effective field theory (EFT). In the Wilsonian formulation, EFTs arise from integrating out UV physics. The dispersion relations for the Wilson coefficients then make manifest the way each UV state enters into the IR observables. From these dispersive representations, one can derive two-sided bounds on the Wilson coefficients, thereby bounding the  space of EFTs. These are ``dual'' bounds in the sense that the parameter space outside the bounds is rigorously excluded:  
 any EFT that arises as the low-energy expansion of a weakly-coupled, unitary, and local UV theory has to have Wilson coefficients that lie within the bounds. 
These ideas have been widely applied in many different contexts, and bootstrap bounds have been derived for EFTs of real and complex scalars, pions, photons, and gravitons~\mbox{\cite{Bellazzini:2020cot,Arkani-Hamed:2020blm,Sinha:2020win,Huang:2020nqy,Henriksson:2021ymi,Li:2021lpe,Caron-Huot:2021rmr,Chiang:2021ziz,Chowdhury:2021ynh,Caron-Huot:2022ugt,Albert:2022oes,Fernandez:2022kzi,Albert:2023jtd,CarrilloGonzalez:2023cbf,Berman:2023jys,Bhat:2023puy,Albert:2023seb,Haring:2023zwu,Berman:2024wyt,Albert:2024yap,Beadle:2024hqg,Bhat:2024agd}}. A reverse analysis is also possible: from the dispersive representation of the Wilson coefficients, constraints can be derived on the allowed spins and masses in the spectrum of unitary theories~\cite{Berman:2024kdh}. 

In this paper, we revisit the massless Abelian scalar EFT bootstrap first studied in~\cite{Caron-Huot:2020cmc,Arkani-Hamed:2020blm} with a new perspective that emphasizes understanding how the UV spectrum manifests itself in IR quantities. The most general low-energy expansion for the four-point amplitude of identical massless Abelian scalars is permutation symmetric in the Mandelstam variables $s,t,u$ and takes the form
\begin{align}
\label{eq:A4simple}
  A(\phi\phi\phi\phi)
  &=
  -\lambda_{\phi}^2
  \bigg( \frac{1}{s} + \frac{1}{t} + \frac{1}{u} \bigg)
  +
  \kappa^2
  \bigg( \frac{tu}{s} + \frac{su}{t} + \frac{st}{u} \bigg)
\no \\
  & \quad
  + g_0 
  + g_2 (s^2 + t^2 + u^2)
  + g_3 (stu)
  + g_4 (s^2 + t^2 + u^2)^2
  + \ldots
  \, .
\end{align}
The first set of massless poles arise from the usual $\tfrac{1}{3!} \lambda_{\phi} \, \phi^3$ interaction, and the second set of poles are generated by the exchange of a massless graviton with $\kappa$ given in terms of Newton's gravitational constant  by ${\kappa^2 = 8 \pi G_\text{N}}$.\footnote{Massless poles with higher-order residues are disallowed in the ansatz because we cannot have interacting massless spin $>2$  states in flat space \cite{Weinberg:1980kq}. The spin-1 spin exchange would be $(t+u)/s$ + perms.~but that adds to zero due to momentum conservation $s+t+u=0$.} The bootstrap analysis of the Wilson coefficients $g_k$ is naturally split into the non-gravitational case with ${\kappa = 0}$ and the gravitational case with ${\kappa \neq 0}$. Our attention is primarily on the non-gravitational set-up. In this case, the tree-level Dirac-Born-Infeld (DBI) open superstring amplitude plays a key role, both as a target of the bootstrap and as a guide for understanding the allowed space of Wilson coefficients more generally. In the gravitational case, a natural target of the bootstrap is the tree-level four-dilaton amplitude from closed superstring theory (i.e.\ the massless scalar version of the Virasoro-Shapiro amplitude). 

The EFT bootstrap relies on several fundamental  assumptions on the four-point amplitude $A(s,u)$ including unitarity, analyticity, the existence of a mass gap $M_\text{gap}$, the partial wave expansion, and polynomial boundedness in the form a Froissart-Martin-like~\cite{Froissart:1961ux, Martin:1962rt}  bound:
\be
\label{eq:froissartn0}
    \lim_{\substack{ |s| \to \infty \\ \text{fixed } u<0 }}
    \frac{A(s,u)}{s^{n_0}} 
    = 0
    \, .
\ee
With $n_0=2$, this is the ``standard'' Froissart bound  for field theory amplitudes \cite{Froissart:1961ux, Martin:1962rt, Arkani-Hamed:2020blm, Haring:2022cyf}, but in certain cases stronger assumptions can be justified, e.g.~$n_0=1$ for the case of large-$N$ QCD~\cite{Albert:2022oes, Fernandez:2022kzi,Albert:2023jtd, Albert:2023seb} or, as relevant in our work here, $n_0=0$ from maximal supersymmetry. The bound \reef{eq:froissartn0} is really an assumption about polynomial boundedness, or Regge-boundedness, but for simplicity we refer to it as ``the Froissart bound'' also for gravitational amplitudes or string amplitudes. 

We consider both $n_0=2$ and $n_0=0$:
\begin{itemize}
    \item \textbf{2SDR bootstrap:}~We derive bounds on the Wilson coefficients of the low-energy expansion of amplitudes that  satisfy~\eqref{eq:froissartn0} with $n_0=2$ using twice-subtracted dispersion relations (2SDR). When no  spectrum assumption (other than a mass gap) are made, these bounds can be considered the ``universal bounds'' in the sense that no special properties, such as supersymmetry or an enhanced Froissart bound, are assumed. 
    \item \textbf{0SDR bootstrap:}~We derive bounds on the Wilson coefficients of the low-energy expansion of amplitudes that  satisfy~\eqref{eq:froissartn0} with $n_0=0$ using zero-times subtracted dispersion relations (0SDR). More Wilson coefficients can be accessed with 0SDR than with 2SDR, so the 0SDR bounds are stricter than the 2SDR bounds. The 0SDR bootstrap  bounds maximally supersymmetric (non-gravitational) amplitudes with $n_0=2$.
\end{itemize}
The consequences of different subtraction-levels in the dispersion relations were previously studied in \cite{McPeak:2023wmq}, but without the connection to maximal supersymmetry. 

To facilitate the analysis, we assume a mass gap $M_\text{gap}$ such that any massive state contributing to the four-point amplitude must have mass $M \geq M_\text{gap}$. Expressing the Wilson coefficients in units of $M_\text{gap}$ makes all the $g_k$ dimensionless.

In both the 2SDR and 0SDR bootstraps, we further assume that the massless states are weakly-coupled. In the weak-coupling limit, unitarity is simply imposed as positivity of the imaginary part of each partial wave of the four-point amplitude. Hence, there are no absolute bounds on the Wilson coefficients themselves (e.g.~$g_2, g_3, g_4, \dots$ in the 2SDR bootstrap). Instead, we derive two-sided bounds on their ratios (e.g.~$g_3/g_2, g_4/g_2, \dots$). Any linear combination of two amplitudes that lie within the bootstrap bounds also obeys the positivity constraints. Therefore, the space of allowed (ratios of) Wilson coefficients is a projective convex space.

To compute two-sided bounds, we use a numerical approach introduced in~\cite{Caron-Huot:2020cmc} and the semi-definite programming solver SDPB~\cite{Simmons-Duffin:2015qma}. The numerical set-up requires truncation of the low-energy expansion~\eqref{eq:A4simple} to finite Mandelstam order $k_\text{max}$, corresponding to terms with $2k_\text{max}$ derivatives in a low-energy effective action. 
The numerical bounds typically become incrementally stronger with increasing $k_\text{max}$. 

Let us now highlight the key results in this paper, focusing on the non-gravitational case entirely until we discuss the bootstrap of the Virasoro-Shapiro amplitude towards the end of the Introduction. 

\begin{figure}
\centering
\begin{tikzpicture}
\node (image) at (0,0) {\includegraphics[width=0.48\linewidth]{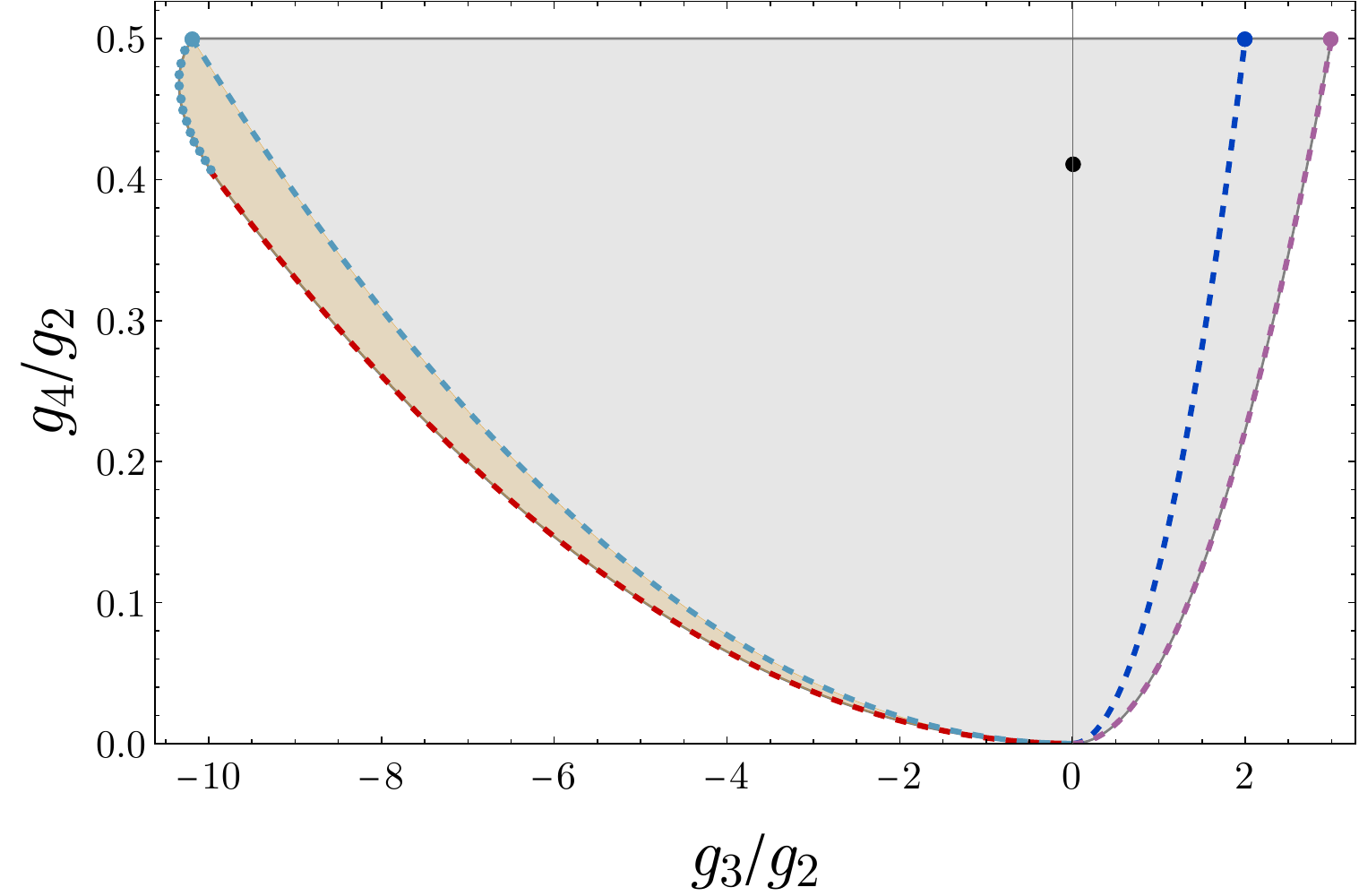}};
\node at (1.7,1.55) {\footnotesize{DBI}};
\end{tikzpicture}
~
{\includegraphics[width=0.48\linewidth]{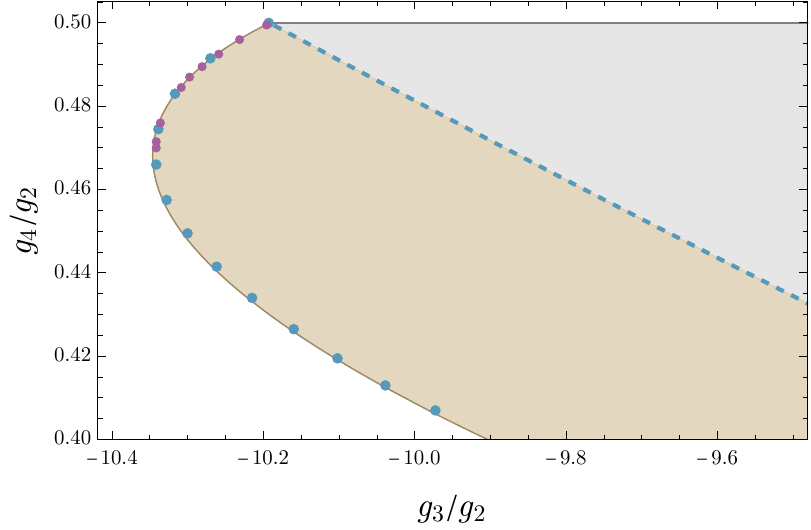}}
\caption{On the left, the universal 2SDR bootstrap bounds on $(g_3/g_2,g_4/g_2)$ in $D=4$. 
The orange extremal sliver is bounded by the scalarless 2SDR extremal theories (light-blue points) that are characterized by having maximal spin 2 coupling for the lowest massive state in the spectrum 
\reef{2sdrExtrSetup}. 
On the right, we zoom-in to show that the numerically contructed 2SDR extremal theories track the upper boundary of the extremal sliver. 
The purple points are examples of 2SDR extremal theories reconstructed from 2SDR maximally supersymmetric extremal theories via a procedure that involves removing a kinematic SUSY factor, subtracting off scalars, and rescaling the spectrum.
}
\label{fig:basicg3g4}
\end{figure}

\subsubsection*{2SDR Bootstrap and its Extremal Theories}
In the 2SDR bootstrap, 
there are dispersion relations for all the Wilson coefficients $g_k$ with $k \geq 2$, but not for $g_0$. 
 Thus, the lowest operator-dimension Wilson coefficients that can be bounded are the ratios $g_3/g_2$ and $g_4/g_2$. Because of positivity, the allowed values of these coefficients form a convex region, first described in~\cite{Caron-Huot:2020cmc} and reproduced on the left of Figure~\ref{fig:basicg3g4}. 
The entire upper bound on $g_3/g_2$ (the purple dashed curve on the right) is determined by the basic model that has a massive scalar state with mass-squared $M^2= \mu M_\text{gap}^2$ as its only massive state. The purple dot corresponds to $\mu=1$ and the purple dashed curve with $\mu \ge 1$ is the model's {\em mass-scaling curve}.

The dark-blue dashed curve in Figure~\ref{fig:basicg3g4} denotes the infinite spin tower (IST) amplitude, which has massive states with all even spins $J=0,2,4,\dots$ at mass-squared $M^2 = \mu M_\text{gap}^2$ for some $\mu \ge 1$. The light-blue dashed curve is obtained by removing the spin-$0$ state from the IST amplitude \cite{Caron-Huot:2020cmc}. We call this the ``scalar-subtracted infinite spin tower'' (ssIST). The numerical bounds computed with SDPB, however, show that the full space is larger: clearly leftover from the convex hull of these identifiable models is a sliver of parameter space shaded orange in Figure~\ref{fig:basicg3g4}. It was noted in~\cite{Caron-Huot:2020cmc} that the amplitudes in this sliver do not contain massive scalar states, but they have otherwise not been identified. Among the goals of this paper is to characterize the four-point amplitudes of these  ``extremal theories"\footnote{We do not know an underlying microscopic theory, but we simply say ``extremal theory'' as synonymous with ``extremal four-point amplitude''.} that define the outer boundary of this orange region. 
We shall further argue that extremal theories play a quintessential role for understanding the full space of allowed Wilson coefficients.

We determine the extremal theories as follows. First, we argue numerically that if there is a finite number of states with mass-squared $M_\text{gap}^2$, they can only have spin 0 or spin 2.\footnote{An analytic derivation will be presented in future work \cite{NickJustin}.} (Odd spins do not contribute to an $stu$-symmetric amplitude.) Next, the extremal theories are in the scalarless region, hence we only need to consider amplitudes that have a spin-$2$ state at $M_\text{gap}^2$. We then assume the absence of other states until a choice of cutoff scale $\mu_c M_\text{gap}^2$, i.e.~
\be
\label{2sdrExtrSetup}
\raisebox{-6.5mm}{\includegraphics[width=5.5cm]{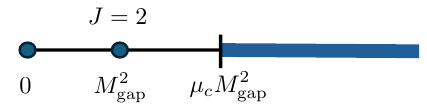}}
~\hspace{1cm}
\raisebox{-8.5mm}{\includegraphics[width=3.5cm]{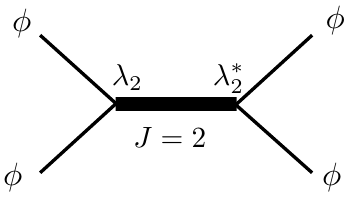}}
\ee
We are agnostic about the spectrum above $\mu_c M_\text{gap}^2$. 
To obtain the extremal theories,  we simply maximize the spin-2 coupling $|\lambda_2|^2$ in units of the Wilson coefficient $g_2$ for a given choice of $\mu_c$. 
Once $|\lambda_2|^2/g_2$ is fixed to its maximal value, we find that each ratio $g_k/g_2$ is uniquely determined in the sense that the SDPB bounds for its maximum and minimum agree numerically.\footnote{The numerical agreement is within the same number of digits that we use to fix the coupling $|\lambda_2|^2/g_2$.}
This means that for each $\mu_c$, the spin-2 coupling maximization selects a unique theory with a particular fixed spectrum above $\mu_c M_\text{gap}^2$ which we do not need to specify to determine the Wilson coefficients numerically. 
Thus, we posit that the extremal theories are a one-parameter family of amplitudes parameterized by $\mu_c$ and that they interpolate between the scalar-subtracted infinite spin tower (ssIST) for $\mu_c =1$ and the pure spin-2 model for  $\mu_c \to \infty$. On the right of Figure~\ref{fig:basicg3g4}, the light-blue points are our results for $g_3/g_2$ and $g_4/g_2$ of the extremal theories with $\mu_c$ in the interval $1 \leq \mu_c \leq 4$. They clearly track the universal boundary of the orange extremal region.

\subsubsection*{Convex Hull Conjecture}
We have discussed the allowed space of the two lowest boundable Wilson coefficients, but the space of couplings comprise a higher-dimensional space,
\be
 \bigg(
 \frac{g_3}{g_2},
 \frac{g_4}{g_2},
 \frac{g_5}{g_2},\dots ,
 \frac{g_{k_{\text{max}}}}{g_2}\bigg)
\ee 
for given $k_\text{max} \ge 3$.  In this space, the one-parameter family of extremal theories lie on a curve, parameterized by $1 \le \mu_c \le \infty$. Our central claim is that 
 \begin{quote}

 {\em the convex hull of the extremal theory curve together with its mass scaling  generate the entire allowed region of scalarless theories with $n_0=2$.} 
  \end{quote}
  Moreover, by the properties of the 2SDR bootstrap, scalars can freely be added to the spectrum of the amplitudes with $n_0=2$. It then follows from the Convex Hull Conjecture that the full space of allowed $n_0=2$ theories (with or without scalars) can be obtained as positive linear combinations of the pure scalar theory and the extremal theories. 

It may seem to be a tall order to claim that a small family of curves generate an entire multi-dimensional space of allowed Wilson coefficients. However, it is quite generally true that a lower-dimensional set of points has a higher-dimensional convex hull. For example, the convex hull of four non-planar points in 3-dimensions is a tetrahedron. 
As another example, the convex hull of a curve of the form
\be
  \label{toymasscurve}
   (a x, b x^2 , c x^3, d x^4, e x^4)
   ~~~\text{for}~~~
   0 \le x \le 1 \,,
\ee
(with non-zero $a,b,c,d,e$) is a four-dimensional region in the five-dimensional embedding space. Actually. this is exactly the situation for any mass-scaling curve. For instance, for the amplitude with only a single scalar at mass-squared $\mu M_\text{gap}^2$ with $\mu \ge 1$, we have
\be
 \label{masscurveex}
  \bigg(
\frac{g_3}{g_2},
\frac{g_4}{g_2},
\frac{g_5}{g_2},
\frac{g_6}{g_2},
\frac{g_6'}{g_2}
  \bigg)
  = 
  \bigg( \frac{3}{\mu}, \frac{1}{2\mu^2}, \frac{5}{2\mu^3} ,\frac{1}{4\mu^4}, 
  \frac{3}{\mu^4} \bigg) \,.
\ee
Here $g_6$ and $g_6'$ denote the Wilson coefficients of the two independent scalar operators at $12^\text{th}$ order in the derivative expansion. As $\mu$ is varied from 1 to infinity, \reef{masscurveex} traces out a curve like that of \reef{toymasscurve} with  $x = \mu^{-1}$. This is the mass-scaling curve of the pure scalar model; its projection to the $(g_3/g_2,g_4/g_2)$-plane is the purple dashed curve in Figure \ref{fig:basicg3g4}. In the space of the first five Wilson coefficients, the convex hull of this mass-scaling curve is four-dimensional. In general, a single mass-scaling curve generates a convex hull of mass-dimension $k_\text{max}-2$ in the space of all Wilson coefficients with $k \le k_\text{max}$. The union of two different mass-scaling curves can generate a convex hull of dimension greater than $k_\text{max}-2$. 

To test the conjecture, we generate the convex hull of the extremal theories and their mass-scaling curves. For the $3$-dimensional $(g_3/g_2,g_4/g_2,g_5/g_2)$ projection, we then pick a point $(g_3/g_2,g_4/g_2)$ and calculate the corresponding max and min of $g_5/g_2$ in two ways: first by optimizing over convex combinations of the extremal theories and second by computing the positivity bounds using SDPB. We find that our results agree to within $10^{-2}$ for $k_\text{max}=12$ with the discrepancy decreasing by about 10\% for $k_\text{max}=14$. A similar analysis is done for  $(g_6/g_2,g_6'/g_2,g_7/g_2)$ with even better agreement.

\subsubsection*{Is Supersymmetry Everything?}

In certain S-matrix bootstrap analyses, maximal supersymmetry is assumed to give a stronger technical handle on the analysis. For example, this is done in the bootstrap of supergravity amplitudes in \cite{Guerrieri:2021ivu, Guerrieri:2022sod, Albert:2024yap} and for the super Yang-Mills bootstrap in \cite{Berman:2023jys, Berman:2024wyt, Albert:2024yap}. A super-skeptic may complain that assuming maximal supersymmetry provides at best a toy model of non-supersymmetric real-world problems. 
To counter this skepticism, we propose that the most general bounds of the non-supersymmetric 2SDR bootstrap may be obtained from a bootstrap with maximally supersymmetry.

Per the Convex Hull Conjecture, we only need to construct the extremal theories of the 2SDR bootstrap to determine the bounds on the entire non-supersymmetric space. The argument that the 2SDR extremal theories can be obtained from the maximally supersymmetric amplitudes goes as follows:
\begin{enumerate}
\item Any non-gravitational, maximally supersymmetric Abelian 4-point amplitude of four identical scalars in $D \ge 4$ can be written as
\be
  A^\text{(SUSY)}(s,t,u) = (s^2 + t^2 + u^2) \, A^\text{(0SDR)}(s,t,u) 
\ee
with $A^\text{(0SDR)}$ a function fully symmetric in $s,t,u$. Assuming the amplitude 
$A^\text{(SUSY)}$ obeys the generic Froissart bound with $n_0 =2$, the $(s^2 + t^2 + u^2)$-``stripped'' amplitude 
$A^\text{(0SDR)}$ must obey Froissart with $n_0=0$. This stripped amplitude can therefore be bootstrapped using 0SDR. Thus, the 0SDR bootstrap can be viewed as a bootstrap of maximally supersymmetric 2SDR theories.
\item In the 0SDR, we show that  a state with mass $M_\text{gap}$ has to be a scalar when there is a gap to the next possible state at $\mu_2 M_\text{gap}^2$. Maximizing the coupling $|\lambda_0|^2/g_0$ of this massive scalar for given $\mu_2$ uniquely determines Wilson coefficients. This gives a one-parameter family of 0SDR extremal theories parameterized by $\mu_2$.
\item The 0SDR extremal theories can be mapped to the 2SDR extremal theories by a two-step process. First, all scalars are subtracted. As a result, the amplitude no longer has $n_0=0$, but does satisfy the $n_0=2$ Froissart bound, so scalar-subtraction maps the 0SDR extremal theories into the 2SDR allowed region. Second, to ensure that the first massive state (which has spin 2) has mass-squared $M_\text{gap}^2$ rather than $\mu_2 M_\text{gap}^2$, we rescale the entire spectrum of the amplitude by $1/\mu_2$.
\end{enumerate}
Numerically, we have shown that this procedure reproduces the 2SDR extremal theories with $\mu_c$ to  $\sim 1.7$; the result is shown in  Figure~\ref{fig:basicg3g4} as the purple dots in the zoomed-in plot on the right.\footnote{It is numerically challenging to produce the extremal theories with gaps larger than $\mu \gtrsim 1.7$.} If the 0SDR extremal theories are indeed in one-to-one correspondence with the 2SDR extremal theories, it would imply that 
{\em the universal bounds on non-supersymmetric theories can be obtained from the bootstrap of maximally supersymmetric theories.}

The extremal theories also correspond to certain ``corner theories'' found in bounds on the Wilson coefficients. This feature is similar to that of the one-parameter family of maximally supersymmetric corner theories found in~\cite{Berman:2024wyt}.
Indeed, it is reasonable to expect that the 0SDR extremal theories found here in the Abelian real scalar case are the Abelianized corner theories from~\cite{Berman:2024wyt}. Hence, it is also plausible that a similar Convex Hull Conjecture holds for the non-Abelian case.  

It is an interesting question if  analytic formulas can be found for the extremal theories.  
Among the analytic candidates in the literature, perhaps the most promising are the deformations of  Veneziano and Virasoro-Shapiro in~\cite{Cheung:2024obl}.\footnote{Similar hypergeometric deformations of the Veneziano and Virasoro-Shapiro amplitudes have been recently studied in~\cite{Cheung:2023adk, Rigatos:2023asb, Geiser:2023qqq, Wang:2024wcc, Cheung:2024uhn, Mansfield:2024wjc}.} However, we argue that to match the $D$-dimensional one-parameter family of extremal solutions, one needs an additional parameter which fixes the critical dimension  independently of the Regge slope --- and one would also need solutions with non-linear Regge trajectories. Neither is the case for the existing families of analytic deformations of the Veneziano and Virasoro-Shapiro amplitudes currently known in the literature.

\subsubsection*{Bootstrapping Super-DBI and Virasoro-Shapiro}
There are two stringy targets of opportunity for our bootstrap: the DBI scalar amplitude and the gravitational dilaton amplitude. The $\mathcal{N}=4$ super-DBI theory in $D=4$ is simply the maximally supersymmetric field theory living on a single flat D3-brane in $D=10$ Minkowski space. Thus, bootstrapping $D=4$ DBI can be interpreted as a bootstrap of the low-energy  D3-brane action. 

The  four-graviton string tree-amplitude amplitude is, up to polarization-factors, given by the Virasoro-Shapiro amplitude. Using supersymmetry to replace the gravitons with gravitational dilatons, it fits directly into the identical scalar bootstrap. Thus, we  bootstrap the four-dilaton amplitude in the context of maximal supersymmetry.

\begin{figure}
    \centering
    \includegraphics[width=0.49\linewidth]{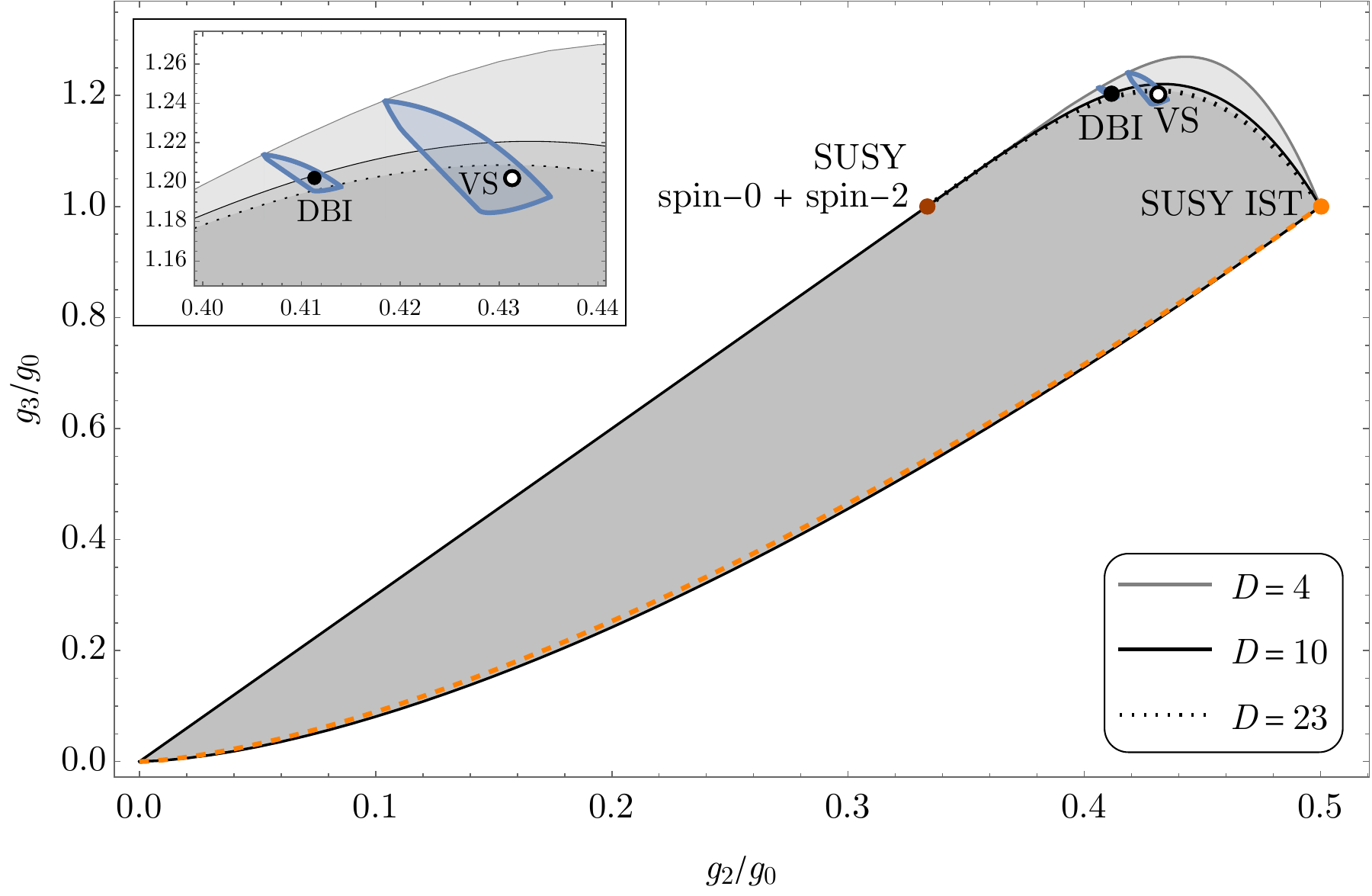}
    \hfill\includegraphics[width=0.49\linewidth]{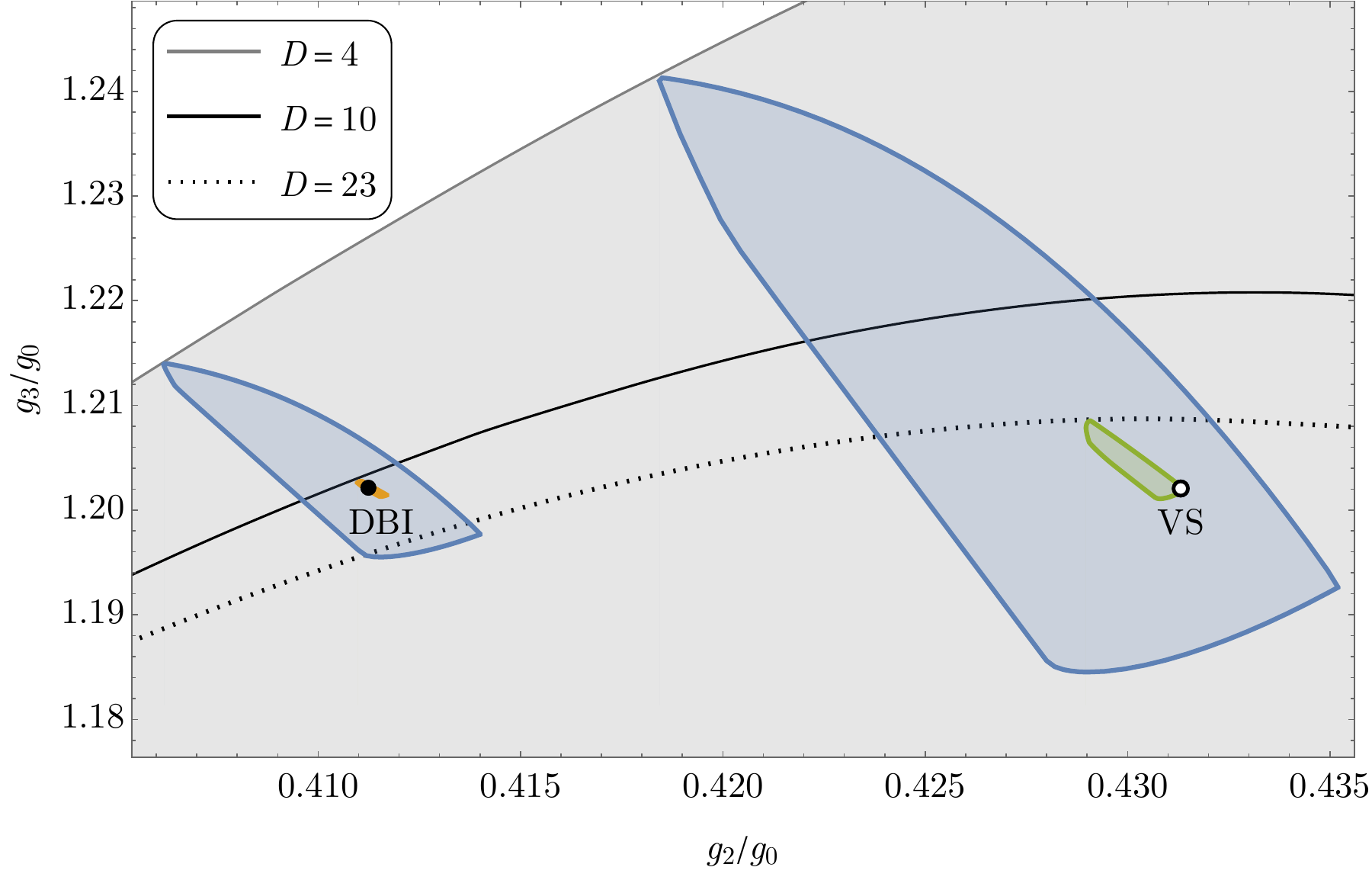}
    \caption{Islands in the allowed parameter space around DBI and Virasoro-Shapiro, shown on the left for $D=4$ in the context of the general bounds for the 0SDR bootstrap. On the right, we zoom in on the $D=4$ islands to contrast their relative immensity with the island for the critical dimensions: $D=10$ for DBI (orange) and $D=23$ for Virasoro-Shapiro (green). The plots are done with $k_\text{max}=16$.}
    \label{fig:twoislands}
\end{figure}

To bootstrap DBI and Virasoro-Shapiro, we are encouraged by the analysis of \cite{Berman:2024wyt} where it was shown for maximally supersymmetric Yang-Mills EFTs that one could obtain a small island shrinking around the Veneziano amplitude by only specifying the value of the coupling of the lowest-mass state to the massless external states (in units of the lowest Wilson coefficient). In this paper, we perform a similar analysis for the DBI and Virasoro-Shapiro amplitudes. The results for $D=4$ with supersymmetry imposed are the blue islands shown on the left of Figure~\ref{fig:twoislands}. As part of the analysis, we discuss the role of supersymmetry and show that without supersymmetry, the resulting islands are significantly bigger.

Examining the DBI and Virasoro-Shapiro islands in  varying the spacetime dimension $D$, it becomes clear that the best bootstrap results are obtained when the target amplitude is as close as possible to violating unitarity without doing so --- and that is exactly the case in the ``formal critical dimension'', i.e.~the highest spacetime dimension $D$ for which the amplitude is unitary. 
In their respective critical dimensions ($D=10$ for DBI and, somewhat bizarrely as we shall explain in the main text, $D=23$ for Virasoro-Shapiro\footnote{In the context of superstring theory, neither of these critical dimension makes much sense. The closed superstring lives in $D=10$, but this constraint does not arise for the tree-level Virasoro-Shapiro amplitude which remains unitary for $D$ up to and including $D=23$. Likewise, it does not make sense to talk about 10-dimensional DBI, because the DBI scalar is the Goldstone mode of the spontaneously broken translation symmetry transverse to the brane, and if the brane is space-filling, as is the case for $D=10$, there are no such scalars, but only the $\mathcal{N}=1$ supersymmetric version of Born-Infeld (BI). Restricting the momenta of the external BI photons to a 4-dimensional subspace and picking their polarizations in the transverse directions, we can nonetheless still use the 4-dimensional characterizations of the external states as scalars. Hence we abuse the language slightly and continue to denote the amplitude as DBI even in $D=10$.}), we do indeed find that the bootstrap islands become significantly smaller around the each of the two theories, as shown on the right in Figure  \ref{fig:twoislands}. 

\subsection*{Outline}
We begin by stating the assumptions of our analysis in Section~\ref{sec:assumptions}. We then proceed to discuss a variety of different example amplitudes that play central roles in the analysis: we consider non-supersymmetric amplitudes in Section~\ref{sec:exampleamps} and then describe the constraints of maximal supersymmetry in Section \ref{sec:imposing_susy} along with examples of such amplitudes, including the open string DBI amplitude and the dilatonic supergravity amplitude. The remainder of Section \ref{sec:2} is dedicated to deriving the dispersion relations (Section \ref{sec:setup}), the null constraints for permutation symmetry (Section \ref{sec:null}), and brief comments on the numerical implementation of the optimization problem in SDPB (Section \ref{sec:num}). Readers familiar with the S-matrix bootstrap may choose to skip or to read only selectively from Section~\ref{sec:2}. 

Section \ref{sec:maxspin} presents numerical evidence for constraints on the possible discrete spectrum of tree-level exchanged states for amplitudes in the 2SDR and 0SDR boostraps. 
In Section \ref{sec:2sdrextr}, the 2SDR extremal theories are characterized, the Convex Hull Conjecture is stated precisely, and evidence in its favor presented. 

In Section \ref{sec:0sdrextr} we describe the maximally supersymmetric (i.e.~0SDR)  extremal theories. We detail the procedure for reproducing the 2SDR extremal theories from those in 0SDR. 
Identifying the extremal theories as ``corner theories'', we comment on the need for non-linear Regge trajectories 
and we make contact with the analytic proposals for hypergeometric deformations of the Veneziano and Virasoro-Shapiro amplitudes \cite{Cheung:2024obl}.

Bootstrapping the DBI amplitude is the focus of Section \ref{sec:susyboot}. We show that a bootstrap without supersymmetry constraints can produce an island around DBI, but imposing SUSY gives much stronger constraints. We emphasize the role of the spacetime dimension by showing that the stronger bounds are obtained in the critical dimension. In Section \ref{sec:gravamps} we carry out a similar supersymmetric bootstrap of the Virasoro-Shapiro closed string amplitude. 

We end in Section \ref{sec:disc} with conclusions and outlook. Some technical details are relegated to the appendices: Appendix \ref{app:SUSY} on supersymmetry constraints, Appendix \ref{app:gegenbauer} with practical formulas for the Gegenbauer polynomials, and Appendix \ref{app:indnulls} with a systematic approach to determining the linearly independent null constraints for the bootstrap.

\section{Assumptions,  Examples, and Dispersion Relations}
\label{sec:2}
In this section, we state our assumptions, discuss some of the key amplitudes that appear in our analysis, and derive the dispersion relations for the  Wilson coefficients of the low-energy expansion.

\subsection{Assumptions} \label{sec:assumptions}

We are studying the scattering of four massless identical  scalars $\phi$. We use all-outgoing conventions, hence Bose symmetry  implies that the Abelian 4-point amplitude is fully symmetric in the Mandelstam variables $s= -(p_1+p_2)^2$, $t=-(p_1+p_3)^2$, and $u=-(p_1+p_4)^2$. Momentum conservation requires $s+t+u=0$.

The assumptions of the bootstrap analysis are:

\begin{enumerate}

\item
\label{prop:massgap}
\textbf{Mass Gap}: There are no massive states below $M_{\gap}$.

\item
\label{prop:partial}
\textbf{Partial wave decomposition}: For physical values of the Mandelstam variables (i.e.\ for $s \geq -u \geq 0$), the amplitude $A(s,u)$ can be written in a basis of partial waves,
\be
\label{eq:partial}
    A(s,u)
    =
    \sum_{j = 0}^\infty
    n_j^{(D)}
    a_j^{\mathstrut}(s) \,
    G_j^{(D)} \Big( 1 + \frac{2u}{s} \Big)
    \, .
\ee
Here $G_j^{(D)}(x)$ is the $D$-dimensional spin-$j$ Gegenbauer polynomial and $n_j^{(D)}$ is a normalization factor defined in Appendix \ref{app:gegenbauer}. The functions $a_j(s)$ are the spin-$j$ partial waves.

\item
\label{prop:weak}
\textbf{Weak coupling}: 
The amplitude is perturbative in the couplings to the massless states such that loops of massless states do not contribute to $a_j(s)$ and there are no branch cuts below $s = M_{\gap}^2$. As a result, the effective couplings do not run and the low-energy expansion is polynomial in $s$, $t$, and $u$, except for possible massless poles.

\item
\label{prop:unitarity}
\textbf{Unitarity}: At  weak coupling, unitarity of the S-matrix implies a simple positivity condition on the imaginary part of the partial waves:
\be
  \label{unitaritypos}
    \Im a_{j}(s) \geq 0 
    \qquad \text{for} \quad
    s > 0
    \, .
\ee

\item
\label{prop:analytic}
\textbf{Analyticity}: In the complex $s$-plane (with fixed $u < 0$), the amplitude $A(s,u)$ is analytic away from the real $s$-axis.

\item
\label{prop:bound}
\textbf{Polynomial boundedness}: The amplitude $A(s,u)$ obeys the following Froissart-Martin-like bound (for $\arg(s) \neq 0, \pi$),
\be
\label{eq:bound}
    \lim_{\substack{ |s| \to \infty \\ \text{fixed } u<0 }}
    \frac{A(s,u)}{s^{n_0}} 
    = 0
    \, ,
\ee
for some integer $n_0$.  
The standard Froissart bound has $n_0=2$. Motivated by maximal supersymmetry, we also explore the case of $n_0=0$.
\end{enumerate}

Assumptions \ref{prop:massgap} and \ref{prop:weak} are physical statements about the spectrum of our theory. The remaining assumptions are technical statements about the S-matrix. For our purposes, we shall simply assume the veracity of each property since they are standard assumptions in the modern S-matrix bootstrap literature. For a more technical discussion on the status of these assumptions, we direct the reader to~\cite{Correia:2020xtr} and the references therein.

The mass-dimension of a four-point amplitude and its Wilson coefficients depends on the spacetime dimension $D$, so to avoid cluttering the formulas with factors of $\text{mass}^{-D}$, we scale our amplitudes by a factor of $M_\text{gap}^{4-D}$, where $M_\text{gap}$ is the assumed mass-gap, so that the four-point amplitude is dimensionless. With this convention, general $D$-dimensional formulas have the same mass-dependence as in $D=4$. 

Permutation symmetry implies  $A(s,u) = A(s,t)$ and, using this together with momentum conservation and the parity property of the Gegenbauer polynomials, $G_{j}^{(D)}(-x) = (-1)^{j} \, G_{j}^{(D)}(x)$, one  finds that only even spins $j$ contribute to~\eqref{eq:partial}. Indeed, only even-spin states  appear in the spectra of the example amplitudes discussed below. 

\subsection{Example Amplitudes}\label{sec:exampleamps}

The most general low-energy expansion of an $stu$-symmetric amplitude can be written as 
\begin{equation}
\label{eq:gmn}
    A(s,t,u)
    =
    A_\text{poles}(s,t,u)
    + \sum_{m,n=0}^\infty
    g_{2m+3n;\,m,n} \, 
    (s^2+t^2+u^2)^m (stu)^n
    \, ,
\end{equation}
where $A_\text{poles}$ contains any massless poles and the polynomial terms 
 are written in terms of $stu$ and $s^2 + t^2 + u^2$ which generate the ring of symmetric polynomials in $s,t,u$ with $s + t + u = 0$. The polynomial terms in \reef{eq:gmn} are in one-to-one correspondence with the local higher-derivative on-shell operators in the EFT action. Specifically,
$g_{2m+3n;\,m,n}$ is the Wilson coefficient of an operator of the form $\partial^{2k}\phi^4$ for $k=2m+3n$. 
The notation $g_{2m+3n;\,m,n}$ for the Wilson coefficients is quite clunky, so we make use of the abbreviated notation
\be 
\label{g0setc}
  g_0 = g_{0;0,0}\,,~~~
  g_2 = g_{2;1,0}\,,~~~
  g_3 = g_{3;0,1}\,,~~~
  \text{etc.}
\ee
Whenever there is more than one Wilson coefficient at a given order, we denote them as $g_k$, $g_k'$, $g_k''$ etc. For example, at sixth order in the Mandelstams, we have
\be
\label{g6s}
  g_6 = g_{6;3,0}\,,~~~
  g_6' = g_{6;0,2}\,.
\ee
For brevity of notation, 
we often refer to the Wilson coefficients collectively as $g_k$, thus leaving it implicit that the property discussed also holds for $g_k'$, $g_k''$, etc.

We now describe a selection of four-point amplitudes with  central roles in our work.

\subsubsection{Single Massive State}
\label{sec:single}

If a spin-$J$ state with mass $M \ne 0$ is exchanged at tree-level in the $s$-channel, the residue of the $s=M^2$ pole is  
\be
  \label{genres}
 \text{Res}_{s=M^2}A(s,t,u)
  \propto  G^{(D)}_J(1+\tfrac{2u}{M^2})\,,
\ee
with $G^{(D)}_J$ the $D$-dimensional spin-$J$ Gegenbauer polynomial which is a degree-$J$ polynomial in~$u$. Because of permutation symmetry, there must be similar terms in the $t$ and $u$ channels, and hence the overall amplitude behaves as $s^J$ at large $|s|$. For this reason, the standard Froissart bound~\reef{eq:froissartn0} with $n_0=2$ excludes theories with a single massive spin $J >2$ state.
Any amplitudes with massive states of spin $J > 2$ must therefore have an infinite set of massive states so that they can resum into an amplitude that obeys the Froissart bound.  

Thus, the two single-state theories that play a role in our analysis are the pure scalar amplitude 
\begin{equation}
\label{eq:A0}
    A^{(0)}(s,t,u)
    =
    -
    \bigg(
      \frac{M^2}{s-M^2}
    + \frac{M^2}{t-M^2}
    + \frac{M^2}{u-M^2}
    \bigg)
    \, 
\end{equation}
and the 
pure spin-2 amplitude 
\begin{align}
\label{eq:AL}
    A^{(2)}(s,t,u)
    =
    - \frac{1}{2}  \,
    \bigg(
    &
    \frac{M^2}{\makebox[\widthof{$u$}][c]{$s$}-M^2}
    \Big[
      G^{(D)}_2 \big( 1+\tfrac{2t}{M^2} \big)
    + G^{(D)}_2 \big( 1+\tfrac{2u}{M^2} \big)
    \Big]
    + \text{perms}
    \bigg)
    \,.
\end{align}
The pure scalar amplitude \reef{eq:A0} goes to a constant as $|s| \to \infty$ for constant $u <0$, so it is a borderline case for the 0SDR bootstrap but within the 2SDR bootstrap.  Being a borderline case means that a small deformation \cite{Albert:2022oes} makes the amplitude satisfy the needed $n_0=0$ Froissart bound, so that in the limit of taking the deformation to zero, these amplitudes are included in the bounds of the 0SDR. 
The pure spin-2 theory \reef{eq:AL} grows as $s^2$ for large $s$, so it is a borderline case for the 2SDR bootstrap.

The Wilson coefficients (in units of $M_{\gap}^2$) in the low-energy expansion of the two amplitudes \reef{eq:A0} and \reef{eq:AL} are tabulated in Table \ref{tab:exWCs}. 
The mass gap $M_\text{gap}$ is a choice of scale for the bootstrap problem. While it is assumed that there are no states below the mass gap, it is natural but not required for the lowest massive state to have mass $M_\text{gap}$. For example, if we assume the pure scalar theory to have mass-squared $M^2 = \mu M_\text{gap}^2$ rather than $M_\text{gap}^2$ for some $\mu \ge 1$, we find  
\be
  \label{spin0parabola}
 \frac{g_3}{g_2}=  \frac{3}{\mu}\,
 ~~~\text{and}~~~
 \frac{g_4}{g_2}=  \frac{1}{2\mu^2}
 ~~~~\implies~~~~
 \frac{g_4}{g_2} = \frac{1}{18}
 \bigg(\frac{g_3}{g_2} \bigg)^2 \,.
\ee
This traces out the {\em mass-scaling curve} in the $(g_3/g_2,g_4/g_2)$-plane  for the pure scalar model. This curve is simply a parabola interpolating between the origin (where $\mu \to \infty$) and $(3,1/2)$ (where $\mu=1$), and it forms one of the universal boundaries of the 2SDR bootstrap, as shown in Figure \ref{fig:basicg3g4}.

\begin{table}[t]
\be\no
  \begin{array}{c|ccccccccccc}
     & \,\tilde{g}_0\, & \,\tilde{g}_2\, & \,\tilde{g}_3\, & \,\tilde{g}_4\, & \,\tilde{g}_5\, & \,\tilde{g}_6\, & \,\tilde{g}_6'\, & \,\tilde{g}_7\,\\[1mm]
     \hline & \\[-4mm]
     \text{spin 0 \reef{eq:A0}} 
     & 3 
     & 1 
     & 3 
     & \frac{1}{2}
     & \frac{5}{2} 
     & \frac{1}{4}
     & 3
     & \frac{7}{4}
      \\[2mm]
     \text{spin 2 \reef{eq:AL}} 
     & 3
     & ~~\frac{3D-4}{D-2}~~
     & -\frac{3(3D-2)}{D-2}~
     & \frac{1}{2}
     & -\frac{2+3D}{2(D-2)}
     & \frac{1}{4}
     & -\frac{3(3D-2)}{D-2}
     & -\frac{6+D}{4(D-2)} \\[2mm]
     \text{IST \reef{eq:basicIST}} 
     & 1
     & \frac{1}{2}
     & 1
     & \frac{1}{4}
     & 1
     & \frac{1}{8}
     & 1
     & \frac{3}{4}
     \\[2mm]
     \text{SUSY DBI \reef{eq:dbi}} 
     & 0
     & \frac{\pi^2}{2}
     & 0
     & \frac{\pi^4}{48}
     & \frac{\pi^2}{2}\zeta_3
     & \frac{\pi^6}{960}
     & 0
     & \frac{\pi^4}{48}\zeta_3+\frac{\pi^2}{4}\zeta_5 \\[2mm]
     \text{closed string \reef{susygravitonAmp}} & 0 & 0 & 0 & 2\zeta_3 & 0 & \zeta_5 & 0 & 2\zeta_3^2
  \end{array}
\ee
\caption{\label{tab:exWCs} Values of the first few Wilson coefficients 
$\tilde{g}_k = |\lambda|^{-2} \mu^{k} g_k$ for our example theories. The coefficients are made dimensionless by scaling them with powers of $M_\text{gap}^{2}$, and $\mu = M^2 / M_\text{gap}^{2}$ is the mass of the lowest mass state relative to the choice of mass gap scale $\mu=M^2/M_{\gap}^2$. In the case of the stringy UV completions, $\lambda$ is proportional to the string coupling. Here ``SUSY DBI'' is the Abelian supersymmetric DBI amplitude \eqref{eq:dbi}, and ``closed string'' is the four-dilaton closed superstring amplitude \eqref{susygravitonAmp} (i.e.\ $(s^2+t^2+u^2)^2$ times Virasoro-Shapiro). For the string amplitudes, the Wilson coefficients include an additional factor of $\alpha' M_\text{gap}^2$ if the mass gap is chosen to be different from $1/\alpha'$.} 
\end{table}

\subsubsection{Infinite Spin Tower}

An example of an amplitude with an infinite set of higher-spin states
is the Infinite Spin Tower (IST),\footnote{This kind of amplitude has been argued to be non-local \cite{Caron-Huot:2016icg, Cheung:2023uwn, Cheung:2024uhn}, but mathematically it satisfies all of the bootstrap assumptions and is thus needed for understanding the universal bounds of allowed UV-completions \cite{Caron-Huot:2020cmc}. Recent work \cite{Wang:2024jhc} argues that IST-type amplitudes may arise from chiral string theory at tree-level.} where all spin states have the same mass 
$M^2$:
\begin{equation}
\label{eq:basicIST}
    A^{\text{IST}}(s,t,u)
    =-|\lambda|^2\frac{M^{6}}{(s-M^2)(t-M^2)(u-M^2)}\,.
\end{equation}
The IST amplitude exchanges states with spins 0, 2, 4, 6, \dots, all at mass $M^2$ and it obeys the Froissart bound with $n_0=-1$. Its lowest Wilson coefficients are given in Table \ref{tab:exWCs}. 

\subsubsection{Scalar-Subtracted Amplitudes}\label{sec:scalarsub}

Given any amplitude $A(s,u)$ that obeys the Froissart bound with $n_0=2$, we can subtract off any scalar contributions without violating the Froissart bound. Specifically, if $A(s,u)$ exchanges a scalar with coupling $\lambda_0$ at mass $M^2 = \mu_0 M_\text{gap}^2$, we define the ``scalar-subtracted'' amplitude by
\begin{align}
    A^\text{ss}(s,t,u)
    =
    A(s,t,u)
    - |\lambda_0|^2 \, 
    A^{(0)}( s,t,u)
    \, ,
\end{align}
where $A^{(0)}$ is the pure scalar amplitude \reef{eq:A0}. 
Unitarity is also still satisfied since we have subtracted exactly the amount of scalar which would make the spectral density reach zero, but not become negative. 
If there are more scalars in the  spectrum, all of them can be subtracted in this way to get a theory that satisfies all of our bootstrap assumptions. Subtracting off spin-$2$ (or higher) states is not allowed since the spin-$2$ exchange~\eqref{eq:AL} violates the $n_0=2$ Froissart bound. Likewise, for the $n_0=0$ bootstrap, scalar subtraction is not allowed. 

As an example, let us subtract the scalar contribution from the Infinite Spin Tower amplitude~\eqref{eq:basicIST} to get the scalar-subtracted IST (ssIST) which only exchanges states with spins $J \geq 2$:
\be
\label{eq:ssIST}
    A^{\text{ssIST}}(s,t,u)
    =
    - |\lambda|^2
    \frac{M^6}{(s-M^2)(t-M^2)(u-M^2)}
    + \frac{4}{9} \,
    _2F_1 \big( \half, 1, \tfrac{D-1}{2}; \tfrac{1}{9} \big) \,
    A^{(0)}(s,t,u)
    \, .
\ee
As described in \cite{Caron-Huot:2016icg}, the ssIST lies at the top left corner of the $(g_3/g_2,g_4/g_2)$ region shown in Figure \ref{fig:basicg3g4}. Its mass-scaling curve is the ``inner'' bound of the extremal sliver. 

\subsection{Amplitudes with Maximal Supersymmetry}
\label{sec:imposing_susy}
In this section, we describe the constraints of maximal supersymmetry and give examples of SUSY amplitudes. We start in $D=4$ dimensions and extend the results to $D>4$ at the end of the section.

\subsubsection{Maximal SUSY Constraints} \label{sec:maxsusyconst}
In $D = 4$, maximal supersymmetry in a non-gravitational theory is $\mathcal{N}=4$ extended SUSY, and the massless Abelian supermultiplet consists of the photon, four photinos, and six real scalars collected into three complex scalars. The leading two-derivative theory is free, but the amplitudes can be non-vanishing when higher-derivative corrections are included. The $\mathcal{N}=4$ SUSY Ward identities imply that all 4-point amplitudes are propertional to each other, and following the steps in 
\cite{Berman:2023jys}, maximal supersymmetry in $D=4$ tells us that the complex scalar amplitude can be written 
\be
  \label{eq:ztophi}
   \text{Abelian $\mathcal{N}=4$ SUSY:}~~~~~
   A(zz\bar{z}\bar{z}) = s^2 f(s,t,u)\,,
\ee
where $f$ has to be fully permutation symmetric in $s,t,u$ for an Abelian amplitude.\footnote{For the non-Abelian case, as in \cite{Berman:2023jys}, $f$ is only crossing symmetric: $f(u,t,s)=f(s,t,u)$.}
If $\phi$ is the real part of $z$, we have (up to an overall normalization)
\be 
  \label{zbarztophisec2}
  A^\text{Abelian}(\phi\phi\phi\phi)
  = 
  A^\text{Abelian}(zz\bar{z}\bar{z})
  + A^\text{Abelian}(z\bar{z}z\bar{z})
   +A^\text{Abelian}(\bar{z}zz\bar{z})\,.
\ee
It follows that any  real scalar amplitude in a maximally supersymmetric theory can be parameterized as 
\be \label{eq:comptoreal}
  \text{Abelian $\mathcal{N}=4$ SUSY:}~~~~~
  A(\phi\phi\phi\phi) 
  =(s^2 + t^2 +u^2)  f(s,t,u)
\ee
with $f$ fully symmetric in $s,t,u$.

\paragraph{SUSY Constraints on Wilson Coefficients.}
The low-energy expansion of an amplitude \reef{eq:gmn} takes the form \reef{eq:comptoreal} when it has no pure $(stu)^n$ terms. This means that supersymmetry can be imposed in the bootstrap by setting 
 \be\label{eq:susynulls}
   g_{3n;\,n,0} = 0 
 \ee
for each $n = 1,2,\ldots,$ up to the desired order in the derivative expansion. We call these the ``SUSY null constraints.''

If a supersymmetric 4-point real scalar amplitude \reef{eq:comptoreal} satisfies the Froissart bound with $n_0=2$, we can think of it as  $(s^2+t^2+u^2)$ times a 4-point  amplitude with $n_0=0$. Conversely, multiplying any amplitude in the 0SDR bootstrap by the factor $(s^2+t^2+u^2)$ gives a maximally supersymmetric amplitude in the 2SDR bootstrap.

Thus, there are two approaches to bootstrapping maximally supersymmetric amplitudes in the 2SDR bootstrap. The first is to work with the twice-subtracted dispersion relations and impose \reef{eq:susynulls} as additional null constraints. The second approach is to bootstrap the ``${(s^2+t^2+u^2)}$-stripped'' amplitude using zero-times subtracted dispersion relations. We  compare these two approaches in Section \ref{sec:susyboot}.


\paragraph{SUSY Constraints on Couplings.} 
\label{sec:susycouplings}
Let us construct a SUSY amplitude by multiplying the scalar amplitude \reef{eq:A0} by  $(s^2+t^2+u^2)$:
\begin{equation}
\label{eq:A0susy}
 A^{\text{SUSY-}0}(s,t,u)
    =
    -|\lambda|^2 \,
    (s^2+t^2+u^2)
    \bigg(
      \frac{M^2}{s-M^2}
    + \frac{M^2}{t-M^2}
    + \frac{M^2}{u-M^2}
    \bigg)
    \, .
\end{equation}
This is no longer simply a pure scalar exchange amplitude; rather it is a combination of spin 0 and spin 2 of the same $\mathcal{N}=4$ massive supermultiplet. The  relationship between the couplings can be directly distilled from the $s=M^2$ residue of \reef{eq:A0susy} by comparing it to a linear combination of Legendre polynomials $P^{(4)}_J$ for $J=0$ and $J=2$, 
\be
  -\mathop{\mathrm{Res}}_{s = M^2}A^{\text{SUSY-}0}(s,t,u)
  = |\lambda_0|^2 P^{(4)}_0 \big( 1+\tfrac{2u}{M^2} \big)
  +
  \frac{|\lambda_2|^2}{2} P^{(4)}_2 \big( 1+\tfrac{2u}{M^2} \big)
  + (t \leftrightarrow u) \,,
\ee
and the result in  $D=4$ is
\begin{align} \label{eq:susymgrel4d}
    \frac{|\lambda_2|^2}{g_2}=\frac{1}{5}
    \frac{|\lambda_0|^2}{g_2}  \,.
\end{align}
The spin 0 and 2 states are just two of the states of the massive $\mathcal{N}=4$ supermultiplet. The other states, the massive spin 1/2, 1, and 3/2, show up in 4-point amplitudes with other external states. The statements we make here are specifically about which states show up in the $s$-channel for the four-$\phi$ amplitude. Similar coupling relations arise for higher-spin supermultiplets and are straightforward to derive.

\paragraph{From 4 to $D$ Dimensions.} 
We extend to $D$  dimensions as follows. If $D > 4$, we consider the external massless states to be restricted to a 4-dimensional subspace, and then we can still use the $D=4$ quantum numbers to label the states. In particular, if the external states are selected to be $D=4$ scalars, the SUSY Ward identities then  
again restrict the amplitudes to take the form \reef{eq:ztophi} and \reef{eq:comptoreal}, respectively. The spin of the internal states will ``know'' they are in $D$ dimensions via the Gegenbauers in the partial wave expansion despite being restricted to the 4-dimensional
subspace by momentum conservation. We slightly abuse notation and continue to use their 4-dimensional quantum numbers. Then the relationship between the couplings of the spin-0 and spin-2 massive states to the external massless states can be found using \reef{eq:A0susy}, which requires them to obey
\begin{align} \label{eq:susymgrel}
    \frac{|\lambda_2|^2}{g_2}=\frac{D-2}{3D-2}
    \frac{|\lambda_0|^2}{g_2} 
    \,.
\end{align}
The supersymmetric DBI amplitude discussed below in Section \ref{sec:dbi} has couplings that obey this relationship. 

\paragraph{SUSY Spin Tower.} The maximally supersymmetric Infinite Spin Tower is obtained by multiplying \reef{eq:basicIST} by 
$(s^2+t^2+u^2)$ to get
\begin{equation}
    \label{susyIST}
    A^{\text{SUSY IST}}(s,t,u)=-|\lambda|^2(s^2+t^2+u^2)\frac{M^6}{(s-M^2)(t-M^2)(u-M^2)}\,.
\end{equation} 
The SUSY IST amplitude goes to a constant for $|s| \to \infty$ so it is a marginal case for the $n_0=0$ Froissart bound. 
Curiously, the SUSY IST amplitude is a particular linear combination of the basic non-SUSY spin tower \reef{eq:basicIST} and the pure scalar exchange amplitude \reef{eq:A0}:
\begin{align}
    A^{\text{SUSY IST}}(s,t,u)
    &=3A^\text{IST}(s,t,u)-A^{(0)}(s,t,u)\;.
\end{align}
This means that it will also be located on the line between the IST model and the scalar model in the allowed region.

\subsubsection{Dirac-Born-Infeld Amplitude}
\label{sec:dbi}
The supersymmetric DBI amplitude can be obtained from the Veneziano amplitude. Details are given in Appendix \ref{app:SUSY}, but it is useful to outline the procedure here. First, the complex scalar color-ordered Veneziano amplitude 
\be
  \label{veneziano}
   A(zz\bar{z}\bar{z}) = -(\a's)^2 \frac{\Gamma(-\a's)\Gamma(-\a'u)}{\Gamma(1+\a't)}\,
\ee
is Abelianized by summing over color-orderings \reef{u1decoup}
to 
get
\be\label{complexdbi}
  A(zz\bar{z}\bar{z})
  = s^2 A^\text{stripped}(s,t,u)\,,
\ee
where the $s,t,u$-symmetric ``stripped'' amplitude is
\be
  \label{Astripped}
 A^\text{stripped}(s,t,u)
 = -{\a'}^2\bigg(
    \frac{\Gamma(-\a's)\Gamma(-\a'u)}{\Gamma(1+\a't)}
    + 
    \frac{\Gamma(-\a't)\Gamma(-\a's)}{\Gamma(1+\a'u)}
    + 
    \frac{\Gamma(-\a'u)\Gamma(-\a't)}{\Gamma(1+\a's)}
    \bigg)\,.
\ee
The real scalar amplitude 
is obtained directly from 
 $A(zz\bar{z}\bar{z})$ via \reef{zbarztophisec2}, so that
\be
  \label{eq:dbi}
  A(\phi\phi\phi\phi)= 
  \big(s^2+t^2+u^2\big)\,A^\text{stripped}(s,t,u)\,.
\ee
This is the real scalar DBI amplitude. It obeys the Froissart bound with $n_0=2$.  
Table \ref{tab:exWCs} lists the leading Wilson coefficients in the low-energy expansion of \reef{eq:dbi}. 
 
The DBI amplitudes \reef{complexdbi} and \reef{eq:dbi} have simple poles at each $s=n/\alpha'$ for odd integers ${n = 1,3,5,\ldots}$. In contrast, the Veneziano amplitude \reef{veneziano} has poles for all positive integer $n$, with only odd-spin states exchanged at the even-$n$ mass levels and only even-spin states at the odd-$n$ mass levels. When Abelianizing Veneziano, the odd spins are ``projected out'' since the Abelian amplitude cannot have any odd spins. Thus, only the odd-$n$ mass levels remain.

 The highest-spin states at each mass level form a linear Regge trajectory with slope $1$.
At the two lowest mass levels of the real scalar DBI amplitude \reef{eq:dbi}, the couplings $|\lambda_{n,J}|^2$ of the massive  spin states to the massless external states of \reef{eq:dbi} are listed in the following table:
\be
  \label{DBIcouplings}
   \begin{array}{|c|cccc|}
     \hline
     \text{mass level \textbackslash\hspace{1mm}spin} & 0 &2 &4 &  ~~6~~~\\
     \hline &&&& ~\\[-3.5mm]
     1 
     & \frac{3D-2}{D-1} 
     & \frac{D-2}{D-1} 
     & - & - \\[2mm]
     \hline &&&& ~\\[-4mm]
     3 
     & {\color{orange}~\frac{3(56+26D-3D^2)}{8(D-1)(D+1)}~}
     &  ~\frac{3(13D+66)(D-2)}{4(D-1)(D+3)} ~
     & ~\frac{27D(D-2)}{8(D+1)(D+3)}~ 
     & -\\[2mm]
     \hline
   \end{array}
\ee
At mass level 1, 
the couplings of the massive spin-0 and spin-2 states satisfy the $D$-dimensional SUSY coupling relation \reef{eq:susymgrel}. At mass level 3, the states are a combination of a supermultiplet with spin 0 up to spin 2 as well as a supermultiplet with spins between 2 and 4. An analysis similar to the one that lead to \reef{eq:susymgrel} can be used to disentangle the contributions from each supermultiplet using the explicit couplings in  \reef{DBIcouplings}.

Importantly, the DBI amplitude \reef{eq:dbi} is unitary only in $2 < D \leq 10$. Explicit violation of unitarity is clear from coupling $|\lambda_{3,0}|^2$ (highlighted in \reef{DBIcouplings}) which is negative for $D \gtrsim 10.45$. 
Thus the highest \textit{integer} dimension for which the DBI amplitude is unitary is $D=10$.

On the other hand, the stripped DBI amplitude \reef{Astripped} has the same mass levels $n=1,3,5,\ldots$, but the highest spin at each mass level is two lower than that of the unstripped DBI amplitude \reef{eq:dbi}. The couplings at the lowest two mass levels are:
\be
  \label{strippedDBIcouplings}
   \begin{array}{|c|ccc|}
     \hline
     \text{mass level \textbackslash\hspace{1mm}spin} & 0 &2 &4  ~~~~\\
     \hline &&& ~\\[-3.5mm]
     1 
     & 1 
     & - 
     & - \\[2mm]
     \hline &&& ~\\[-5mm]
     3 
     & {\color{orange}~\frac{10-D}{24(D-1)}~}
     &  \colorbox{white}{$~\frac{3(D-2)}{8(D-1)} ~$}
     & - \\[2mm]
     \hline
   \end{array}
\ee
The highlighted coupling is zero in exactly $D=10$ and negative in higher $D$. Hence, the stripped DBI amplitude has critical dimension \textit{exactly} equal to 10. The same is true for the complex scalar DBI amplitude \eqref{complexdbi}. Importantly, the complex scalar SUSY factor of $s^2$ does not change the critical dimension, but the real scalar SUSY factor of $s^2+t^2+u^2$ does.

\subsubsection{Closed Superstring}
\label{sec:exclosed}
With gravitons in the spectrum, maximal supersymmetry in $D=4$ is $\mathcal{N}=8$ supergravity. Considering a pair of conjugate scalars in the  $\mathcal{N}=8$ theory, 
the 4-point amplitude takes the form
\be
    \text{$\mathcal{N}=8$ SUSY:}~~~~~ A(zz\bar{z}\bar{z}) = s^4 f(s,t,u) \,,
\ee
where $f$ is fully symmetric in $s$, $t$, and $u$. The minimally coupled graviton exchange is captured by $f(s,t,u) = - \kappa^2/(stu)$.
It follows from \reef{zbarztophisec2} and \reef{s4tos22} that a real scalar amplitude compatible with $\mathcal{N}=8$ supersymmetry is
\be
  \text{$\mathcal{N}=8$ SUSY:}~~~~~ A(\phi\phi\phi\phi) = 
  \big(s^2+t^2+u^2\big)^2 f(s,t,u)\,.
\ee
Just as we discussed in Section \ref{sec:maxsusyconst}, the results are straightforwardly generalized to $D>4$ dimensions. 

The quintessential example of a maximally supersymmetric gravitational amplitude is the  closed superstring tree-amplitude
\begin{equation}
   \label{susygravitonAmp}
    A^\text{grav}(\phi\phi\phi\phi)=(s^2+t^2+u^2)^2 A^\text{VS}(s,t,u)\;,
\end{equation}
where\footnote{In the string theory, the VS amplitude is written with $\alpha'\to \alpha'/4$.}
\begin{equation}\label{eq:VS}
    A^\text{VS}(s,t,u)=-\alpha'^2\frac{\Gamma(-\alpha's)\Gamma(-\alpha't)\Gamma(-\alpha'u)}{\Gamma(1+\alpha's)\Gamma(1+\alpha't)\Gamma(1+\alpha'u)}\,
\end{equation}
is the Virasoro-Shapiro (VS) amplitude. (See Appendix \ref{app:SUSY} for the relation to the graviton string amplitude.) The VS amplitude satisfies a stronger Froissart bound,
\begin{align}\label{eq:VSfroissart}
    \lim_{\substack{ |s| \to \infty \\ \text{fixed } u<0 }}
    s^2A^\text{VS}(s,t,u)
    = 0
    \, ,
\end{align}
i.e.~$n_0 = -2$. Hence, the real scalar closed string amplitude \reef{susygravitonAmp} satisfies Froissart with $n_0=2$. The Wilson coefficients for the low-energy expansion of \eqref{susygravitonAmp} are given in Table \ref{tab:exWCs}.

The VS amplitude has simple poles at $s=n/\alpha'$ for all integers $n=0,1,2,3,\ldots$ and the highest-spin states at each level lie on a linear Regge trajectory with slope $2$. At the lowest mass-level $n=1$, the VS amplitude has only a massive scalar with coupling $|\lambda_{1,0}|^2 = 1$. However, the factor $\big(s^2+t^2+u^2\big)^2$ implies that the real scalar amplitude \reef{susygravitonAmp} has the spin 0, 2, and 4 states of the massive $\mathcal{N}=8$ supermultiplet exchanged in the $s$-channel at mass-level 1. The  SUSY relations between the spin 0, 2, and 4 couplings can be derived similarly to the relationship \reef{eq:susymgrel}.

Even though Type II closed superstring theory is unitary and consistent only in $D=10$, the VS amplitude \reef{eq:VS} is by itself unitary for $D\leq23$ \cite{Cheung:2024obl}. Specifically, the spin-4 coupling at mass-level $n=4$ is $|\lambda_{4,4}|^2= \tfrac{(23-D)(D-2)D}{18(D+1)(D+3)(D+7)}$. It is zero for $D=23$ and negative for $D>23$. Thus, albeit a bit bizarre, the formal critical dimension for bootstrapping the VS amplitude is $D=23$ rather than $D=10$.

\subsection{Dispersion Relations}
\label{sec:setup}
We now derive the dispersive representations of the Wilson coefficients. It is useful to break the low-energy expansion of the permutation symmetric  amplitude $A(s,t,u)$ into two parts, one containing the massless poles and the other containing polynomials terms:
\begin{equation}
    A_\text{low-E}(s,t,u)
    =
    A_\text{poles}(s,t,u)
    +
    A_\text{local}(s,t,u)
    \, .
\end{equation}
The most general ansatz for the massless poles is 
\begin{equation}
\label{eq:masslesspoles}
    A_\text{poles}(s,t,u)
    =
    -\lambda_{\phi}^2
    \bigg( 
      \frac{1}{s} 
    + \frac{1}{t}
    + \frac{1}{u}\bigg)
    +
    \kappa^2 
    \bigg( 
      \frac{tu}{s} 
    + \frac{su}{t}
    + \frac{st}{u}
    \bigg)
    \, .
\end{equation}
The first set of pole terms in this equation arises from a $\frac{1}{3!}\lambda_{\phi}^2\phi^3$ interaction in the effective Lagrangian, and the second set of poles arise from a graviton exchange with coupling $\kappa^2 = 8\pi G_\text{N}$. In order to derive the most general set of dispersion relations, we consider non-zero $\kappa^2$ throughout this sub-section. However, for the majority of this paper, we study non-gravitational EFTs with $\kappa^2=0$.

Eliminating $t=-s-u$, the local terms are conveniently parameterized as 
\begin{equation}\label{eq:akq}
    A_\text{local}(s,u)=
    \sum_{k=0}^\infty\sum_{q=0}^k 
    a_{k,q} s^{k-q}u^q\,.
\end{equation}
Comparing equation \reef{eq:akq} order-by-order in the Mandelstam expansion 
to the manifestly permutation symmetric low-energy expansion in \reef{eq:gmn}-\reef{g6s}, one finds
\be
   \label{gfromakq}
   g_0 = a_{0,0}\,,~~
   g_2 = \frac{1}{2}a_{2,0}\,,~~
   g_3 = -a_{3,1}\,,~~
   g_4 = \frac{1}{4}a_{4,0}\,,
   ~~\text{etc.}
\ee
along with conditions 
\be
  \label{naivenull}
  \begin{split}
  &
  a_{1, 1}=a_{1, 0} = 0\,, ~~~~
  a_{2, 2}= a_{2, 1} = a_{2, 0}\,,  ~~~~
  a_{3, 3}=a_{3, 0} = 0\,, ~~~~
  a_{3, 2} = a_{3, 1}\,, ~
  \\
  &
  a_{4, 4} 
  = \tfrac{1}{2}a_{4, 3}
  = \tfrac{1}{3}a_{4, 2}
  = \tfrac{1}{2}a_{4, 1}
  = a_{4, 0}\,,
  ~~\text{etc.}
  \end{split}
\ee
that fix the redundancies in the representation \reef{eq:akq} of the polynomials terms. 

\begin{figure}[t]
	\centering
    \begin{subfigure}[t]{0.45\textwidth}
     \centering
	\raisebox{2mm}{\scalebox{0.73}{
  \begin{tikzpicture}
  [
    decoration={%
      markings,
      mark=at position 0.5 with {\arrow[line width=1pt]{>}},
    }
  ]
  \draw [help lines,->] (-4,0) -- (4,0) coordinate (xaxis);
  \draw [help lines,->] (0,-3) -- (0,3) coordinate (yaxis);
  \node [right] at (xaxis) {Re};
  \node [left] at (yaxis) {Im};
  \path [draw, line width=0.8pt, postaction=decorate,blue] (0.2,0) arc (0:180:.2);
  \path [draw, line width=0.8pt, postaction=decorate,blue] (-0.2,0) arc (-180:0:.2);

  \node at (4/3,0) {$|$};
  \node at (-0.83,0) {$|$};

  \node[red] at (0,0) {$\times$};
  \node[red] at (0.5,0) {$\times$};
  \node[below=0.25cm] at (0.4,0) {$-u$};

  \draw[line width=0.8pt, red] (4/3,0) [decorate, decoration=zigzag] --(4,0);
  \draw[line width=0.8pt, red] (-0.83,0) [decorate, decoration=zigzag] --(-4,0);

  \node[above=0.25cm] at (4/3,0) {$M_{\gap}^2$};
  \node[above=0.25cm] at (-1.1,0) {$-M_{\gap}^2-u$};

 \node[draw] at (4,3) {$s'$};
 \node at (3.8,2.5) {fixed $u<0$};
\end{tikzpicture}
}}
    \end{subfigure}\hspace{0mm}\begin{subfigure}[t]{0.45\textwidth}
\centering
\scalebox{0.73}{
\begin{tikzpicture}
  [
    decoration={%
      markings,
      mark=at position 0.5 with {\arrow[line width=1pt]{>}},
    }
  ]
  \draw [help lines,->] (-4,0) -- (4,0) coordinate (xaxis);
  \draw [help lines,->] (0,-3) -- (0,3) coordinate (yaxis);
  \node [right] at (xaxis) {Re};
  \node [left] at (yaxis) {Im};

  \node at (4/3,0) {$|$};
  \node at (-0.83,0) {$|$};

  \node[red] at (0,0) {$\times$};
  \node[red] at (1/2,0) {$\times$};

  \draw[line width=0.8pt, red] (4/3,0) [decorate, decoration=zigzag] --(4,0);
  \draw[line width=0.8pt, red] (-0.83,0) [decorate, decoration=zigzag] --(-4,0);
  
  \path [draw, line width=0.8pt, postaction=decorate,blue] (0.3,0) arc (180:0:.2);
  \path [draw, line width=0.8pt, postaction=decorate,blue] (0.7,0) arc (0:-180:.2);
  
  \node[above=0.25cm] at (4/3,0) {$M_{\gap}^2$};
  \node[above=0.25cm] at (-1.1,0) {$-M_{\gap}^2-u$};
  \node[below=0.25cm] at (0.4,0) {$-u$};
  
  \path [draw, line width=0.8pt, postaction=decorate,blue,bend angle=90] (4,0.2) to[bend right] (-4,0.2);
 \path [draw, line width=0.8pt, postaction=decorate,blue,bend angle=90] (-4,-0.2) to[bend right] (4,-0.2);
  
  \path [draw, line width=0.8pt, postaction=decorate,blue] (4,-0.2) -- (4/3,-0.2);
  \path [draw, line width=0.8pt, postaction=decorate,blue] (4/3,-0.2) arc (270:90:0.2) -- (4,0.2);
  \path [draw, line width=0.8pt, postaction=decorate,blue] (-0.83,-0.2) -- (-4,-0.2);
  \path [draw, line width=0.8pt, postaction=decorate,blue] (-4,0.2) -- (-0.83,0.2 ) arc (270:90:-0.2);
  
  \node[draw] at (4,3) {$s'$};
 \node at (3.8,2.5) {fixed $u<0$};
\end{tikzpicture}
}
   \end{subfigure} 
    \caption{Contour deformation used to derive dispersion relations for Wilson coefficients. 
    }
    \label{fig:contour}
\end{figure}
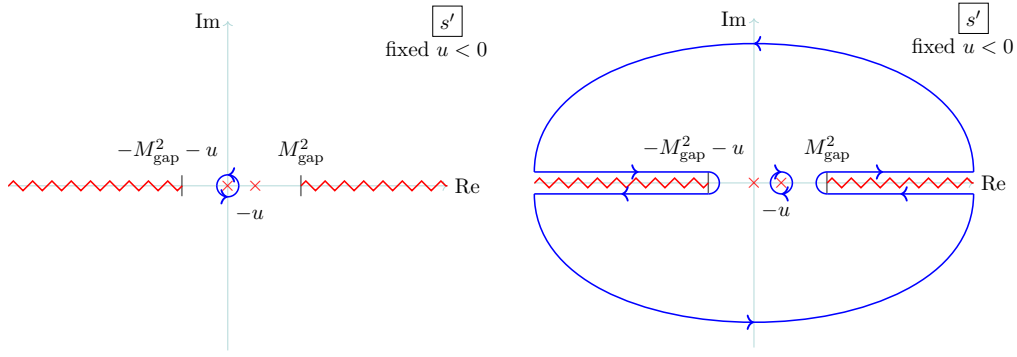

The starting point of the dispersion relations is the contour integral
\be
  \label{contour0}
   \oint_{\mcC_{s=0}}\frac{\d s'}{2\pi i}\frac{A(s',u)}{s'^{k-q+1}}\,,
\ee
where $\mcC_{s=0}$ is a small contour encircling the origin $s=0$ of the complex $s$-plane. 
When deforming this contour as shown in Figure \ref{fig:contour},  contributions from  $|s| \to \infty$ vanish as long as we take  
\be  
  \label{kqn0}
   k-q \ge n_0\,,
\ee
where $n_0$ is the integer from the assumed Froissart-Martin-like bound \reef{eq:bound}. 
In the low-energy regime, the integral \reef{contour0} picks up contributions from the massless poles \reef{eq:masslesspoles} as well as from the local terms \reef{eq:akq}. From the latter, we isolate $a_{k,q}$
by taking $q$ derivatives with respect to $u$ and then setting $u=0$. The contour deformation illustrated in Figure \ref{fig:contour} then gives, for $k-q\ge n_0$,
\be
   \label{cdeform0}
   \begin{split}
   a_{k,q} 
   &= \frac{1}{q!}\frac{\partial^q}{\partial  u^q}
   \oint_{\mcC_{s=0}}\frac{\d s'}{2\pi i}\frac{A_\text{local}(s',u)}{s'^{k-q+1}} \Bigg|_{u=0}\\
   &=
   \frac{1}{q!}\frac{\partial^q}{\partial  u^q}
   \Bigg(
   -\oint_{\mcC_{s=0}}\frac{\d s'}{2\pi i}\frac{A_\text{poles}(s',-s'-u,u)}{s'^{k-q+1}}
   +\oint_{\mcC_{s=-u}}\frac{\d s'}{2\pi i}\frac{A_\text{poles}(s',-s'-u,u)}{s'^{k-q+1}}~\\
   &~\hspace{2cm}
     +\text{$s$-channel cuts}
     +\text{$u$-channel cuts}
   \Bigg)\Bigg|_{u=0}\,.
   \end{split}
\ee
The contour integral around $s=-u$ picks up contributions from the massless $t$-pole. After cancellations between the $s=0$ and $s=-u$ contour integrals in the second line of \reef{cdeform0}, they combine to give
\be
  \label{stpolecontrib}
  -\lambda_{\phi}^2 
  \bigg(
    \delta_{k-q,-1} 
    + \frac{1}{u} \delta_{k-q,0} 
  \bigg)
  - \kappa^2
  \bigg(
    u^2 \delta_{k-q,-1} 
    + u \delta_{k-q,0} 
    + \delta_{k-q,1} 
    + \frac{1}{u}\delta_{k-q,2} 
  \bigg) \,.
\ee
Whether these terms contribute depends on the range of values of $q$ for given $k$ allowed by \reef{kqn0}. 
When present, the $1/u$ poles obstruct the limit 
$u \to 0$. This can be handled by including averages as done in \cite{Albert:2024yap}, but here we avoid the issue as follows:
\begin{enumerate}
 \item For $n_0 =2$, we decouple gravity completely by  setting $\kappa = 0$. 
 \item For $n_0 = 0$, we assume  $\lambda_{\phi} = 0$ in addition to $\kappa = 0$. This is justified for a non-gravitational amplitude where a symmetry, for example a basic $\phi \to -\phi$ symmetry, disallows the $\phi^3$-interaction. 
This, for example, is the case the DBI amplitude.  
\end{enumerate}
For the gravitational case with $\kappa \neq 0$, we take a slightly different approach (see  Section \ref{sec:gravamps}).

Under these assumptions, equation \reef{cdeform0} reduces to an expression for $a_{k,q}$ in terms of the cuts. For brevity, the cuts in \reef{cdeform0} stand for both poles and discontinuities on the real $s$ axis for both $s \ge M_\text{gap}^2$ and $s \le -u - M_\text{gap}^2$ respectively. Using permutation symmetry, the $s$- and $u$-channel contributions combine into one dispersive integral, so for $k-q \ge n_0$ we get
\be
   a_{k,q}
    = 
\frac{1}{q!}\frac{\partial^q}{\partial  u^q}
    \int\limits_{M^2_{\text{gap}}}^{\infty}
    \frac{ \mathrm{d}s' }{\pi}
    \bigg[
    \frac{1}{ (s')^{k-q+1} }
    +
    \frac{ (-1)^{k-q} }{ (s'+u)^{k-q+1} }
    \bigg]
    \Im A(s', -s'-u,u)
     \Bigg|_{u=0}
    \, .
\ee
Using the partial wave expansion \reef{eq:partial} and performing the derivatives, we find 
\begin{equation}\label{eq:akqsdr}
\boxed{
a_{k,q}=\sum_{j=0}^\infty\int_{1}^\infty\frac{\d y}{y^{k+1}
    }\;\rho_{j}(yM_{\gap}^2)\, w_{j;\,k,q}^{(D)}\;,\quad k-q\geq n_0\;,}
\end{equation}
where the sum over $j$ includes only even spins, we have changed integration variable to the dimensionless $y=s'/M_{\gap}^2$, and defined 
$\rho_j(s) = \frac{1}{\pi} n^{(D)}_j s^{(D-4)/2} \Im a_j(s)$. 
The constants $w_{j;\,k,q}^{(D)}$ are defined as 
\begin{equation}\label{eq:wkq}
    w_{j;\,k,q}^{(D)}=\sum_{n = 0}^{ q }
    \binom{k-n}{q-n}
    \big[ \delta_{n,q} + (-)^{k-n} \big]
    v_{j,n}^{(D)}\,
\end{equation}
in terms of the $v^{(D)}_{j,n}$ which are the  coefficients in the expansion of Gegenbauer polynomials; explicit formulas are given in Appendix \ref{app:gegenbauer}. To keep our expressions compact, it is useful to introduce the notation
\begin{align}
    \< Q \>_\mu \equiv 
    \sum_{j=0}^\infty
    \int_\mu^\infty \frac{dy}{y}\rho_{j}(y M_{\gap}^2)\,Q \,.
\end{align}
Then the dispersion relation \reef{eq:akqsdr} is simply 
$a_{k,q}=\< y^{-k} \, w_{j;\,k,q} \>_1$.

The dispersive representation \reef{eq:akqsdr}
links the low-energy Wilson coefficients $a_{k,q}$ to the high-energy spectrum encoded in the spectral density $\rho_j(s)$. By unitarity, \reef{unitaritypos}, it satisfies the positivity condition  $\rho_j(s) \ge 0$, which is essential for the bootstrap. As noted in the Introduction, we refer to the $n_0=0$ and $n_0=2$ cases as the ``0SDR'' and ``2SDR'' bootstrap, respectively. In the 0SDR bootstrap, all coefficients $a_{k,q}$ with $k \ge q$ have a dispersive representation \reef{eq:akqsdr}, but in the 2SDR bootstrap the dispersive representation is only valid for the $a_{k,q}$ with $k -q\ge 2$. 
Thus, not only does the 0SDR bootstrap bound more Wilson coefficents, its bounds on the $a_{k,q}$ with $k -q\ge 2$ are also stronger than in the  2SDR bootstrap because it has access to more null constraints from imposing the $stu$-symmetry. Let us now discuss the null constraints.

\subsection{Null Constraints}
\label{sec:null}

It was convenient to derive the  dispersive representation \reef{eq:akqsdr} for the Wilson coefficients $a_{k,q}$, which ares related to the $g_k$ coefficients of the manifestly permutation symmetric form of the amplitude via \reef{gfromakq}.  Importantly, the redundancy of the  $a_{k,q}$ representation of the low-energy amplitude leads to relations among the $a_{k,q}$'s, such as those listed for the lowest orders in \reef{naivenull}. Writing these relations in terms of the dispersion relations for the $a_{k,q}$'s imposes constraints on the spectral density $\rho_j(s)$; those are the {\em null constraints}. 
However, some of these relations are rendered trivial due to the particular values of the $w^{(D)}_{j;\,k,q}$ in \reef{eq:wkq}. For example, 
because
\begin{equation}
    w^{(D)}_{ j;\,4,1} = 2w^{(D)}_{ j;\,4,0}  = 4\, ,
\end{equation}
the relation $a_{4,1} = 2 a_{4,0}$ from \reef{naivenull} is automatically true and thus does not constrain $\rho_j(s)$. In fact, in the 2SDR bootstrap, where we can only access Wilson coefficients with $k-q \ge 2$, the only non-trivial relation remaining from \reef{naivenull} is 
$a_{4, 2}= 3a_{4, 0}$. This then becomes the lowest-order 2SDR null constraint, 
\be
   0 =  3a_{4,0} - a_{4,2} = 
   \Big\langle y^{-4}\big(3w^{(D)}_{j;\,4,0}-w^{(D)}_{j;\,4,2}\big) \Big\rangle_1 \,.
\ee
In the 0SDR bootstrap, we can access all $a_{k,q}$ coefficients, but some relations in \reef{naivenull} are trivial. For example, the relations $a_{1,0} = 0$ and $a_{3,1} = a_{3,2}$  from \reef{naivenull} do not impose any constraints on the spectral density because $w^{(D)}_{j;\,1,0} = 0$ and 
$w^{(D)}_{j;\,3,1} = w^{(D)}_{j;\,3,2}$. Appendix \ref{app:indnulls} presents a general method for deriving the set of independent null constraints. 

\subsection{Numerical Implementation}
\label{sec:num}
Following the procedure outlined in \cite{Caron-Huot:2020cmc,Albert:2022oes,Berman:2023jys}, we use the semidefinite optimization package SDPB \cite{Simmons-Duffin:2015qma} to obtain numerical bounds on Wilson coefficients. The numerical implementation requires a truncation in the Mandelstam order $k$ of the infinite set of null constraints and additionally a truncation in the infinite sum over even spins that enter the sum in \eqref{eq:akqsdr}. 

The former truncation can be dealt with simply:  one imposes all non-trivial null constraints with $k-q \ge n_0$ up to some maximum Mandelstam order $k_\text{max}$. As we increase $k_{\max}$, the problem becomes strictly more constrained such that if a choice of Wilson coefficients is ruled out by our bounds for any finite choice of $k_{\max}$, those coefficients could not come from a UV amplitude which satisfies our assumptions. In general we find that the bounds converge with $k_{\max}$ in the sense that the change in the allowed region between orders in $k_{\max}$ decreases as $k_{\max}$ increases. However, the computational time also increases rapidly with $k_{\max}$, so finding bounds including all null constraints up to values of $k_{\max} \gtrsim 20$ is computationally intensive. Often we settle for  $k_\text{max}$ lower than 20 when the bounds do not change significantly by going to higher $k_\text{max}$. 

Truncating the sum over spins, on the other hand, must be dealt with more carefully. Including more spins actually increases the allowed region, so the numerical bounds are only  rigorous when all spins are included. However, that is not computationally possible, so we must make a choice of some maximal spin 
such that the bounds are well converged. 
We find in most optimization problems that we obtain well-converged bounds by using a list of spins that includes all even spins up to some $\mcO(100)$ number and then also has a very large spin $J_{\infty} = 10^{a}$ for some positive integer $a$ chosen for the particular optimization problem. Generally, we increase the spin list until the resulting bounds do not change beyond the numerical precision needed. 

\section{Maximal Spin Constraint}\label{sec:maxspin}
Consider a spectrum 
\be
\label{our1statespec}
\raisebox{-8.5mm}{\includegraphics[width=6cm]{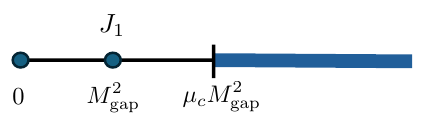}}
~~~\hspace{5mm}
\ee
with a spin-$J_1$ state of mass-squared $M^2  = M_\text{gap}^2$ exchanged at tree level
\be
\label{spinJexch}
\raisebox{-1cm}{\begin{tikzpicture}
	\node at (0.95,1/2) {$\lambda_{ J}^*$};
	\node at (-0.95,1/2) {$\lambda_{ J}^{\phantom{*}}$};
	\node at (0,3/8) {$ J$};
	\node at (0,-1/2) {$M^2$};
	
	\node (image) at (0,0) {\includegraphics[width=4cm]{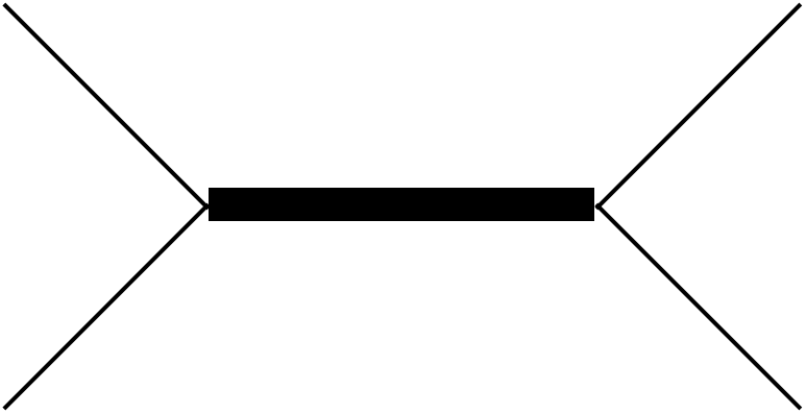}};
\end{tikzpicture}}
\ee
in the massless four-scalar amplitude. 
We assume that there are no other states with masses below the cutoff $\mu_c M_\text{gap}^2$ for some $\mu_c > 1$, but are agnostic about the spectrum above the cutoff, as indicated by the solid blue bar.

In the case of $s\leftrightarrow u$ crossing symmetry (e.g.~for scalars in the adjoint representation of a large-$N$ gauge group), it was first noted numerically in \cite{Berman:2024wyt} and later analytically proven in~\cite{Berman:2024kdh} that the dispersive representations of the Wilson coefficients imply an upper
bound on the spin $J_1$. 
For the fully $stu$-symmetric case, similar numerical bounds were found in \cite{Albert:2024yap} for the case of scalars in $\mathcal{N}=8$ supergravity. In this section we give numerical evidence for a maximal spin constraint in the general $stu$-symmetric case, leaving an analytic derivation for future work \cite{NickJustin}. We consider the 2SDR and 0SDR cases in turn. 

\begin{figure}
\centering
\includegraphics[width=0.48\linewidth]{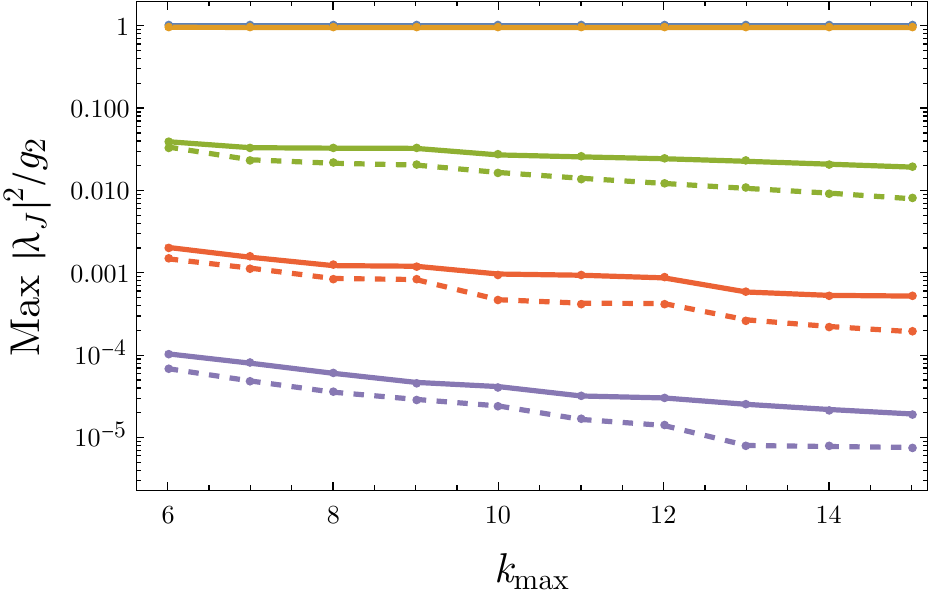}
~
{\includegraphics[width=0.48\linewidth]{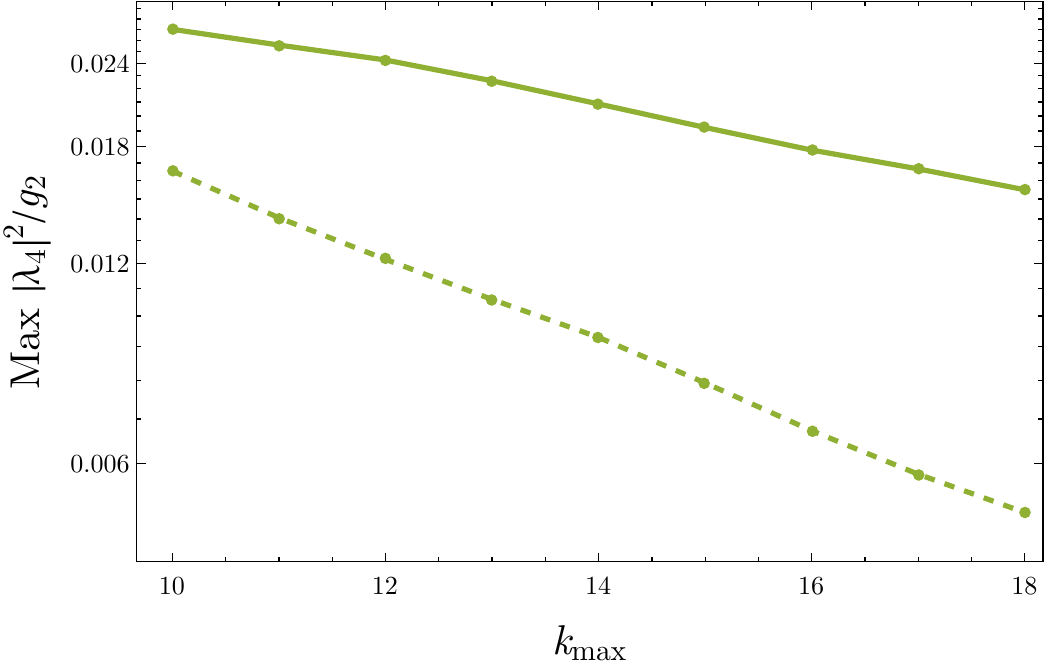}}
\caption{\label{fig:spinsuppression} Maximum normalized coupling $|\lambda_J|^2 / g_2$ for spin-$J$ states with mass-squared $M_\text{gap}^2$. The dashed lines are computed in the setup \reef{our1statespec} with $J=J_1$
and $\mu_c = 1.1$. The solid lines are computed in same setup while also allowing all lower even spins $J=0,2,\ldots,J_1$.
On the left, we show the results for $\max(|\lambda_{J_1}|^2 / g_2)$ as a function of the truncation level $k_\text{max}$ for  $J = 0$ (blue), $2$ (orange), $4$ (green), $6$ (red), and $8$ (purple). On the right, we select the $J=4$ case to better illustrate the exponential suppression with increasing $k_\text{max}$. We conclude that only spin-$0$ and spin-$2$ states can be exchanged at tree-level at the lowest mass state in the spectrum. 
}
\end{figure}

\vspace{2mm}
\noindent {\bf 2SDR bootstrap.} The key idea is to bound the couplings $|\lambda_{J_1}|^2$ of the massive state to the massless external states. Since the positivity bounds are projective, the couplings are bounded relative to the lowest accessible Wilson coefficient, which in the 2SDR bootstrap is $g_2$. 
Assuming that the spin\nobreakdash-$J_1$ state is the only state at the mass gap, as in \reef{our1statespec},
we compute the maximum of  $|\lambda_{J_1}|^2/g_2$ 
for fixed $\mu_c=1.1$ for $J_1 = 0,2,4,\ldots$. 
As shown by the dashed lines in Figure~\ref{fig:spinsuppression}, we find for $J_1=4,6,8$ that max($|\lambda_{ J_1,q}|^2/g_2$) becomes exponentially suppressed with increasing $k_\text{max}$. This suppression indicates that in the $k_\text{max}\to\infty$ limit, spins greater than 2 are not allowed for the lowest mass states of the spectrum. In contrast, the spin-$0$ and spin-$2$ couplings are not suppressed with increasing $k_\text{max}$. 

Instead of a single state at $M_\text{gap}$, suppose we allow all even spins up to spin $J_1$. Then, when we bound the coupling, $|\lambda_{ J_1}|^2/g_2$, of the highest spin state,
we find that its maximum value $\max(|\lambda_{ J_1}|^2/g_2)$ is slightly larger at finite $k_\text{max}$ than it was in the single-state case. This is shown as the solid lines in Figure \ref{fig:spinsuppression}.  However, we still see exponential suppression in $k_\text{max}$ for spins $J_1 \ge 4$.

We interpret these results to mean that 
only spin-$0$ and spin-$2$ states are allowed at the mass gap. 
The result that spins $J_1>2$ are disallowed at the mass gap is independent of the particular choice of the cutoff, so long as $\mu_c > 1$.  
(The Infinite Spin Tower \reef{eq:basicIST} is allowed when $\mu_c=1$.) For larger values of $\mu_c$, the higher-spin couplings are even more suppressed, but the spin-$0$ and spin-$2$ couplings are still singled out by not exhibiting exponential suppression with increasing $k_\text{max}$.

Repeating the analysis at the second mass-level, i.e.
\be
\label{our2statespec}
\raisebox{-8.5mm}{\includegraphics[width=6cm]{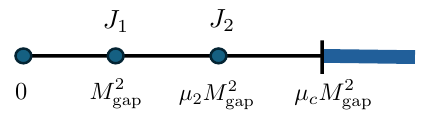}}
\ee
for some choices of $\mu_c>\mu_2>1$, we find that the only non-trivial options for the spins are
\be
 \label{spin12choice}
  \text{2SDR bootstrap:}~~~
  J_1 =0, 2 
 ~~~~\text{and}~~~~
  J_2 =0,2,4
 \,,
\ee
with $ J_2=4$ only allowed if $ J_1=2$ appears at the mass gap.

\vspace{2mm}
\noindent {\bf 0SDR bootstrap.} 
An equivalent analysis for the 0SDR bootstrap of $stu$-symmetric theories, leads to the conclusion that the only allowed spins are
\be
 \label{spin12choice0SDR}
  \text{0SDR bootstrap:}~~~
  J_1 =0 
 ~~~~\text{and}~~~~
  J_2 =0,2
 \,.
\ee

\vspace{2mm}
These results suggest that the leading Regge trajectory  has sequentially increasing spins, similar to the result proved in~\cite{Berman:2024kdh}. For our purposes, this constraint is practical because it greatly reduces the possible freedom in the spectrum of relevant UV-complete tree-level amplitudes.

\section{Extremal Theories of the 2SDR Bootstrap} \label{sec:2sdrextr}

The space of allowed Wilson coefficients is projective and convex. 
This fact implies that any allowed theory can in principle be written as a positive linear combination of the set of theories that live on the boundary of the full convex region. Keeping in mind that the allowed region is, in the $k_\text{max} \to \infty$ limit, an infinite-dimensional bounded space, this may not seem to be a useful characterization of amplitudes. However, in this section, we conjecture that there is a one-parameter family of extremal theories which, together with the pure scalar amplitude, completely determine the full space of allowed Wilson coefficients in the 2SDR bootstrap. We begin by characterizing the 2SDR extremal theories. Then we state and test the conjecture.

\subsection{2SDR Extremal Theories}
\label{sec:extr2sdr}

\begin{figure}
    \centering
    \begin{subfigure}[b]{0.48\textwidth}
    \centering
    \includegraphics[height=5cm]{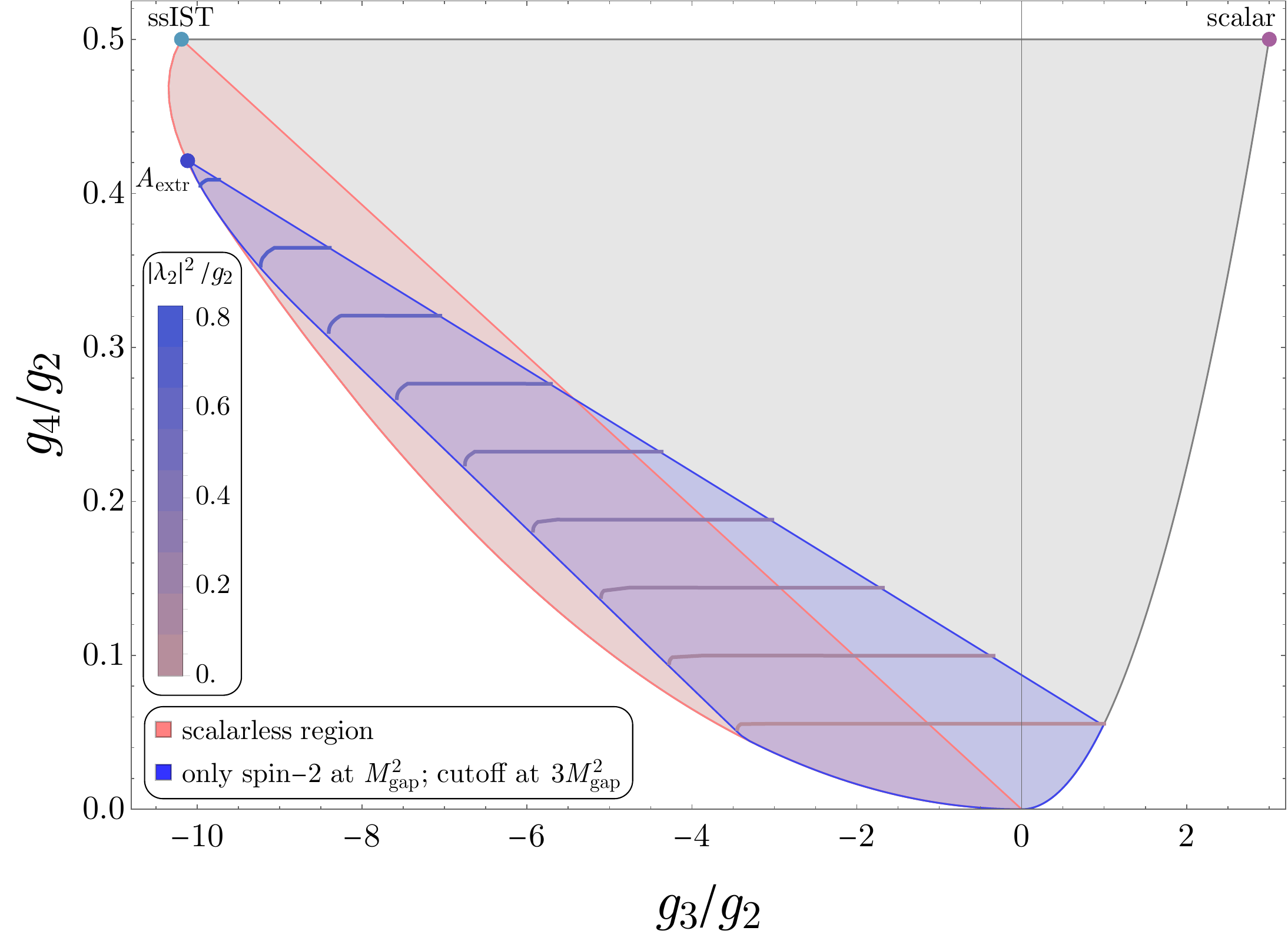}
    \caption{{}}
    \end{subfigure}
    \begin{subfigure}[b]{0.48\textwidth}
    \centering
    \begin{tikzpicture}
    \node (image) at (0,0) {\includegraphics[height=5cm]{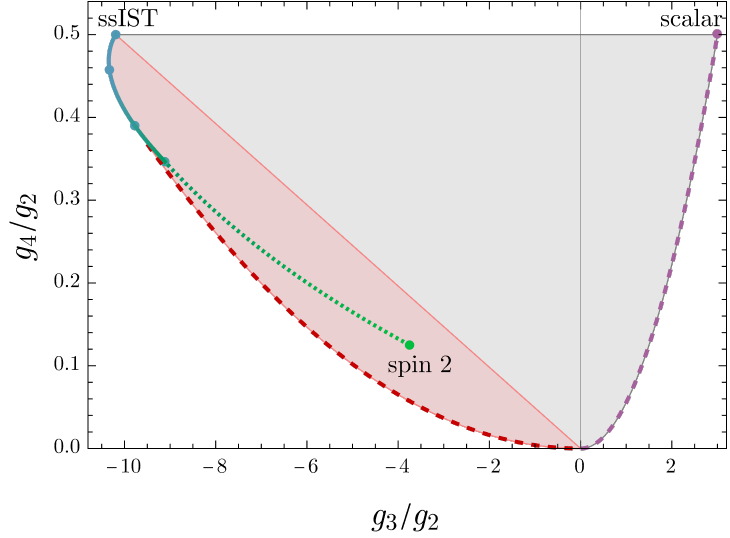}};
    \draw[->] (-0.65,1) -- (-1.8,1);
    \draw[->] (-0.9,1.35) -- (-2.05,1.35);
    \draw[->] (-1.15,1.85) -- (-2.3,1.85);
    \node at (-0.15,1) {\footnotesize{$\mu_c = 6$}};
    \node at (-0.4,1.35) {\footnotesize{$\mu_c = 4$}};
    \node at (-0.65,1.85) {\footnotesize{$\mu_c = 2$}};
    \end{tikzpicture}
    \caption{{}}
    \end{subfigure}
    \caption{The $D=4$ allowed region of the $(g_3/g_2,g_4/g_2)$-plane. The scalarless region is shown in pink. The purple dot at the top-right is the scalar amplitude \reef{eq:A0}. The light-blue dot at the top-left is the scalar-subtracted infinite spin tower \reef{eq:ssIST}.
    {\bf (a)}
    The blue region is the allowed region assuming the spectrum \reef{onlyspin2atgap} with cutoff $\mu_c = 3$. Within this region, ``contours'' of constant $|\lambda_2|^2/g_2=0,0.1,\ldots,0.8$ are shown: these curves indicate the maximal $g_4/g_2$ allowed for each choice of the coupling. 
    At the sharp corner, marked with the dark-blue dot, we find the 
     $\mu_c=3$ extremal theory $A_\text{extr}$ for which $|\lambda_2|^2/g_2\approx0.828$.
     {\bf (b)} The extremal theories characterized by the spectrum \reef{onlyspin2atgap} with the spin-2 coupling $|\lambda_2|^2/g_2$ maximized for given cutoff $\mu_c$ shown as the blue-to-green curve for $ 1 \le \mu_c \le 5$, starting  at $\mu_c=1$ with the ssIST model (light-blue dot) and asymptoting to the pure spin-2 theory (green dot)  as $\mu_c \to \infty$. The dotted green curve is obtained as an interpolation function based on the numerical points for $1 \le \mu_c \le 5$ and the spin 2 theory as the end point. The red dashed curve is a mass-scaling curve of the extremal theory with $\mu_c\approx 4$ and  
    the purple dashed curve on the right is the mass-scaling curve of the pure scalar theory.
    }
    \label{fig:extremal}
\end{figure}
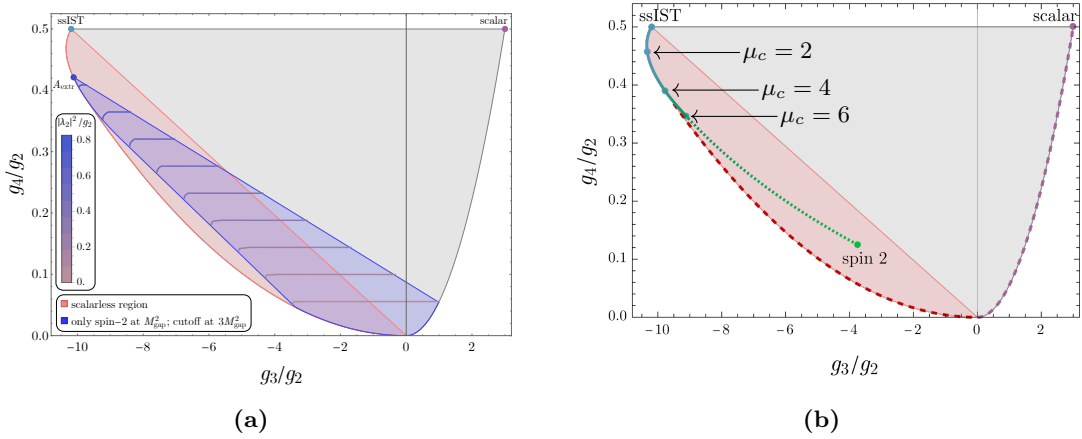

In the 2SDR bootstrap, we compute bounds on the ratios of Wilson coefficients $g_k/g_2$. For any finite $k_\text{max}$, this procedure determines a convex region of allowed ratios $g_k/g_2$ embedded within the full space of Wilson coefficients $g_k/g_2$ with $k\le k_\text{max}$; e.g.~for $k_\text{max}=12$, the full space is 17-dimensional. To visualize it, we project the higher-dimensional space into $2$-dimensional planes, such as the $(g_3/g_2,g_4/g_2)$-plane shown in Figure~\ref{fig:basicg3g4} for $D=4$. However, it is important to keep in mind that any such plot is only a projection of a much more complex higher-dimensional space. 

Figure \ref{fig:extremal}(a)  shows the $(g_3/g_2,g_4/g_2)$ region, now with the {\em scalarless region} shown in pink; this is the region where no spin-0 contributions are needed in the dispersion relations. The scalarless region shares the entire lower bound on $g_3/g_2$ with the general allowed region, so we anticipate that the extremal theories that determine the bound also have no scalars. 

The spectrum analysis in Section \ref{sec:maxspin} showed that the lowest tree-exchanged massive state can only have spin $0$ or spin $2$. 
Hence, in the absence of scalars,  we consider a spectrum with a spin-$2$ state at $M_\text{gap}^2$ and a gap  up to the next possible state at $\mu_c M_\text{gap}^2$:
\be
\label{onlyspin2atgap}
\raisebox{-8.5mm}{\includegraphics[width=6cm]{figures/onlyspin2atgap.pdf}}
\ee
We are agnostic about any states at or above $\mu_c M_\text{gap}^2$. 
For $D=4$ and $\mu_c =3$, the resulting allowed region is shown as the blue region in Figure \ref{fig:extremal}(a). There is a sharp corner at its maximum $g_4/g_2$ value.  
As $\mu_c$ takes values in the range $1<\mu_c \lessapprox 4$, the region resulting from the spectrum assumption \reef{onlyspin2atgap} similarly has a sharp corner that tracks part of the $g_3/g_2$ boundary of the allowed region. This was shown as the blue points in Figure \ref{fig:basicg3g4}.

Within the $\mu_c=3$ blue region of Figure \ref{fig:extremal}(a), the values of the spin-2 coupling $|\lambda_2|^2/g_2$ varies, as shown by the ``contour-like'' curves which indicate the upper bound on $g_4/g_2$ for given value of $|\lambda_2|^2/g_2$. When $|\lambda_2|^2/g_2 = 0$, there is no state at the mass gap, and all possible spectra are allowed at and above $\mu_c =3$. As a result, the $|\lambda_2|^2/g_2 = 0$ region is the same shape as the general region but scaled down by a factor of $\mu_c=3$ in the $g_3/g_2$-direction and by $\mu_c^2=9$ in the $g_4/g_2$-direction; this is the blue region below the $|\lambda_2|^2/g_2 = 0$ contour. 
As $|\lambda_2|^2/g_2$ increases, the  $|\lambda_2|^2/g_2$ ``contours'' move towards the sharp corner where $|\lambda_2|^2/g_2$ takes its maximal value. This motivates the characterization: 
\begin{quote}
{\em The extremal theories are the theories with spectrum \reef{onlyspin2atgap} that for given $\mu_c$ has the maximum possible coupling $|\lambda_2|^2/g_2$.} 
\end{quote}
For given $\mu_c$, fixing $|\lambda_2|^2/g_2$ to the maximum allowed value indeed determines a unique amplitude, in the sense that the computed values of the maximum and minimum of any ratio $g_k/g_2$ give the same result within the numerical precision to which $|\lambda_2|^2/g_2$ is fixed. 
This is the one-parameter family of ``extremal theories''. Parameterized by $1 \le \mu_c \le  \infty$, the extremal theories interpolate between the ssIST \reef{eq:ssIST}, reached as $\mu_c \to 1$, and is expected to approach the pure spin-2 model  \reef{eq:AL} asymptotically in the limit $\mu_c \to \infty$.\footnote{It may be possible to take the $\mu_c \to \infty$ limit in different ways, depending on what is scaled or held fixed in the spectrum above $\mu_c$, but, based on the locations of the extremal theories in the higher dimensional geometry, the extremal theories appear to have the pure spin 2 amplitude as their limit point.}

The range of maximal values of $|\lambda_2|^2/g_2$ covered as $\mu_c$ varies between 1 and $\infty$ depends on the spacetime dimension $D$, for example
\be
  \label{extrspin2range}
  \text{$D=4$:}~~~
  \frac{1}{4} \le \frac{|\lambda_2|^2}{g_2}
  \le 0.959789\,,
  \hspace{1cm}
    \text{$D=10$:}~~~
  \frac{4}{13} \le \frac{|\lambda_2|^2}{g_2}
  \le 0.927552\,.
\ee
The lower bound $|\lambda_2|^2/g_2 = (D-2)/(3 D-4)$ is given by the pure spin-2 theory and the upper bound determined by the spin 2 coupling of the ssIST model.

We show the extremal theories in Figure \ref{fig:extremal}(b) as the blue-to-green points. For $1 \le \mu_c \le 5$, the points are obtained numerically by the spin-2 coupling maximization method and the dotted green curve is an extrapolated of these points to the spin 2 model. 
For $1 \le \mu_c \lesssim 4$, the difference between the value of $g_3/g_2$ for the extremal theories computed with coupling-maximization and the general region bounds with the same value of $g_4/g_2$ is less than 0.01\% and it decreases with $k_\text{max}$.

For values 
$\mu_c \gtrsim 4$, the extremal points lie in the interior of the $(g_3/g_2,g_4/g_2)$ region shown in Figure \ref{fig:extremal}(b), as necessary in order for them to asymptote the pure spin 2 model. Thus, it may appear strange to call the $\mu_c \gtrsim 4$ theories ``extremal'' since they appear to lie in the interior of the allowed space. First of all, they are extremal in the sense of maximizing the spin-2 coupling. Secondly, the fact that they lie in the interior of the allowed region of the  $(g_3/g_2, g_4/g_2)$-plane is an artifact of this particular projection of the higher-dimensional region. As we discuss in Section \ref{sec:0sdrextr}, in other projections the full range of extremal theories track the outer boundary of the scalarless region.

The part of the boundary that is not covered by the extremal theories with $1 \le \mu_c \lesssim 4$ is shown as the dashed red curve in Figure \ref{fig:extremal}(b). This is simply the mass-scaling curve\footnote{It was noted in \cite{Chiang:2021ziz} that this segment of the universal bounds is indeed just a parabola, as expected for a mass-scaling curve, and a closed-form expression for it was found based on EFT-hedron type bounds for $k_\text{max}=5$. Those values agree well with the SDPB bounds for $k_\text{max}=5$, but are significantly different for higher $k_\text{max}$.} of the extremal theory with  $\mu_c \approx 4$.\footnote{As far as we can tell, there is nothing special about the extremal theory with $\mu_c \approx 4$; this just happens to be the value for the $(g_3/g_2,g_4/g_2)$ projection. We have observed similar behavior in other projections where this family of extremal theories tracks the outer boundary of the region as a function of $\mu_c$ until some point where the boundary turns into a mass-scaling curve, but there is no consistent value for that cutoff mass.} Let us recall that a mass-scaling curve is obtained by scaling the entire spectrum by $\mu$, 
such that the lowest-mass state is no longer at $M_\text{gap}^2$ but at $\mu M_\text{gap}^2$,
i.e.~
\begin{align}
A(s,t,u) ~\to~A(s/\mu,t/\mu, u/\mu)\,
\end{align}
gives 
\be
\label{gkmasscaled}
\frac{g_{k}}{g_{2}} \to \frac{1}{\mu^{k-2}}\frac{g_{k}}{g_{2}} \, .
\ee
Hence, every point $(g_3/g_2,g_4/g_2,g_5/g_2,\ldots )$ in the allowed space, has a mass-scaling curve
\be
\label{massSCcurve}
\bigg(\frac{1}{\mu}\frac{g_3}{g_2},\frac{1}{\mu^2}\frac{g_4}{g_2},\frac{1}{\mu^3}\frac{g_5}{g_2},\ldots \bigg)
\ee
parameterized by $1 \le \mu < \infty$ that is also in the allowed space. In Figure~\ref{fig:extremal}(b), the $\mu_c \approx 4$ mass-scaling curve turns out to be on the boundary of the allowed space, but in other projections it can be the mass-scaling curve of extremal theories with different  $\mu_c$. Hence, the mass-scaling curves  are also important for understanding the full allowed region.

\subsection{Convex Hull Conjecture}
\label{sec:CHC}

Based on the discussion in the previous section, we can visually convince ourselves that in the $(g_{3}/g_2,g_{4}/g_{2})$-plane, taking positive linear combinations of the extremal theories along with their mass-scaling curves (e.g.~\reef{massSCcurve}) gives the entire scalarless region shown in pink in Figure~\ref{fig:extremal}(a). In other words, the scalarless region in the $(g_{3}/g_2,g_{4}/g_{2})$-plane is the convex hull of the one-parameter family of extremal theories and their mass-scaling curves. Moreover, the full universal region in the $(g_{3}/g_2,g_{4}/g_{2})$-plane (the union of the gray and pink regions in Figure~\ref{fig:extremal}(b)) can clearly be obtained as the convex hull of the scalarless region and the pure scalar mass-scaling curve (purple in Figure \ref{fig:extremal}(b)). 

More generally, recall from Section \ref{sec:scalarsub} that any theory with a scalar in the 2SDR bootstrap can simply be scalar-subtracted to bring it into the scalarless region. Or, stated inversely, the full allowed region can be obtained as the positive linear combination of amplitudes in the scalarless region and the pure scalar amplitudes \reef{eq:A0} with any masses $\ge M_\text{gap}$. Hence, to characterize the full space of theories, we only have to focus on the scalarless region. 

The scalarless region in the $(g_{3}/g_2,g_{4}/g_{2})$-plane is merely a projection of a much more intricate higher-dimensional convex space. To characterize the bounds of the full multi-dimensional (or really, as $k_\text{max} \to \infty$, infinite-dimensional) scalarless region, we have to ask ourselves what could the boundary theories of this higher-dimensional region possibly be? While we cannot a priori exclude amplitudes with branch cuts reaching  down to $M_\text{gap}^2$, we find it sufficient to focus on the case with a tree-level exchange at $M_\text{gap}^2$. In absence of scalars, the maximal spin bound in Section \ref{sec:maxspin} leaves as the only possibility a state with spin 2 at $M_\text{gap}^2$. With a gap $\mu_c$ to the next state, the only property that makes this class of theories special is maximization of the spin-2 coupling $|\lambda_2|^2/g_2$. We know from Section \ref{sec:extr2sdr} that this gives the one-parameter family of extremal theories. This logic motivates the ``Convex Hull Conjecture'':
 
\begin{quote}  {\em In the $k_\text{max} \to \infty$ limit, the scalarless allowed 2SDR region is the convex hull of the  one-parameter family of extremal theories and their mass-scaling curves.}
\end{quote} 
As discussed in the Introduction, it is surprising that a multi-dimensional convex space would be exactly the convex hull of a one-parameter family of curves!

To test the conjecture, we compute the Wilson coefficients 
\be
  \bigg(
\frac{g_3}{g_2},
\frac{g_4}{g_2},
\frac{g_5}{g_2},
\frac{g_6}{g_2},
\frac{g_6'}{g_2},
\frac{g_7}{g_2}
  \bigg)
\ee
for the extremal theories. With these coefficients and their mass-scalings curves \reef{massSCcurve}, we can produce their 6-dimensional convex hull (this of course is just a $6$-dimensional projection of an even higher-dimensional geometry).
Minimizing/maximizing a Wilson coefficient over the convex hull  
can easily be done in Mathematica,
and the resulting values can be compared with the positivity bounds from SDPB. 
If the convex hull truly is the entire region allowed by the bootstrap, then the convex hull bounds and SDPB bounds should agree to better and better precision with increasing $k_\text{max}$.

\begin{figure}
    \centering
    \includegraphics[height=6.5cm]{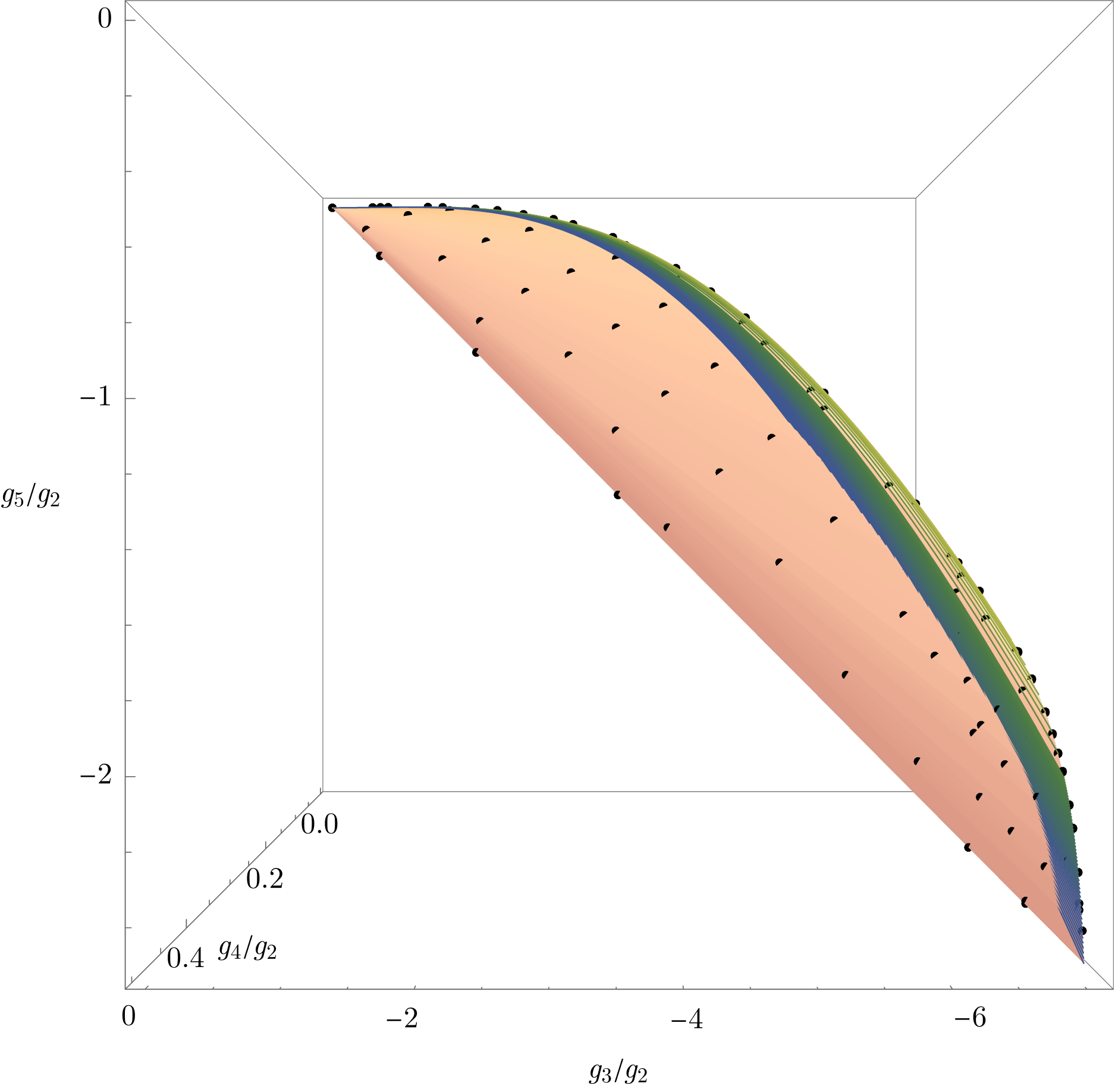}\hfill
    \includegraphics[trim={8cm 0cm 8.2cm 0cm},clip,height=4cm]{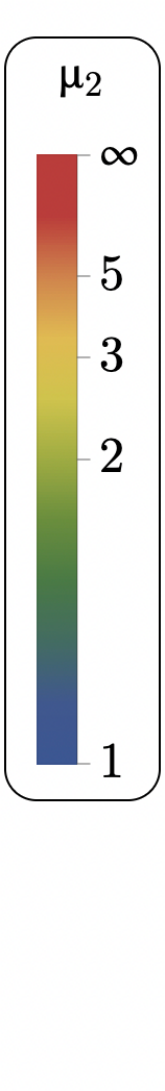}\hfill
    \includegraphics[height=6.5cm]{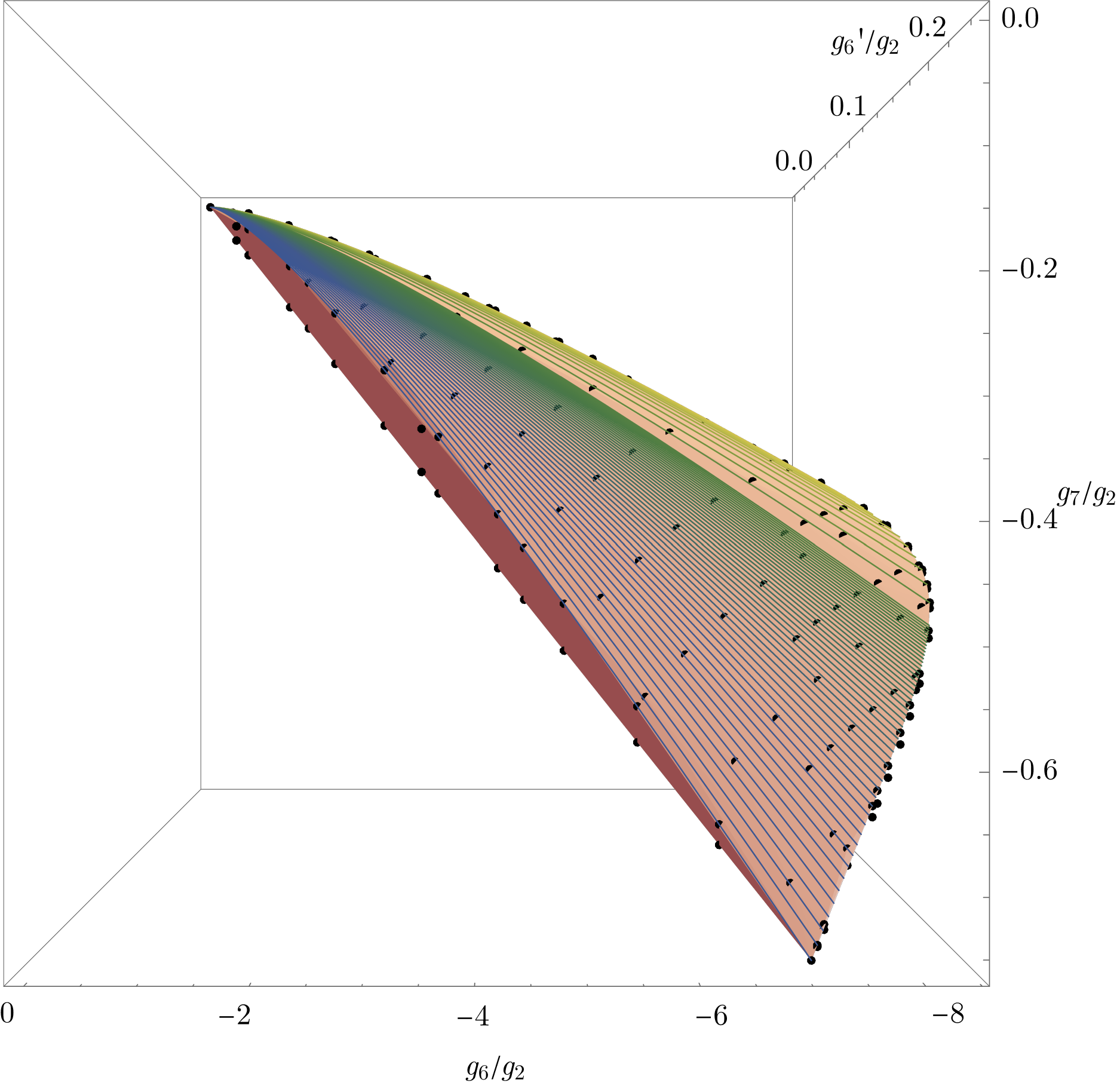}
    \caption{Three-dimensional projections of the convex hull computed from the one-parameter family of extremal theories (determined here at $k_\text{max}=12$) and their mass-scaling curves. Increasing values of $\mu_c$ for the extremal theories and their mass-scaling curves are color-coded from blue for $\mu_c=1$ to red for $\mu_c \to \infty$ (the latter limit being the pure spin-2 model). The black dots are examples of bounds obtained from SDPB at $k_\text{max}=12$ assuming no scalars in the spectrum. The plots illustrate the agreement between the convex hull and the general SDPB bounds since none of the black points lie significantly away from the convex hull. The greatest discrepancy is of order $10^{-2}$.
    }
    \label{fig:convexhull}
\end{figure}

We find very good agreement between the two types of bounds, as illustrated in Figure \ref{fig:convexhull}, which show the 3d projections  $(g_3/g_2,g_4/g_2,g_5/g_2)$
and 
$(g_6/g_2,g_6'/g_2,g_7/g_2)$ of the extremal convex hull. The black points are the SDPB results for the scalarless positivity bounds. As expected, these points lie on the outside of the convex hull, but generally differ from it by very little. (We display the figures from angles that show the biggest discrepancy.) 
Comparing the results obtained for  $k_\text{max}=10,12,14$, we find that the convex hull and positivity bounds indeed tend to converge toward each other with increasing $k_\text{max}$. 
At $k_\text{max}=14$, we find the discrepancies to be of order 
${\sim10^{-2}}$ for $(g_3/g_2,g_4/g_2,g_5/g_2)$ and ${\sim10^{-3}}$ for $(g_6/g_2,g_6'/g_2,g_7/g_2)$, which is comparable with how well the extremal theories with $1< \mu_c \lesssim 4$ match the universal bounds at finite $k_\text{max}$. 
The convex hull method gets progressively better as more extremal theories are included in the hull as well as with increasing $k_\text{max}$. The improvement from including more extremal theories outpaces the improvement from increasing $k_\text{max}$, so the main bottleneck to convergence comes from the fact that extremal theories with high $\mu_c$ are more difficult to evaluate numerically. 

\section{The 0SDR Bootstrap} 
\label{sec:0sdrextr}

In this section, we characterize 
the {\em maximally supersymmetric} extremal amplitudes and devise an algorithm that maps them to the 2SDR extremal amplitudes. 
At the end of this section, we make connections to the analytic deformations of the string amplitudes studied in \cite{Cheung:2024obl}.

\subsection{0SDR and Supersymmetry} \label{s:SUSYandOSDR}

Any amplitude of identical real scalars that can be embedded in an $\mathcal{N}=4$ supersymmetric Abelian theory can be written as the factor $(s^2+t^2+u^2)$ times a function that is fully symmetric in $s,t,u$. (See Section~\ref{sec:maxsusyconst}.) If the amplitude obeys the Froissart bound with $n_0=2$, then this symmetric function must obey the Froissart bound with $n_0=0$. Thus, we can think of the maximally supersymmetric amplitude in the 2SDR real scalar bootstrap as the factor $(s^2+t^2+u^2)$ times an amplitude\footnote{One might question why the function obtained from stripping off the $(s^2+t^2+u^2)$-factor is a valid amplitude. This is simply because $s^2$ times the stripped function is  the complex scalar amplitude $A(zz\bar{z}\bar{z})$ from \reef{eq:ztophi}. When  $A(zz\bar{z}\bar{z})$ has a positive partial wave expansion, so too does the function without the $s^2$ factor.} in the 0SDR real scalar bootstrap:
\begin{align}
    A^{\text{SUSY}}(s,t,u) = (s^2+t^2+u^2) \, A^{0\text{SDR}}(s,t,u)\, .
\end{align}
Reversing the logic, any 0SDR amplitude can be understood as a maximally supersymmetric 2SDR amplitude via multiplication by $(s^2+t^2+u^2)$. This establishes a one-to-one map between the amplitudes of the 0SDR bootstrap and maximally supersymmetric amplitudes in the 2SDR bootstrap. 

\subsection{Extremal 0SDR Theories}
\label{sec:extr0sdr}

\begin{figure}
    \centering
    \includegraphics[width=.7\linewidth]{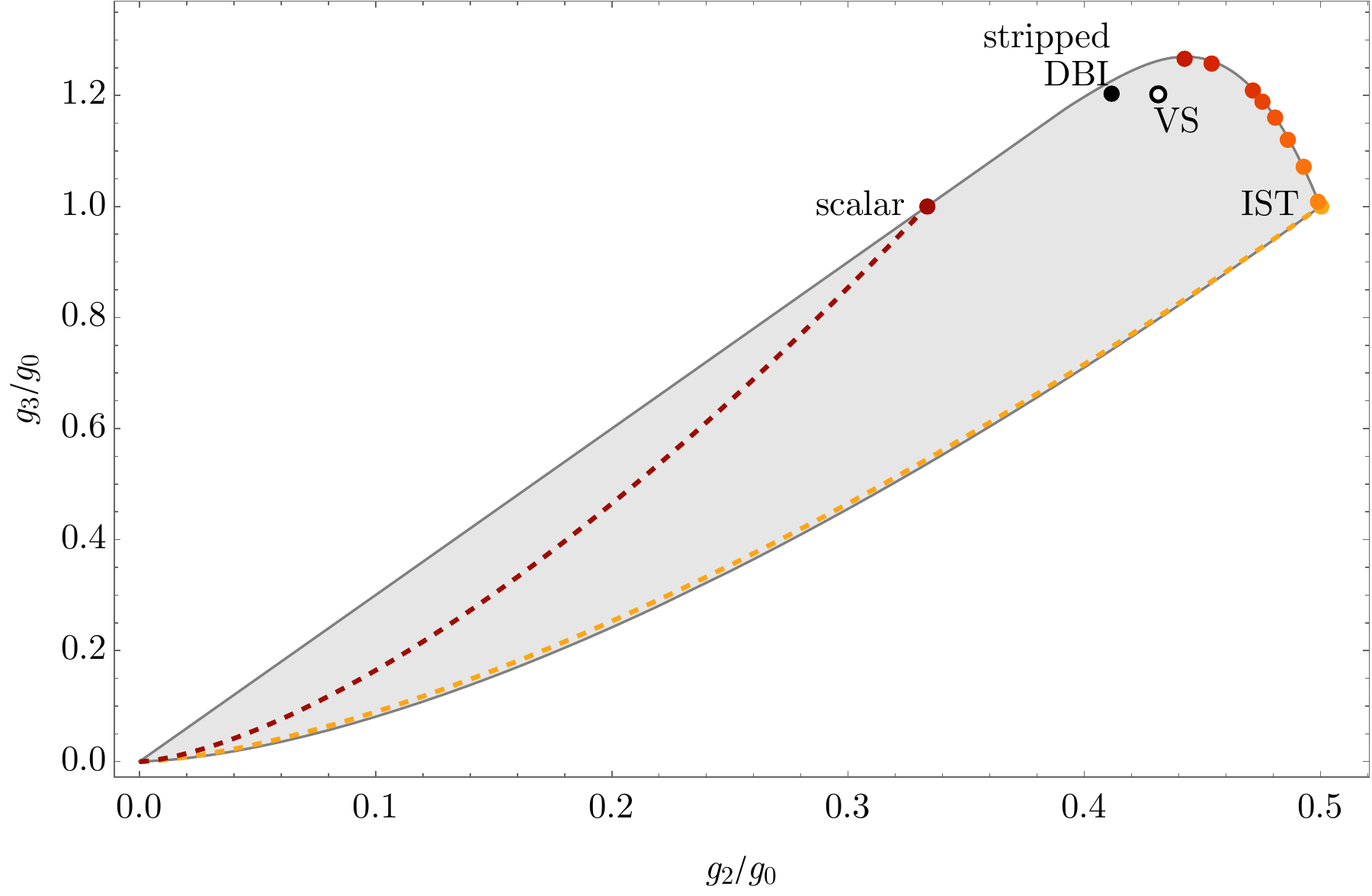}
    \caption{
    The $D=4$ general allowed region (gray) of the 0SDR bootstrap for the $(g_2/g_0,g_3/g_0)$ projection, computed at $k_\text{max}=20$. The IST is shown in orange and appears to determine the lower bound on the allowed region. The pure scalar exchange, marginal in this bootstrap, is shown in dark red. 
    The 0SDR extremal theories are shown in orange and were computed for $\mu_2<1.5$ with $k_\text{max}=16$,
    for $1.5\leq\mu_c<3$ with $k_\text{max}=19$, and for $\mu_2 = 3$ with $k_\text{max}=22$.{\protect\footnotemark}  The extremal theories interpolate between the IST and the pure scalar model as $\mu_2$ increases from 1 to $\infty$. We additionally show the locations of the relevant stringy amplitudes: stripped DBI \eqref{Astripped} and Virasoro-Shapiro \eqref{eq:VS}.
    }
    \label{fig:0sdr_extr}
\end{figure}

In the 0SDR bootstrap, we have dispersion relations for all the Wilson coefficients $g_k$, including $g_0$.\footnote{Recall that we have turned off the cubic self-interaction of the $\phi$'s in order to avoid the $1/u$ pole issue described around \reef{stpolecontrib}. The $\phi^3$ interaction is not allowed by maximal supersymmetry which automatically takes care of the $1/u$ pole issue.} It is therefore natural to bound the ratios
\be
  \frac{g_k}{g_0}
\ee
rather than $g_k/g_2$ in the 2SDR boostrap. Thus, the lowest-order coefficients that can be bounded are $g_2/g_0$ and $g_3/g_0$, and the corresponding general allowed region is shown in Figure~\ref{fig:0sdr_extr}. The curved upper bound connecting the pure scalar model and the Infinite Spin Tower (IST) is the 0SDR equivalent of the extremal sliver for the 2SDR bootstrap. In analogy with the 2SDR analysis, we now proceed to characterize the 0SDR extremal amplitudes that lie on the boundary of the extremal sliver. 

We know from \reef{spin12choice0SDR} that in the 0SDR bootstrap, a state exchanged at tree-level with mass $M_\text{gap}$ can only be a scalar. Thus, in parallel with the numerical determination of the 2SDR extremal theories in Section \ref{sec:extr2sdr}, we
assume the spectrum
\be
\label{onlyspin0atgap}
\raisebox{-8.5mm}{\includegraphics[width=6cm]{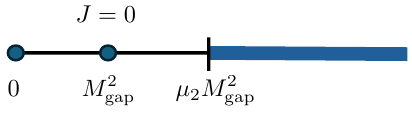}}
\ee
and maximize the spin-0 coupling $|\lambda_0|^2/g_0$ for given $\mu_2$. 
This procedure selects a unique theory (within numerical agreement between the maximum and minimum of each Wilson coefficient). These theories, as a function of $\mu_2$, are the one-parameter family of 0SDR extremal theories. In the limit $\mu_2 \to 1$, they approach the IST, and as $\mu_2 \to \infty$, we expect them to asymptote to the pure scalar theory. The orange-to-red points along the extremal boundary curve in Figure \ref{fig:0sdr_extr} are extremal theories with $\mu_2$ varying between 1 and 3. 
Analogously to the 2SDR case, we expect that the one-parameter family of 0SDR extremal theories and their mass-scaling curves generate a convex hull that is equivalent to the full allowed space of the 0SDR bootstrap in the limit $k_\text{max} \to \infty$.

While both characterizations involve coupling maximization, the 0SDR and 2SDR extremal theories nonetheless have different spectra and appear unrelated. However, we now propose a map between them.\footnotetext{For larger values of $\mu_c$, the extremization results of SDPB are much more sensitive to $k_{\max}$ than they are for smaller values of $\mu_c$, so to find sensible results for the 0SDR extremal theories at larger $\mu_c$, it is necessary to go to larger $k_{\max}$ than it is for smaller~$\mu_c$.}
%

\subsection{Reconstructing 2SDR Bounds Using the 0SDR Bootstrap}\label{sec:2sdrfrom0sdr}

\begin{figure}
    \centering
    \hspace{-5mm}
    \raisebox{3.7cm}{
    \begin{tabular}{l}
    \includegraphics[width=0.3\linewidth]{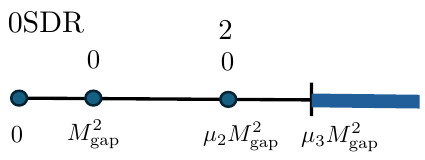}
    \\[0mm] 
    \hspace{1.2cm}
    {\color{BurntOrange} {\Large $\downarrow$} 
    {\small Step 1}
    {\Large $\downarrow$}}
    \\[-1mm]
\includegraphics[width=0.3\linewidth]{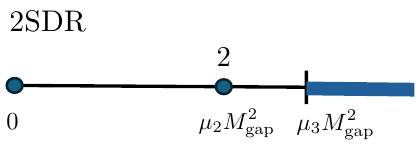}
    \\[0mm]
    \hspace{1.2cm}
    {\color{ForestGreen} {\Large $\downarrow$} 
    {\small Step 2}
    {\Large $\downarrow$}}
    \\[-1mm]
\includegraphics[width=0.3\linewidth]{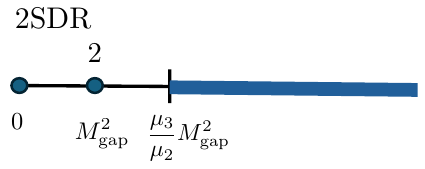}
    \end{tabular}}
    \hspace{-2mm}
    \raisebox{4mm}{\includegraphics[width=0.65\linewidth]{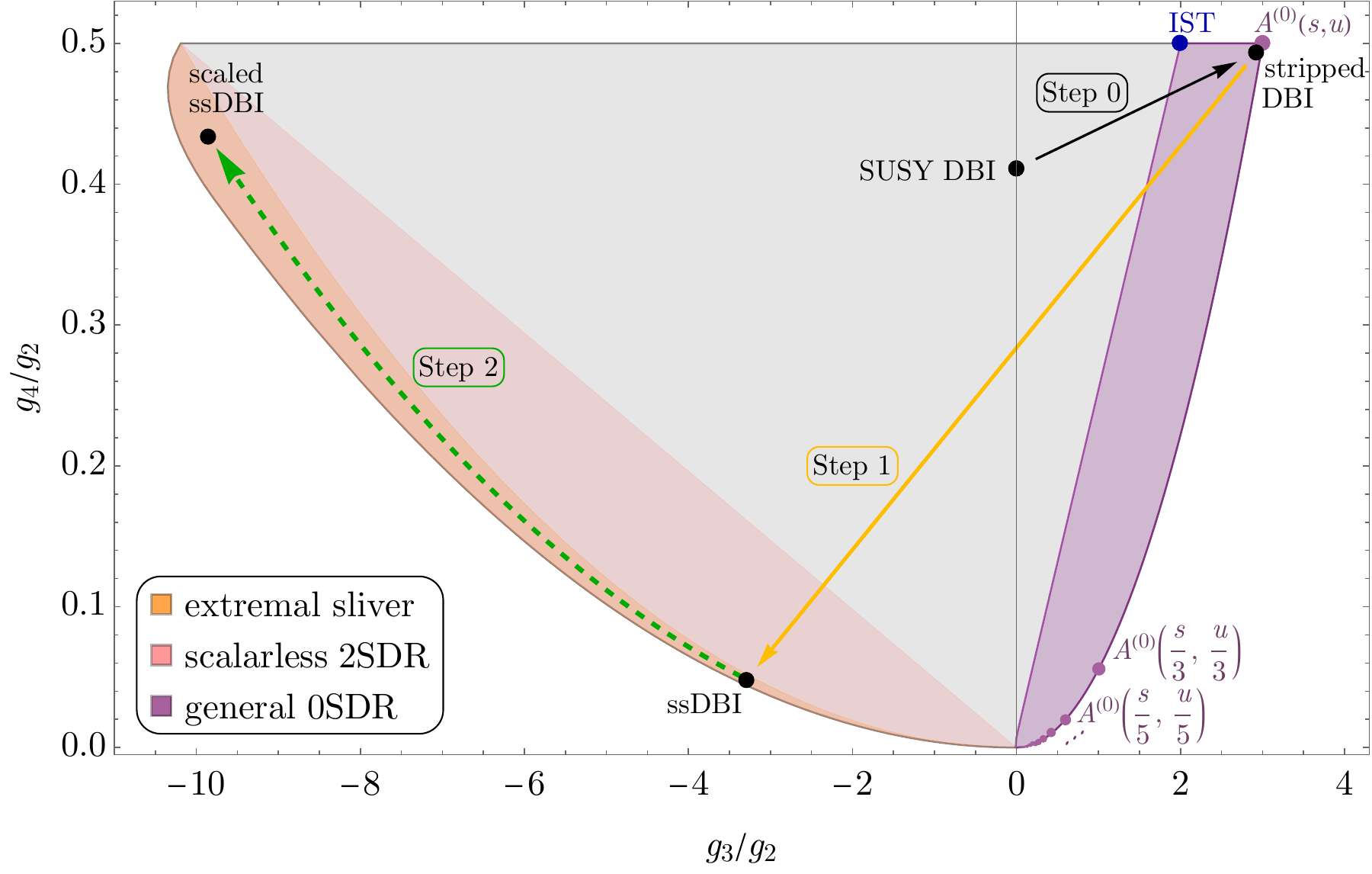}}
    \caption{Left: The procedure for relating 0SDR and 2SDR extremal theories. Right: Example with $\mu_2 = 3$ and $\mu_3 = 5$ for $D=4$ DBI in the general 2SDR allowed region in the $(g_3/g_2,g_4/g_2)$-plane.
    Step 0 removes the SUSY factor $(s^2+t^2+u^2)$ from the SUSY DBI amplitude to get the stripped amplitude  \eqref{eq:dbisubamp} that obeys the $n_0=0$ Froissart bound. Steps 1 and 2 are  performed on the stripped DBI amplitude to generate an amplitude in the 2SDR extremal sliver.}
    \label{fig:dbiprocedure}
\end{figure}

The 0SDR bounds are more restrictive than the 2SDR bounds. This difference can be seen by comparing the bounds on $g_k/g_2$ obtained by the respective bootstraps:
Figure \ref{fig:dbiprocedure} shows the 0SDR region as the purple subregion within the 2SDR bounds in the $(g_3/g_2,g_4/g_2)$-plane. The top corners are the IST and the pure scalar theory. Also shown in Figure \ref{fig:dbiprocedure} is the procedure to map 0SDR theories to the scalarless region (pink) that we now describe in detail.

Consider a 0SDR amplitude with the most general allowed tree-level spectrum (cf.~\reef{spin12choice0SDR}):  a scalar at $M_\text{gap}^2$ and then a scalar and a spin 2 at $\mu_2 M_\text{gap}^2$ for $\mu_2 >1$. We assume a cutoff $\mu_3 M_\text{gap}^2$ beyond which we remain agnostic about the spectrum. 
As illustrated on the left of Figure \ref{fig:dbiprocedure}, in
 Step 1, we subtract all scalars. The amplitude no longer obeys the $n_0=0$ Froissart bound, but it still lies within the 2SDR bootstrap bounds. 
Instead of having the lowest massive state at $M_\text{gap}^2$, the amplitude has its lightest state at $\mu_2 M_\text{gap}^2$. In Step 2, we rescale the spectrum by $1/\mu_2$ so that the lowest spin-2 state is indeed at $M_\text{gap}^2$. These are the two steps to go from an amplitude in the 0SDR bootstrap to one in the 2SDR bootstrap with both having a state at the mass gap $M_\text{gap}^2$.

Let us illustrate the procedure with a concrete example. The DBI amplitude \reef{eq:dbi} obeys the Froissart bound with $n_0=2$. Stripping off the overall $(s^2 + t^2 + u^2)$ factor, however, gives a unitary amplitude that obeys the Froissart bound with $n_0=0$:
\begin{align}\label{eq:dbisubamp}
    A_\text{0SDR}^{\text{DBI}}(s,t,u)=-\bigg(\frac{\Gamma(-s)\Gamma(-u)}{\Gamma(1+t)}+\frac{\Gamma(-t)\Gamma(-s)}{\Gamma(1+u)}+\frac{\Gamma(-u)\Gamma(-t)}{\Gamma(1+s)}\bigg)\;.
\end{align}
(For simplicity, we have set $\alpha'=M_\text{gap}^2=1$.)
The spectrum of this amplitude is just like that of the top-left of Figure \ref{fig:dbiprocedure} with $\mu_2=3$, $\mu_3=5$, etc. As shown in Figure \ref{fig:dbiprocedure}, the resulting ``stripped'' DBI amplitude now lies in the 0SDR region.

The scalar couplings of \reef{eq:dbisubamp} can be computed explicitly and then subtracted as discussed in  Section~\ref{sec:scalarsub}. This is Step 1, which maps the stripped DBI amplitude from the 0SDR region to the 2SDR scalarless region. In Step 2, the spectrum is rescaled by taking 
$s,t,u\mapsto 3s,3t,3u$, resulting in 
\be
\label{eq:dbiprocedure}
  A_\mathrm{2SDR}^{\text{DBI}}(s,t,u)=-\bigg(\frac{\Gamma(-3s)\Gamma(-3u)}{\Gamma(1+3t)}+\frac{\Gamma(-3t)\Gamma(-3s)}{\Gamma(1+3u)}+\frac{\Gamma(-3u)\Gamma(-3t)}{\Gamma(1+3s)}\bigg)
  -\sum\mathrm{scalars}\,.
\ee
This rescaled, scalar-subtracted, stripped DBI amplitude lies in the extremal sliver of the 2SDR scalarless region.

The DBI example illustrates the map explicitly, but we can also implement the map numerically to draw a correspondence between the 0SDR and 2SDR extremal theories. The procedure for doing so is as follows. 
Assume the spectrum \reef{onlyspin0atgap} with spin 0 at $M_\text{gap}^2$ and a gap to  $\mu_2 M_\text{gap}^2$ beyond which we are agnostic about the spectrum. For each $\mu_2$, the resulting allowed region in the $(g_2/g_0,g_3/g_0)$-plane from Figure \ref{fig:0sdr_extr} has a sharp corner at the extremal boundary. (This is analogue to the blue region on the left of Figure \ref{fig:extremal} for the 2SDR analysis.) This allows us to identify the 0SDR extremal theories as the unique theories that maximize $g_2/g_0$.\footnote{This characterization of the 0SDR extremal theories is different from the spin-0 coupling maximization of Section \ref{sec:extr0sdr}, but it works better for the numerical reconstruction procedure because it gives a better determination of the $g_2/g_0$ coefficient.}. The reconstruction procedure can be summarized as follows:

\begin{enumerate}
\item 
For given $\mu_2$, we compute $\text{max}(g_2/g_0)$ in the 0SDR bootstrap. 

\item Fixing $g_2/g_0$ to the maximum value, we compute the unique (up to numerical precision) values of $|\lambda_0|^2/g_0$, $g_3/g_0$, $g_4/g_0,\ldots$ for the extremal theories.  

\begin{figure}
\centering
\begin{tikzpicture}
    \node (image) at (0,0) {\includegraphics[width=0.7\linewidth]{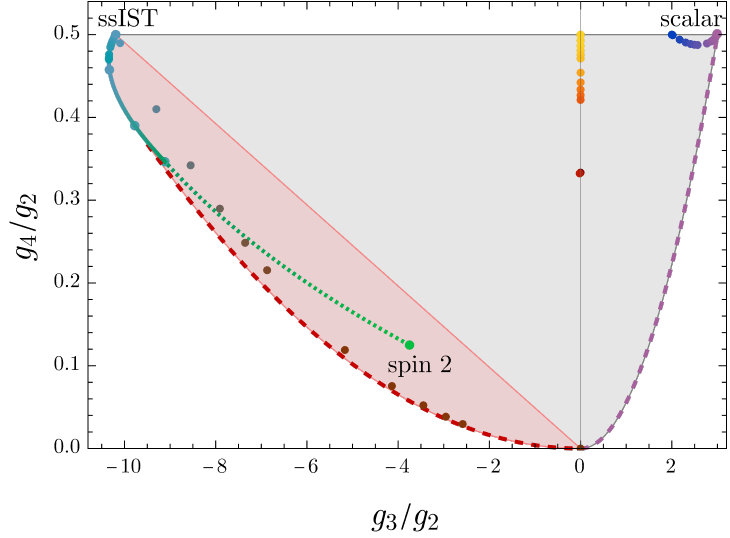}};
    \draw[-{Latex[length=2mm]}] (3.13,2.65)   -- (4.45,3.05);
    \node[rotate = 13] at (3.7,2.95) {\tiny{Step 0}};
    \draw[-{Latex[length=2mm]}] (4.7,2.95)   -- (-0.38,-0.92);
    \node[rotate = 35] at (2.1,1.2) {\tiny{Step 1}};
    \draw[-{Latex[length=2mm]}] (-0.46,-0.9)   -- (-3.8,3);
    \node[rotate = -50] at (-2,1.2) {\tiny{Step 2}};
\end{tikzpicture}
\caption{Mapping the 0SDR extremal theories to the 2SDR extremal theories in $D=4$.}
\label{fig:2sdrextr}
\end{figure}
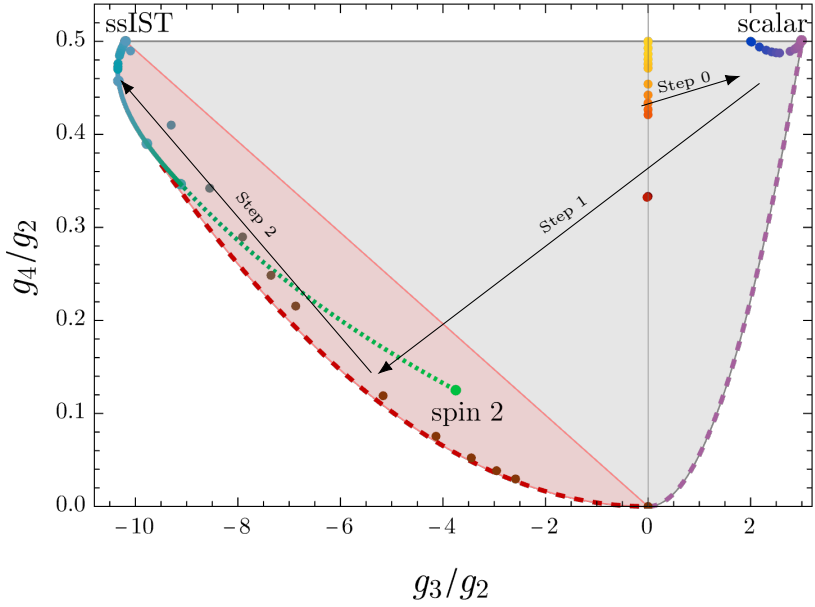

The blue-to-violet points in the purple 0SDR region of Figure \ref{fig:2sdrextr} are obtained from these values and illustrate the location of the 0SDR extremal theories in the $(g_3/g_2,g_4/g_2)$-plane. 

\item Scalar subtraction acts directly on the Wilson coefficients via the dispersion relations, e.g.~
\be
  \label{g2g3scalarsub}
  g_2 \to g_2^{(2)} = g_2 - |\lambda_0|^2 \frac{1}{2}w_{0,2,0}^{(D)}\,,
  ~~~~
  g_3 \to g_3^{(2)} = g_3 - |\lambda_0|^2 \big(-w_{0,3,1}^{(D)}\big)\,,
  ~~~
  \dots
\ee
Here we have used the relation \reef{gfromakq} between the $g_k$ and the $a_{k,q}$ along with the dispersion relations \reef{eq:akqsdr} with a delta-function 
$|\lambda_0|^2\delta_{j,0}\,
\delta(s-M_\text{gap}^2)$ 
in the spectral density.  We also use the superscript ${}^{(2)}$ to indicate that the resulting Wilson coefficients are in the 2SDR bootstrap. 
The location of the scalar subtracted 0SDR extremal theories in the 2SDR region is then, e.g.
\be
  \frac{g_3^{(2)}}{g_2^{(2)}}
  = \frac{g_3^{(2)}/g_0}{g_2^{(2)}/g_0}
  \,.
\ee
We have divided both the numerator and denominator by $g_0$ in order to make the expressions from \reef{g2g3scalarsub} depend just on the ratios $|\lambda_0|^2/g_0$ and  $g_k/g_0$ from step 2. 

This procedure maps the blue-to-violet points of the 0SDR extremal theories to the blue-to-brown points in the 2SDR scalarless region, as shown in Figure \ref{fig:2sdrextr}.
 
\item The last step is the rescaling of the spectrum. This is implemented simply as the mass-scaling $g_k^{(2)} \to g_k^{(2)} \mu_2^k$.

The result is shown in Figure~\ref{fig:2sdrextr} as the blue-to-green dots. The results  agree to $\mathcal{O}(10^{-3})$ with the general 2SDR bounds. The zoom-in on the extremal bulge boundary was presented in Figure~\ref{fig:basicg3g4},  which compares the extremal theories obtained in the 2SDR bootstrap (blue points) and reconstructed from the 0SDR extremal theories (purple points) with the general 2SDR bounds.
\end{enumerate} 

It is important to note here that in Step 3 of this procedure, we subtract only a single scalar: the one at $M_\text{gap}^2$ for which we have numerically fixed the coupling. One might think that the resulting subtracted theory would not necessarily be entirely scalarless, as it could have scalar contributions above the mass gap. However, as has been noted in several previous works \cite{Caron-Huot:2021rmr,Albert:2023seb,Berman:2024wyt,Albert:2024yap}, amplitudes found to be extremal in numerical  bootstraps often appear to have only a single massive scalar. Therefore, we expect that there is only a single scalar in the numerical spectrum of the extremal theories. This point comes up again in Section \ref{sec:bifrost}, and we discuss it further in Section \ref{sec:disc}.

\begin{figure}
     \centering         \includegraphics[width=0.45\textwidth]{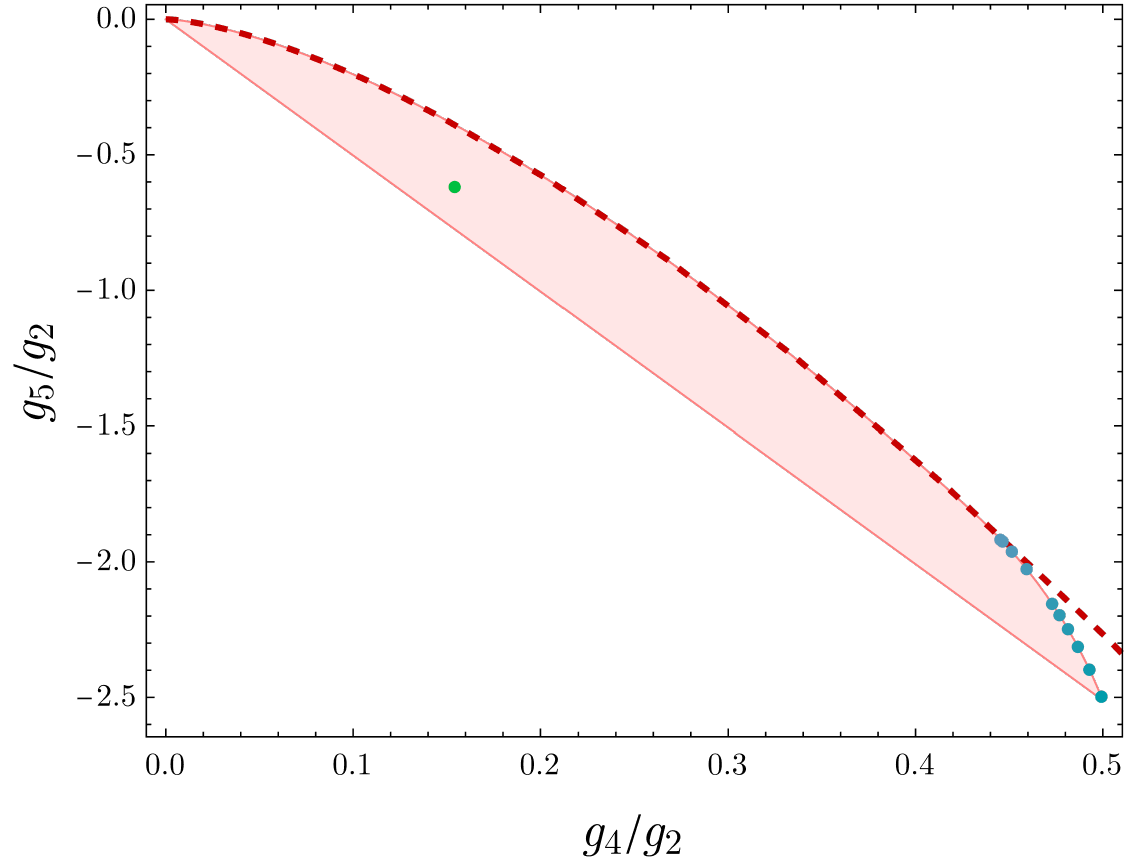}
 ~~~\includegraphics[width=0.45\textwidth]{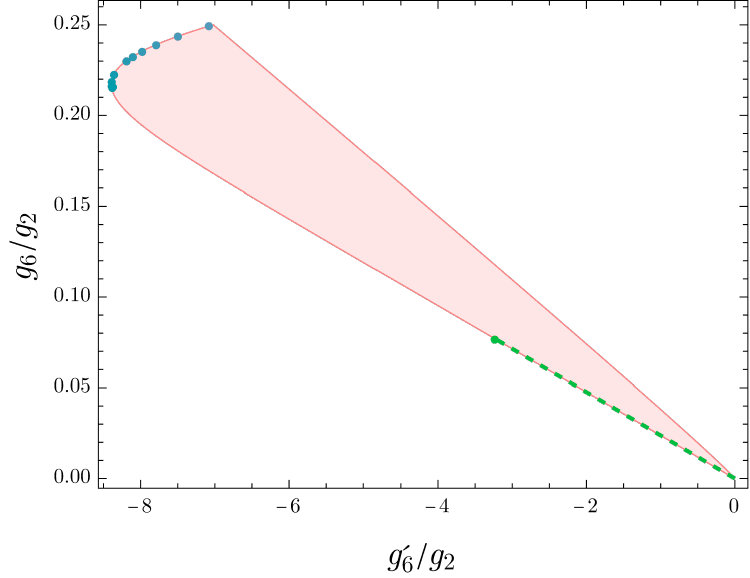}
     \caption{
     The scalarless 2SDR allowed region in two different projections. The blue-green points are the points reconstructed from the 0SDR extremal theories. The green dot is the pure spin-2 amplitude that the extremal theories should approach for large cutoff. Note that in the $(g_6/g_2,g_6'/g_2)$-plane on the right, the entire family of extremal theories lies on the boundary of the space along with the spin-2 model.
     (Details: for $\mu_2<1.5$ using $k_\text{max}=16$, for $1.5\leq\mu_2<3$ using $k_\text{max}=19$, and for $\mu_2 = 3$ using  $k_\text{max}=22$.)
     }
    \label{fig:scalarless}
\end{figure}

With this subtlety in mind, we tested the construction of the 2SDR extremal theories from the 0SDR extremal theories in several other projections. Two sample plots are shown in Figure \ref{fig:scalarless}. They show that the 0SDR extremal theories map to the boundary of the 2SDR scalarless regions. The plots in Figure \ref{fig:scalarless} are in $D=10$, where the numerics converge at lower $k_\text{max}$ than in $D=4$.

With our numerical reconstruction of the 2SDR extremal theories from the 0SDR extremal theories in $D=4$ and $D=10$, we have been able to reach gap values $\mu_c^{(2)} \approx 1.7$. 
This requires $\mu_2$ for the 0SDR bootstrap to be above 3, and for these values, the extremal theories only converge to the boundary for very high $k_\text{max}$. For example, in Figure~\ref{fig:0sdr_extr} the 0SDR extremal theories with $\mu_2 > 3$ were computed with $k_\text{max} =22$. At such high $k_\text{max}$, the numerics are also increasingly sensitive to the spin list, and this means a very high cost in computational time. 

This leaves the question open if the entire 2SDR extremal bound can be reconstructed from the 0SDR theories. If that were the case, then the convex hull conjecture would imply that the 0SDR and 2SDR bootstraps are equivalent and that one can recover the more ``generic'' bootstrap bounds of the 2SDR bootstrap from the more constrained 0SDR bootstrap. Moreover, since the 0SDR bootstrap can be viewed as a bootstrap of maximally supersymmetric amplitudes, this would mean that a maximally supersymmetric bootstrap could, in principle, determine the entire non-supersymmetric bootstrap!

To gain more insights, let us now study the 0SDR extremal theories from a different perspective.

\subsection{Bifrost: Extremal Theories as ``Corner Theories''}
\label{sec:bifrost}

\begin{figure}
    \centering
\includegraphics[width=0.7\linewidth]{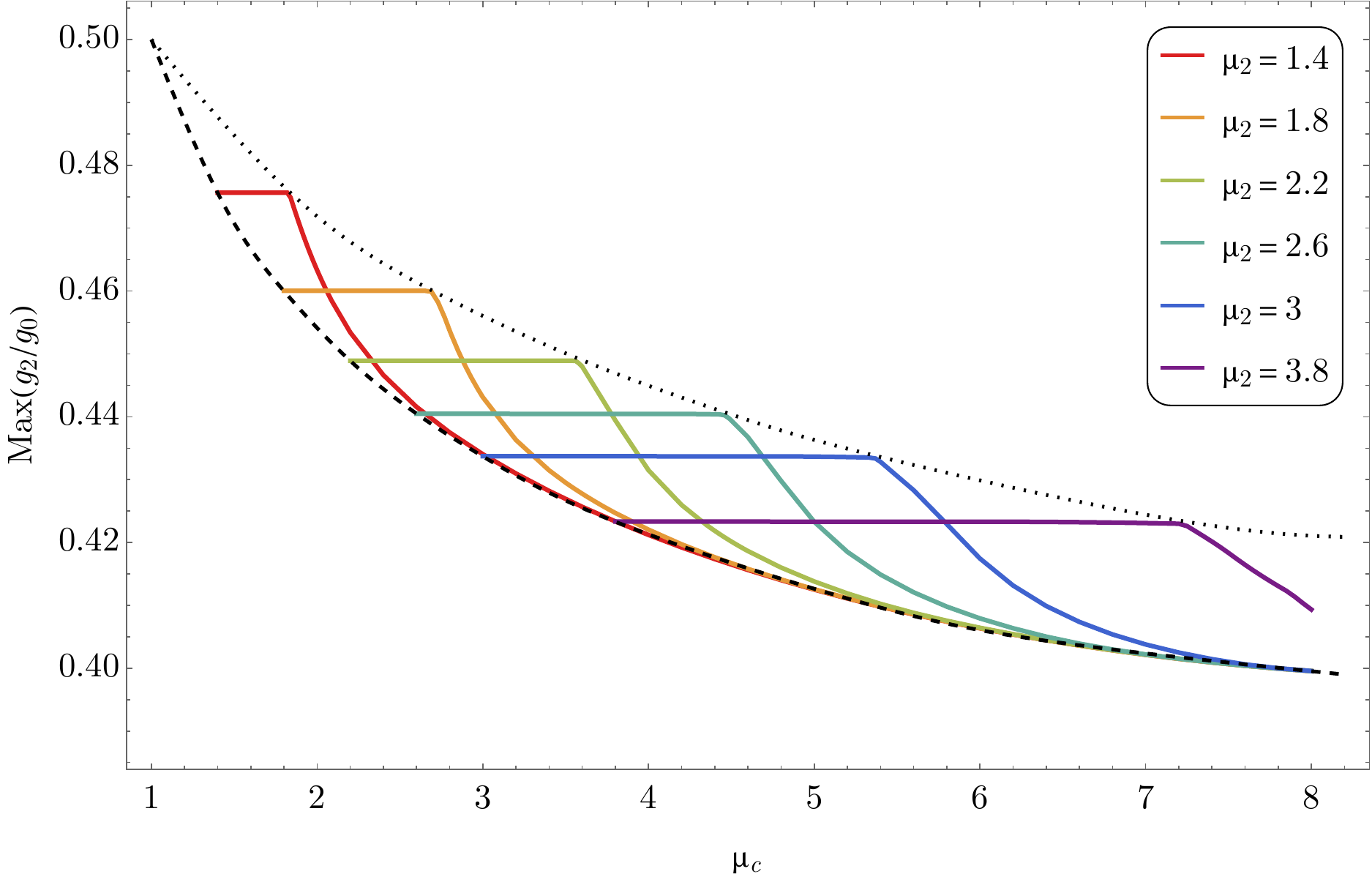}
    
    \caption{Max($g_2/g_0$) as a function of the cutoff $\mu_c$ computed at $k_\text{max}=14$ in the 0SDR bootstrap for spectrum \reef{bifrostspec} in $D=4$. 
    The sharp corners identify a one-parameter family of ``corner theories'' parameterized by the mass-squared ratio $\mu_2$ of the  two lowest massive states. The corner theories are expected to  interpolate between the IST ($\mu_2=1$) and the scalar-only theory ($\mu_2=\infty$). To indicate the proposed connection to the non-Abelian corner theories, the bounds are color-coded the same as the equivalent curves in Figure 3 of \cite{Berman:2024wyt}.}
    \label{fig:bifrost}
\end{figure}

In the context of the 0SDR bootstrap, consider the spectrum 
\be
\label{bifrostspec}
\raisebox{-7.5mm}{
\includegraphics[width=5cm]{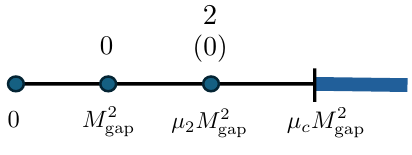}}
\ee
with couplings of the massive spin states left unfixed.
For given choice of $\mu_2$, compute the maximum of $g_2/g_0$
as a function of $\mu_c > \mu_2$ in the 0SDR bootstrap. Figure \ref{fig:bifrost} shows that for each $\mu_2$, $\text{max}(g_2/g_0)$ has a plateau that ends at a corner with $\mu_c= \mu_c^\text{corner}(\mu_2)$, which marks a sudden sharp reduction in the allowed space for values of $\mu_c$ greater than $\mu_c^\text{corner}(\mu_2)$.

As in other bootstraps \cite{Berman:2024wyt,Albert:2024yap},  a kink in the bounds is a clear indication that there is a theory near the corner. Specifically in this case, for each $\mu_2$, there should be a 4-point amplitude whose $g_2/g_0$ value equals the plateau value and has a massive state near, or perhaps a bit below, the corner value of $\mu_c$, i.e.~with mass-squared $\approx \mu_c^\text{corner}(\mu_2) M_\text{gap}^2$. It is natural to expect that these ``corner theories''  are nothing but the 0SDR extremal theories. This is confirmed  by computing the value of $g_2/g_0$ for the 0SDR extremal theory for each of the six cases $\mu_2=1.4, 1.8, 2.2, 2.6, 3, 3.8$ in Figure \ref{fig:bifrost} 
and finding the result to be within numerical precision of the plateau value of $g_2/g_0$.

For values of $\mu_2$ sufficiently close to 1 (e.g.~$\mu_2 \lesssim 1.2$), the corner locations are approximately $\mu_c^\text{corner}(\mu_2)=2\mu_2-1$. If this is the mass-squared of the third massive state, it would be compatible with a linear leading Regge trajectory
\be 
  \label{Regge0sdr}
  \text{low-$\mu_2$ 0SDR extremal theory:}
  \hspace{6mm}
  \frac{M^2_J \hfill}{M_\text{gap}^2} = \frac{1}{2}(\mu_2-1)J + 1 \,.
\ee
When mapped to the 2SDR extremal theories, the scalars are subtracted off and the spectrum  rescaled by $\mu_2$ so that the lowest spin-2 state has mass-squared $M_\text{gap}^2$. 
If the 0SDR extremal theories have  linear leading Regge trajectory \reef{Regge0sdr}, we would also obtain a linear
leading  Regge trajectory for the 2SDR extremal theories, but with a different slope and intercept:
\be 
  \label{Regge2sdr}
  \text{low-$\mu_2$ 2SDR extremal theory:}
  \hspace{6mm}
  \frac{M^2_J \hfill}{M_\text{gap}^2} = \frac{1}{2\mu_2}(\mu_2-1)J + \frac{1}{\mu_2} \, .
\ee
This says that the lowest mass state which has $J=2$ has mass-squared $M_\text{gap}^2$ and the second level with spin $J=4$ is at $(2-\frac{1}{\mu_2})M_\text{gap}^2$. Thus, if the 
0SDR extremal theories have a linear spectrum \reef{Regge0sdr}, they could only reproduce 2SDR extremal theories with $\mu_c^{(2)}$ in the range between 1 and 2. 
There are two possible conclusions one can draw from this analysis. Either 
\begin{enumerate}
\item The map from the 0SDR extremal theories to the 2SDR is only valid for gap $1 \le \mu_c^{(2)} < 2$ in the 2SDR extremal spectrum. This would be the case if all 0SDR extremal theories have linear spectra \reef{Regge0sdr} for all $\mu_2$,
\item[] or
\item The map from the 0SDR extremal theories can reach  2SDR extremal theories with any value of $\mu_c^{(2)} \ge 1$. This requires 0SDR extremal theories to have non-linear spectra, at least for $\mu_2$ above a certain value.
\end{enumerate}

Based on the location of the corners in Figure \ref{fig:bifrost}, one would however not necessarily expect a linear spectrum of the 0SDR theories. For example, for $\mu_2=3.8$, we find the corner to be around $\mu_c \approx 7.2$, a significant departure from the prediction $\mu_c=6.6$ of the linear Regge behavior \reef{Regge0sdr}. 
If we had been able to numerically reconstruct the 2SDR extremal theories with $\mu_c^{(2)} > 2$, we would have had additional evidence in favor of a non-linear spectrum and hence the second option above. Unfortunately, 
as discussed in Section~\ref{sec:2sdrfrom0sdr}, the large increase in computation time prevented us from  numerically constructing the 2SDR extremal theories  for $\mu_c^{(2)} > 1.7$.

A one-parameter family of corner theories were previously found in the S-matrix bootstrap of maximally supersymmetric Yang-Mills EFTs in \cite{Berman:2024wyt,Albert:2024yap}.\footnote{The corner theory plots in \cite{Berman:2024wyt} were done in $D=10$. For the Abelian real scalars, we have done the corner analysis in both $D=4$ and $D=10$, and the results are very similar.} These non-Abelian corner theories have similar properties to those in Figure~\ref{fig:bifrost}. In fact, since the 0SDR bootstrap can be thought of as a supersymmetric bootstrap of Abelian complex scalar amplitudes $A^\text{SUSY}(zz\bar{z}\bar{z}) = s^2 A^\text{0SDR}(s,t,u)$, it is quite natural to expect that the 0SDR one-parameter family of corner theories are nothing but the Abelianization of the non-Abelian corner theories found in \cite{Berman:2024wyt}. 

The results in Figure \ref{fig:bifrost} are found to be independent of whether the scalar at the second mass-level $\mu_2 M_\text{gap}^2$ is allowed or not (hence the parenthesis in \reef{bifrostspec}). Furthermore, if we exclude spin 0 and 2 in the entire spectrum above the cutoff $\mu_c$, we find same plateau value of max($g_2/g_0$) and the same  corner location $\mu_c^\text{corner}(\mu_2)$, but the dropoff after the corner does become much sharper. The absence of these lower-spin states at the higher mass-levels indicate an absence of  daughter trajectories, and one may speculate whether the extremal theories could be amplitudes with just a single-Regge trajectory. There are analytic bootstrap arguments that such amplitudes are inconsistent~\cite{Eckner:2024pqt}. However, at finite $k_\text{max}$, it appears that SDPB can extremize the Wilson coefficients by pushing off any daughter trajectories to infinity and instead populating the lower mass-levels as much as possible. We discuss these single-Regge trajectory aspects further in Section \ref{sec:disc}. For now, let us turn to the question of analytic candidates for the family of extremal amplitudes.

\subsection{Possible Analytic Candidates}
\label{sec:deformedamps}
There are currently no unitary analytic candidates for the 0SDR extremal amplitudes with non-linear Regge trajectories, so we study here candidates with linear trajectories of varying slopes. Specifically, we consider the hypergeometric deformations of the Veneziano and Virasoro-Shapiro amplitudes in \cite{Cheung:2024obl} as candidates for the extremal theories.\footnote{Other hypergeometric amplitudes have also been studied in~\cite{Cheung:2023adk, Rigatos:2023asb, Geiser:2023qqq, Wang:2024wcc, Cheung:2024uhn, Mansfield:2024wjc}.} 

The authors of~\cite{Cheung:2024obl} constructed analytic expressions for the two families of deformed string amplitudes 
in terms of a parameter that allowed them to vary the slope $\g$  of the assumed linear leading  Regge trajectory:\footnote{To avoid confusion with couplings, the mass-spacing parameter $\lambda$ of \cite{Cheung:2024obl} is relabeled here to $1/\gamma$ for dDBI and to $1/(2\gamma)$ for dVS. 
} 
\be  
  M^2_J = \gamma J+1 \,.
\ee
The deformed Veneziano amplitude can be Abelianized using \reef{u1decoup} to give a deformation of the DBI amplitude (with the $(s^2+t^2+u^2)$ factor stripped off). Both the deformed DBI (dDBI) and the deformed Virasoro-Shapiro (dVS) amplitudes obey the Froissart bound with $n_0=0$, and, since they are unitary by construction, their low-energy expansions have Wilson coefficients within the 0SDR bootstrap bounds. Although these deformed amplitudes are not known to arise from a microscopic theory, we can still study where they sit within EFT-space. 

As the parameter $\g$ varies in $D=4$ dimensions, the two deformed amplitudes interpolate between the pure scalar amplitude \reef{eq:A0} in the $\g \to \infty$ limit and the IST amplitude \reef{eq:basicIST} in the $\g \to 0$ limit. For $\g=1$, the dDBI amplitude coincides with the stripped DBI amplitude, and for $\g=1/2$
the dVS amplitude becomes the Virasoro-Shapiro amplitude.

\begin{figure}
\centering
\includegraphics[width=0.48\linewidth]{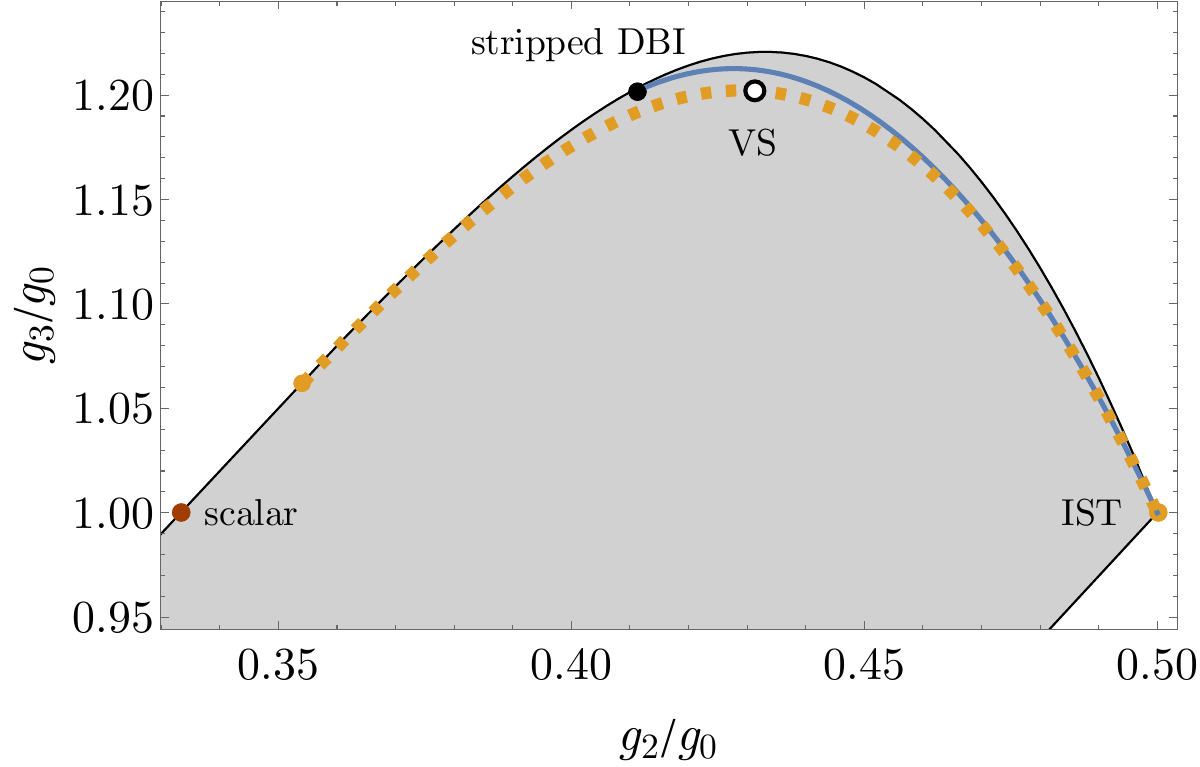}
~
{\includegraphics[width=.48\linewidth]{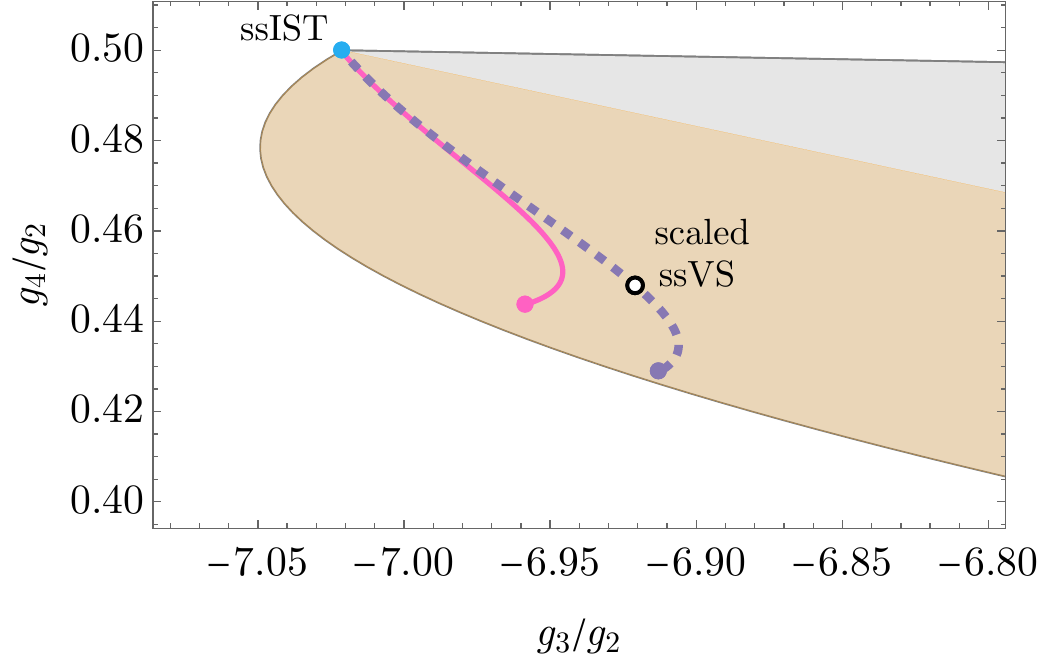}}
\caption{{\bf Left:} The analytic deformations \cite{Cheung:2024obl} of Veneziano, Abelianized to DBI (dDBI, blue), and Virasoro-Shapiro (dVS, orange) in the 0SDR extremal region for $D=10$. For $\gamma=0$ they both start at the Infinite Spin Tower (IST), shown as the orange dot the corner of the gray allowed region. 
The dDBI amplitude reaches DBI at its maximally allowed slope $\gamma=1$. The dVS amplitude reaches VS at $\gamma=1/2$ but in $D=10$ it goes beyond as $\gamma$ is allowed to extend all the way to 6.1; hence the curve goes past VS. (In the formal critical dimension, $D=23$, the maximal slope is $\gamma=1/2$.)
 In $D=10$, neither deformed theory reaches the pure spin-0 amplitude~\reef{eq:A0}, shown as the maroon dot.
{\bf Right:}
Zoom-in of the extremal sliver in $D=10$, with the scalar-subtracted-and-scaled deformed amplitudes  of~\cite{Cheung:2024obl}
shown as the pink solid curve for dDBI and the purple dashed curve for dVS.  The $D=10$ universal boundary is computed at $k_\text{max}=22$.}
\label{fig:gcamps}
\end{figure}

However, in dimensions $D > 4$ the DBI deformation has a $D$-dependent Regge slope $\g_c^{\text{DBI}}(D)$ at which the amplitude is critical, and in $D > 9$ the same is true for the Virasoro-Shapiro deformation, which has critical Regge slope $\g_c^{\VS}(D)$. In $D = 10$, which we use as an example, the critical slopes are
\be\label{eq:critslopes}
\g_c^{\text{dDBI}}(10) = 1 \, ,
\quad \quad 
\g_c^{\text{dVS}}(10) \approx 6.1  \, .
\ee 
The analytic dDBI and dVS theories are plotted for $D=10$ in Figure \ref{fig:gcamps}, where they can be compared with the 
$D=10$ bounds of the 0SDR allowed region in the $(g_2/g_0,g_3/g_0)$-plane. The dots at the end of the curves represent the minimal Regge slope~\reef{eq:critslopes} allowed by unitarity. The amplitudes at those ending points are critical in $D=10$.

To understand whether the two families of deformed amplitudes really are candidates for the extremal theories, we map them into the $(g_3/g_2,g_4/g_2)$ region and subject them to the procedure of scalar-subtraction and rescaling discussed in Section \ref{sec:2sdrfrom0sdr}.  
This gives the dashed-pink (dVS) and purple (dDBI) curves shown on the right  Figure~\ref{fig:gcamps}.
This shows that each curve reaches the scalar-subtracted IST amplitude in the limit $\gamma \to 0$, but then it bends into the space such that for general $\gamma$ values, they do not reproduce the numerical extremal boundary. They come closest at the end-points where the deformed amplitudes become extremal in $D=10$. This correspond to a ratio of the two lowest massive states which is $\mu_c^{(2)}=5/3$ for dDBI and  $\mu_c^{(2)} \approx 1.92$ for dVS.

In order to reproduce extremal theories with linear Regge slopes, we believe that it would be necessary to incorporate an additional parameter in the amplitudes found in \cite{Cheung:2024obl}. The new parameter would control the critical dimension for a particular gap between the two lightest states: in other words, it would allow one to vary the leading Regge trajectory while keeping the critical dimension fixed. While the analytic examples of \cite{Cheung:2024obl} both have linear Regge trajectories, we discussed the possibility that some of the extremal theories may have nonlinear Regge trajectories in Section \ref{sec:bifrost}, so this parameter may also introduce nonlinearity into the leading Regge trajectory.

\section{Bootstrapping DBI}\label{sec:susyboot}

In the context of maximally supersymmetric Yang-Mills EFTs, Ref.~\cite{Berman:2024wyt} presented numerical evidence that fixing the coupling of the lightest massive scalar to the  string value resulted in the Veneziano amplitude being the only allowed 10-dimensional UV completion. In this section, we turn our attention to the Abelian DBI amplitude \eqref{eq:dbi} with the goal of cornering it in an analogous fashion to \cite{Berman:2024wyt}.

The supersymmetric DBI amplitude \eqref{eq:dbi} has a spin~0 and spin~2 with mass-squared $M_\text{gap}^2 = 1/\alpha'$. Their couplings to the massless external scalars~$\phi$ are, when normalized by $g_2 = \pi^2/2$,
\be
 \label{DBIcoup02}
 \text{DBI:}~~~
 \frac{|\lambda_0|^2}{g_2}
 =\frac{2(3D-2)}{\pi^2 (D-1)}
 ~~~~\text{and}~~~~
 \frac{|\lambda_2|^2}{g_2}
 =\frac{2(D-2)}{\pi^2 (D-1)}\,.
\ee
The spin 0 and 2 states at this mass are part of the same supermultiplet, so their couplings \reef{DBIcoup02} satisfy the supersymmetric relationship \reef{eq:susymgrel}.

We now proceed to bootstrap the DBI amplitud, first without imposing supersymmetry and afterwards with supersymmetry.

\subsection{Bootstrap Without SUSY}
\label{sec:DBInosusy}

\begin{figure}
    \centering
    \includegraphics[width=0.55\linewidth]{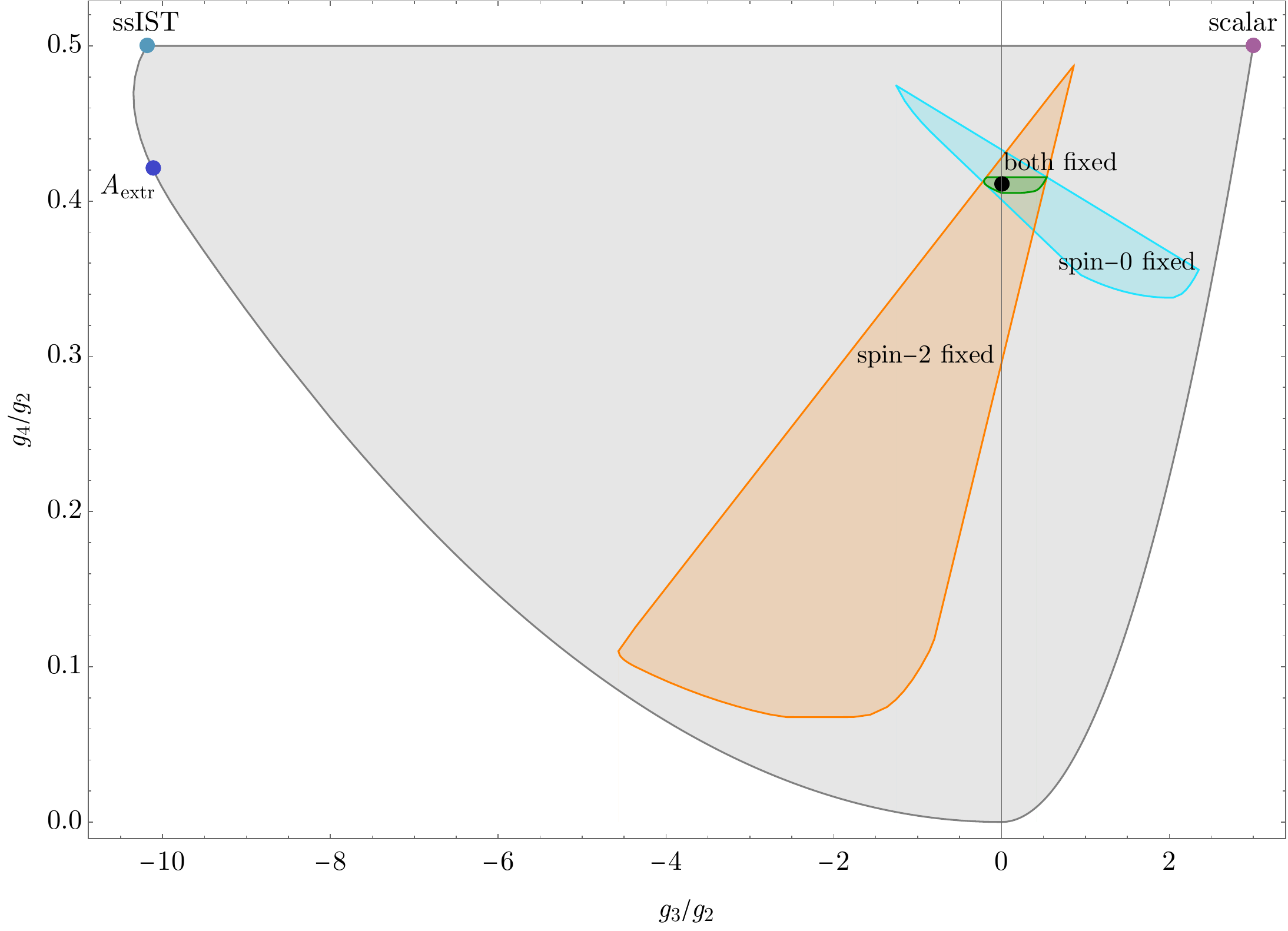}\hfill
    \includegraphics[width=0.39\linewidth]{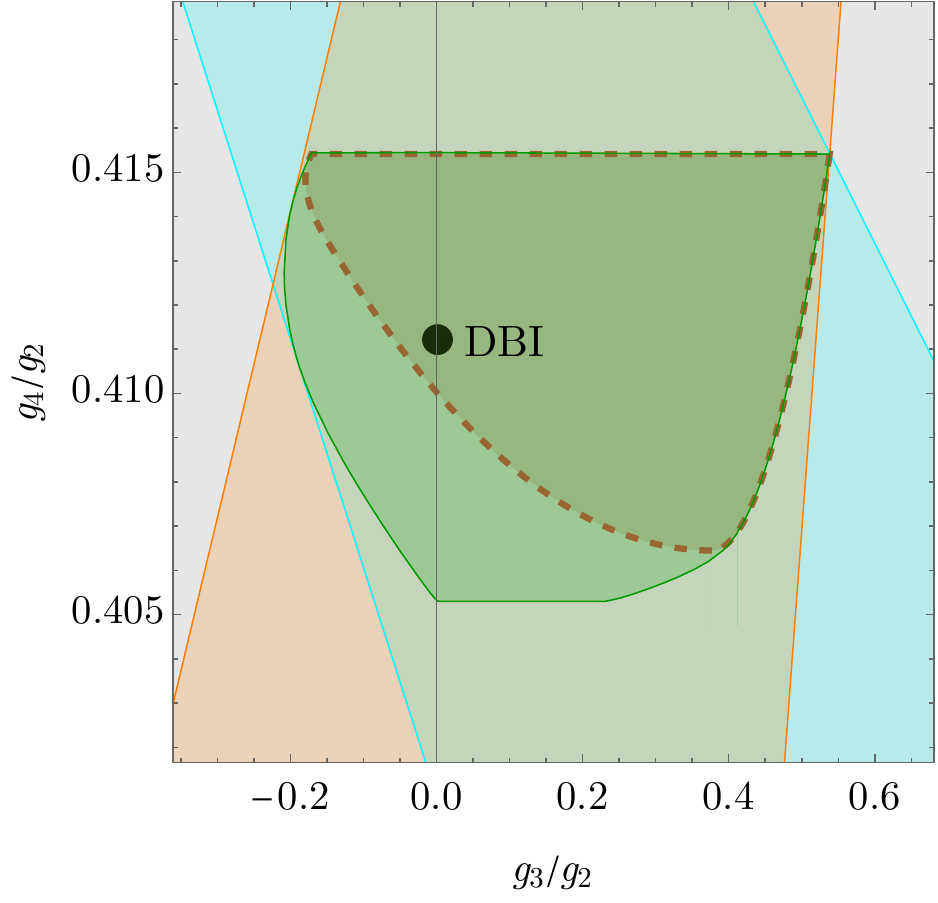}
    \caption{With spin-0 and spin-2 states at $M_\text{gap}^2$, the plot shows the allowed regions when either the spin-0 (cyan) or spin-2 (orange) couplings are fixed to their $D=4$ DBI values \reef{DBIcoup02}. The green island is obtained when \textit{both} the spin-0 and spin-2 couplings are fixed to their DBI values.  
   The dashed-brown curve within the green island denotes the boundary of the region which can be obtained by adding $A_\text{extr}$ and $A^{(0)}$ to any theory with \textit{no states} below $3M_\text{gap}^2$. These bounds are computed at $k_\text{max}=12$.}
    \label{fig:bothfixed}
\end{figure}

Motivated by the DBI spectrum, we assume that apart from the spin 0 and 2 states at $M_\text{gap}^2$, there are not other massive states below $3M_\text{gap}^2$.  
In the absence of supersymmetry, we can freely ignore the coupling relationship \reef{eq:susymgrel}. If we fix the spin 2 coupling to its DBI value in \reef{DBIcoup02}, while leaving the spin 0 coupling free, we obtain the orange region shown in Figure~\ref{fig:bothfixed}, which of course contains SUSY DBI, but is a quite large region. Oppositely, if we fix the spin 0 coupling to its DBI value \ref{DBIcoup02}, we obtain the somewhat smaller cyan region in Figure~\ref{fig:bothfixed}.\footnote{In Appendix~\ref{app:fixcoupling}, we provide a detailed discussion of the shapes of the two fixed coupling regions.
}  
DBI necessarily lies in the overlap of the fixed spin-0 and spin-2 regions of Figure~\ref{fig:bothfixed}. Fixing {\em both} the spin-0 and spin-2 couplings to the DBI values \reef{DBIcoup02} gives the green island in Figure \ref{fig:bothfixed}, which is strictly smaller than the intersection of the two single-fixed-coupling regions. The green island includes the SUSY DBI amplitude, but beyond $k_\text{max}=12$ it barely shrinks.

It is useful to better understand why the non-SUSY DBI island does not shrink with increasing $k_\text{max}$. 
Consider the space of theories that have no massive states below $3 M_\text{gap}^2$. The resulting allowed region has the same shape as the full region, but scaled down by a factor of 3 in the $g_3/g_2$-direction and by a factor of $3^2$ in the $g_4/g_2$-direction. Now take any theory in that region and add to it a linear combination of the pure scalar amplitude $A^{(0)}$ and the 2SDR extremal theory $A_\text{extr}$ with $\mu_c=3$ such that the spin-0 and spin-2 states at $M_\text{gap}^2$ in the resulting theory has couplings $|\lambda_0|^2/g_2$ and $|\lambda_2|^2/g_2$ equal to its DBI values \reef{DBIcoup02}. This procedure yields the dashed-brown ``reconstructed'' region in Figure \ref{fig:bothfixed}. The non-SUSY DBI island must always contain this ``reconstructed'' region, so the green region cannot shrink beyond the dashed-brown boundaries. Moreover, we could also have added an extremal theory with $\mu_c>3$ (at least up to some maximal $\mu_c$ value for which the couplings can still be fixed to \reef{DBIcoup02}) and we expect this to give the rest of the green island. 

It is not surprising that the DBI amplitude is not isolated by fixing the couplings of the two lowest massive states. The basic reason is that the DBI amplitude cannot be interpreted as ``extremal'' in the sense that it is not near any boundary of any of the allowed region nor do the spin-0 or spin-2 couplings maximize the bounds as in the case of the extremal theories discussed in Section \ref{sec:extr2sdr}. To do better, we have to impose supersymmetry in the bootstrap analysis.

\subsection{2SDR SUSY Bootstrap via Null Constraints}\label{sec:2sdrsusy_nulls}
\begin{figure}
    \centering
    \includegraphics[width=0.6\linewidth]{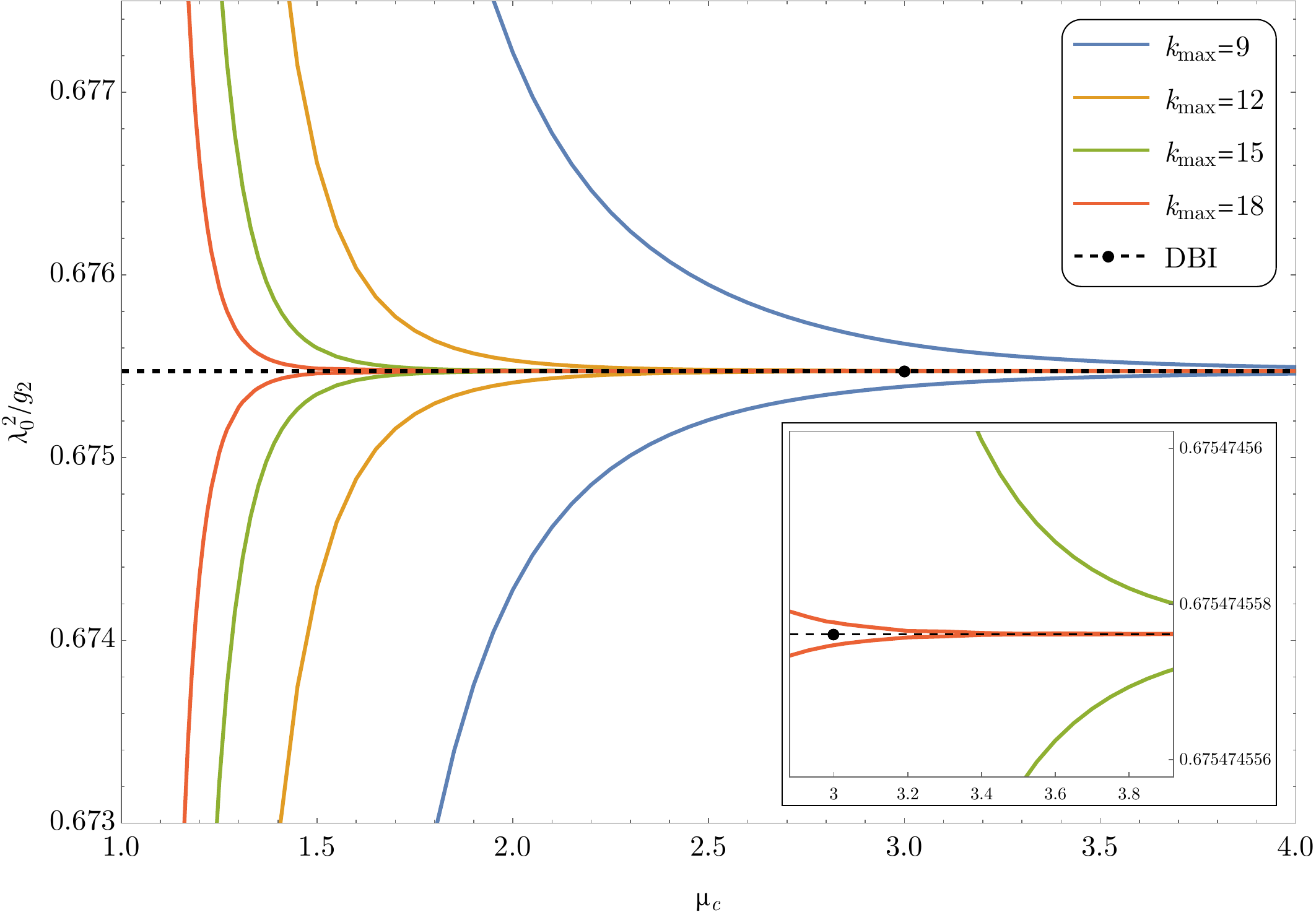}
    \caption{Minimum and maximum allowed spin-0 coupling at the gap when the spin-2 coupling at the gap is fixed to the DBI value. The computation is done in $D=4$ in the 2SDR bootstrap with SUSY imposed via the additional null constraints \reef{eq:susynulls}. As $k_\text{max}$ increases, the spin-0 coupling is forced to obey the SUSY constraint \reef{eq:susymgrel}, except at $\mu_c=1$. The black dot indicates DBI with $\mu_c=3$. 
    }
    \label{fig:2sdr_susyratio}
\end{figure}

In the 2SDR bootstrap, supersymmetry can be imposed with the additional low-energy null constraints $g_{3n;\,{0,n}}=0$ from \reef{eq:susynulls}  for $3n \le k_\text{max}$. It is not obvious if these constraints suffice to fix SUSY coupling relationships such as \reef{eq:susymgrel}. To examine this, 
consider again spin-0 and spin-2 states at $M_\text{gap}^2$, but now assume a gap $\mu_c M_\text{gap}^2$ up to the next possible state. We fix the spin-2 coupling $|\lambda_2|^2/g_2$ to the $D=4$ DBI value \reef{DBIcoup02} and compute the maximum and minimum spin-0 coupling $|\lambda_0|^2/g_2$ as a function of $\mu_c$ in the 2SDR bootstrap with the SUSY null constraints \reef{eq:susynulls} imposed. Figure \ref{fig:2sdr_susyratio}
shows that as $k_\text{max}$ increases, the range of allowed values of $|\lambda_0|^2/g_2$ narrows sharply to the value determined by \ref{fig:2sdr_susyratio}, especially for higher values of $\mu_c$. 
In particular, at the DBI-motivated gap $\mu_c = 3$, the bootstrapped value of $|\lambda_0|^2/g_2$ is for $k_\text{max} = 18$ within $\sim 3\cdot 10^{-10}$ of the DBI value.

Although the numerical agreement at high $k_\text{max}$ is excellent, some freedom remains at finite $k_\text{max}$, not only for the lowest mass states studied here, but also for the remainder of the spectrum above the cutoff. 
Thus, supersymmetry is not completely imposed by the null constraints \reef{eq:susynulls} in the 2SDR bootstrap. Instead, stronger SUSY bounds are obtained at finite $k_\text{max}$ if the supersymmetric coupling relationships are imposed exactly and this is the case in the 0SDR bootstrap. So for the purpose of bootstrapping a supersymmetric amplitude, such as DBI, we use the 0SDR bootstrap.

\subsection{0SDR Bootstrap of DBI}
\label{sec:0sdrsusy}

We have discussed both in Sections~\ref{sec:maxsusyconst} and~\ref{s:SUSYandOSDR} that an Abelian maximally supersymmetric 4-point amplitude that obeys Froissart with $n_0=2$ can have its overall factor $(s^2+t^2+u^2)$ stripped off to give a real scalar amplitude with $n_0=0$. If one amplitude obeys the positivity constraints, so does the other. 
The stripped amplitude can only have a scalar at $M_\text{gap}^2$, hence, as discussed in Section \ref{sec:susycouplings}, multiplication by $(s^2+t^2+u^2)$ immediately gives the correct SUSY relationship \reef{eq:susymgrel} between the scalar coupling and the spin-2 coupling for the 2SDR supersymmetric amplitude. Similar coupling relationships will be automatically obeyed for any other massive supermultiplets at any higher masses. In this sense, we expect the 0SDR bootstrap of the SUSY amplitudes to be strictly stronger at finite $k_{\max}$ than the 2SDR bootstrap with SUSY null constraints described in Section \ref{sec:2sdrsusy_nulls}. 

\begin{figure}
    \centering
\includegraphics[height=4.7cm]{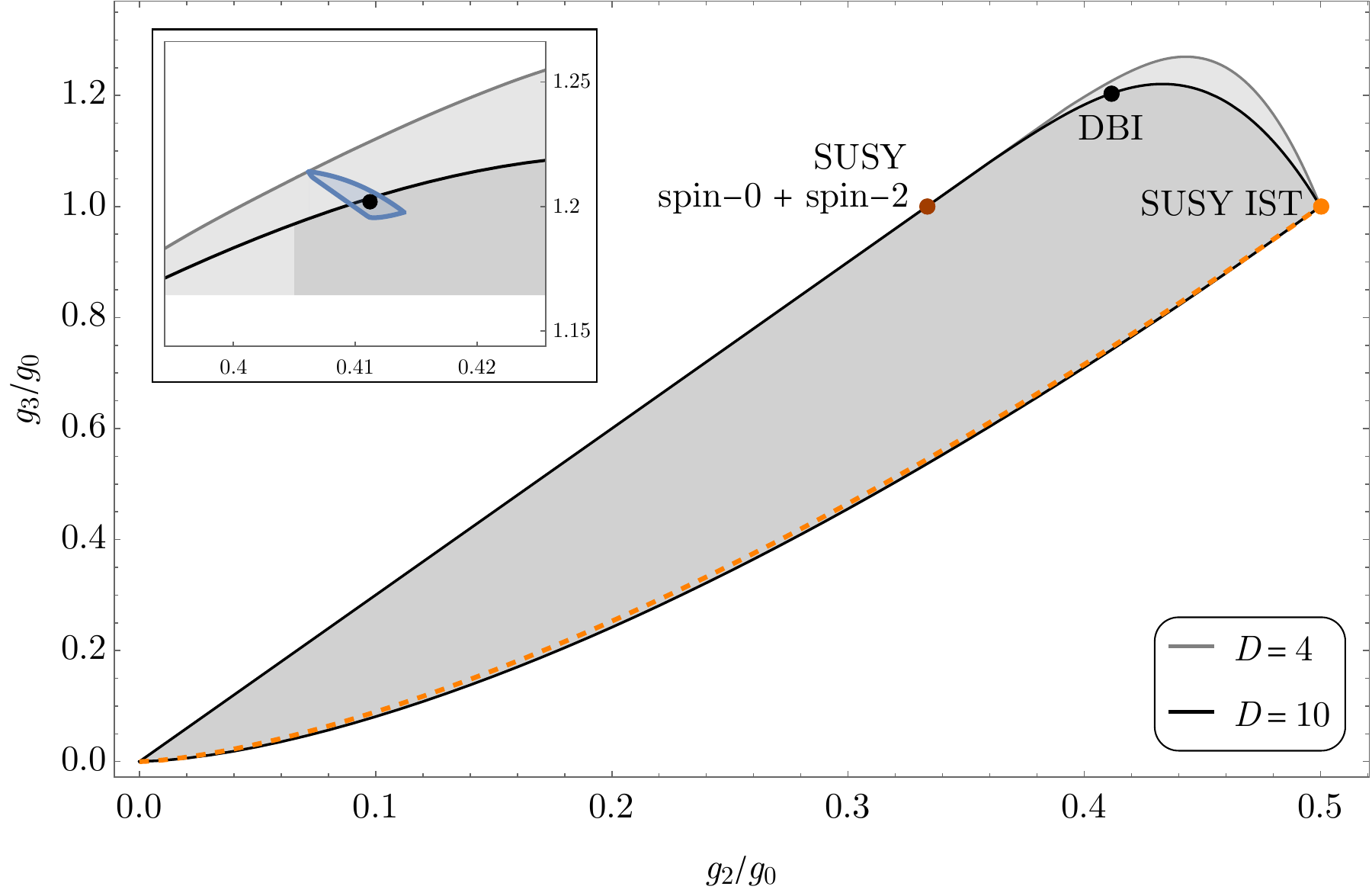}
\includegraphics[height=4.7cm]{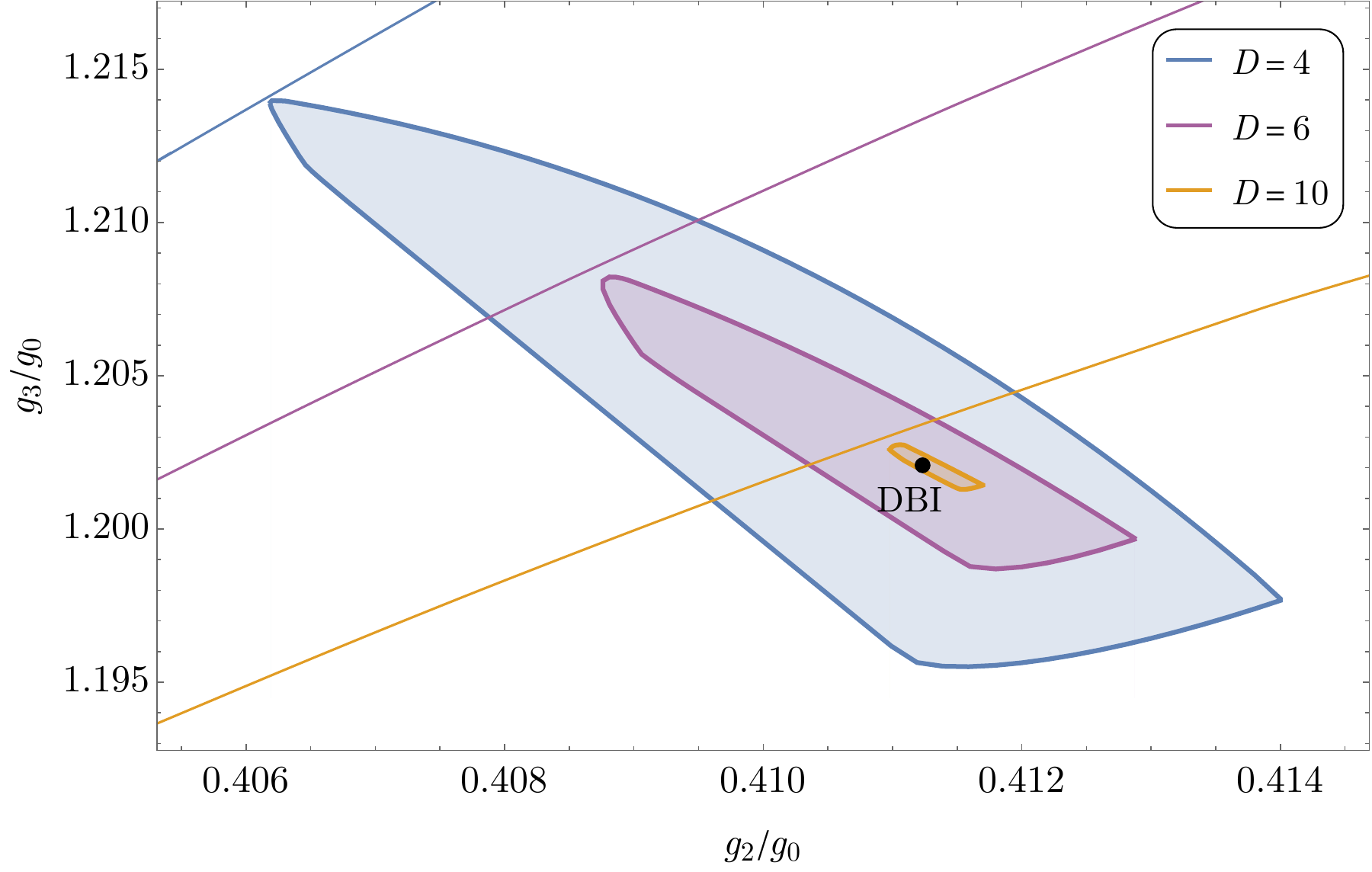}
    \caption{\textbf{Left.} The $(g_2/g_0,g_3/g_0)$ allowed region for 0SDR amplitudes in $D=4$ (light grey) and $D=10$ (dark gray) computed at $k_\text{max}=16$. 
The lower-boundary appears to converge to the mass-scaling curve (dashed orange) of the SUSY IST \reef{susyIST}. The supersymmetric spin-0 and spin-2 amplitude \reef{eq:A0susy} is shown as the maroon dot.  The $(s^2 + t^2 + u^2)$-stripped DBI amplitude (black dot) is very close to the general 0SDR boundary in its critical dimension, $D=10$. 
\textbf{Right.} With the scalar coupling at the gap fixed to the DBI value and the cutoff set to $\mu_c=3$, the allowed 0SDR region (shown for $k_\text{max}=16$) becomes a small island around DBI, shown in blue for $D=4$, orange for $D=6$, and green for $D=10$.
As $D$ gets closer to $10$ (the critical dimension of DBI), the island shrinks. 
    In $D=10$, the bounds continue shrinking slowly as $k_\text{max}$ increases, while in $D<10$ they have stopped visibly shrinking by $k_\text{max}=16$. This may suggests that DBI could be the only theory in the $D=10$ island as $k_\text{max}\to\infty$.}
    \label{fig:g4g5}
\end{figure}

The lowest-dimension Wilson coefficients that can be bounded in the 0SDR bootstrap are $g_2/g_0$ and $g_3/g_0$. For the ``stripped'' DBI amplitude, we have 
\be
 g_0 = \frac{\pi^2}{2}\,,~~~~
 g_2 = \frac{\pi^4}{48}\,,~~~~ 
 g_3 = \frac{\pi^2 \zeta_3}{2}\,,~~~~ \dots
\ee
so that
\be 
  \label{DBIin0SDR}
  \frac{g_2}{g_0} = \frac{\pi^2}{24} \approx 0.41123\,,~~~~
  \frac{g_3}{g_0} = \zeta_3 \approx 1.20206\,.
\ee
The general allowed region of the $(g_2/g_0,g_3/g_0)$-plane  is shown on the left of Figure \ref{fig:g4g5} for $D=4$ and $D=10$. The $D=10$ bounds are stronger and get very close to DBI (black dot).\footnote{In $D>10$, we expect the numerical bounds to rule out DBI for sufficiently large $k_\text{max}$ since we know this amplitude is non-unitary in $D>10$.}

Fixing the coupling of the spin-0 state at $M_\text{gap}^2$ to the DBI value, $|\lambda_0|^2/g_0 = 2/\pi^2$, and setting the cutoff to the DBI-motivated value $\mu_c=3$, we find the allowed region to be a small island around DBI. Figure \ref{fig:g4g5} shows the DBI island in blue for $D=4$, purple for $D=6$, and orange for $D=10$. 
The dimensional dependence makes it clear that the bounds around DBI are significantly tighter for $D=10$ than in lower dimensions. This is closely related to the fact that in $D=10$, DBI lies close to the boundary shown on the left-hand side of Figure \ref{fig:g4g5}. Indeed, being (near-)extremal is key for being able to bootstrap the model well.\footnote{Based on the zoom-in on Figure \ref{fig:g4g5}, it seems likely that the islands converge faster in $k_\text{max}$ than the general bounds do; this accounts for the non-zero separation between the general upper bound and the upper-left corner of the island in each $D$.} 

As noted in Section \ref{sec:dbi}, the DBI amplitude is unitary only for $D\le 10$. In the critical dimension $D=10$, the stripped DBI amplitude is infinitesimally close to violating the bootstrap unitarity assumptions. Hence, we can expect the finite $k_\text{max}$ bounds to be tighter in the critical dimension. As $k_\text{max} \to \infty$, we may expect a critical amplitude to be right on the boundary in the full-dimensional geometry. 

The $D=10$ DBI island shrinks only very slowly with increasing $k_\text{max}$, so while the smallness of the island suggests that DBI is likely the unique UV completion, it is  harder to reach this conclusion  based on the numerics than it was in the bootstrap of the Veneziano amplitude studied in \cite{Berman:2024wyt}. One can try to make additional assumptions to shrink the DBI island further. Assuming that the leading Regge trajectory for higher mass states is at most linear with slope 1 hardly changes the bounds. Adding in the next set of states at $3M_\text{gap}^2$ and fixing their couplings to the known DBI values while increasing the cutoff to $\mu_c=5$ makes the bounds quite a bit tighter, but it does not appear to make them shrink faster with increasing $k_\text{max}$.

\begin{figure}[t]
    \centering
    \begin{subfigure}{0.48\textwidth}
        \includegraphics[width=70mm]{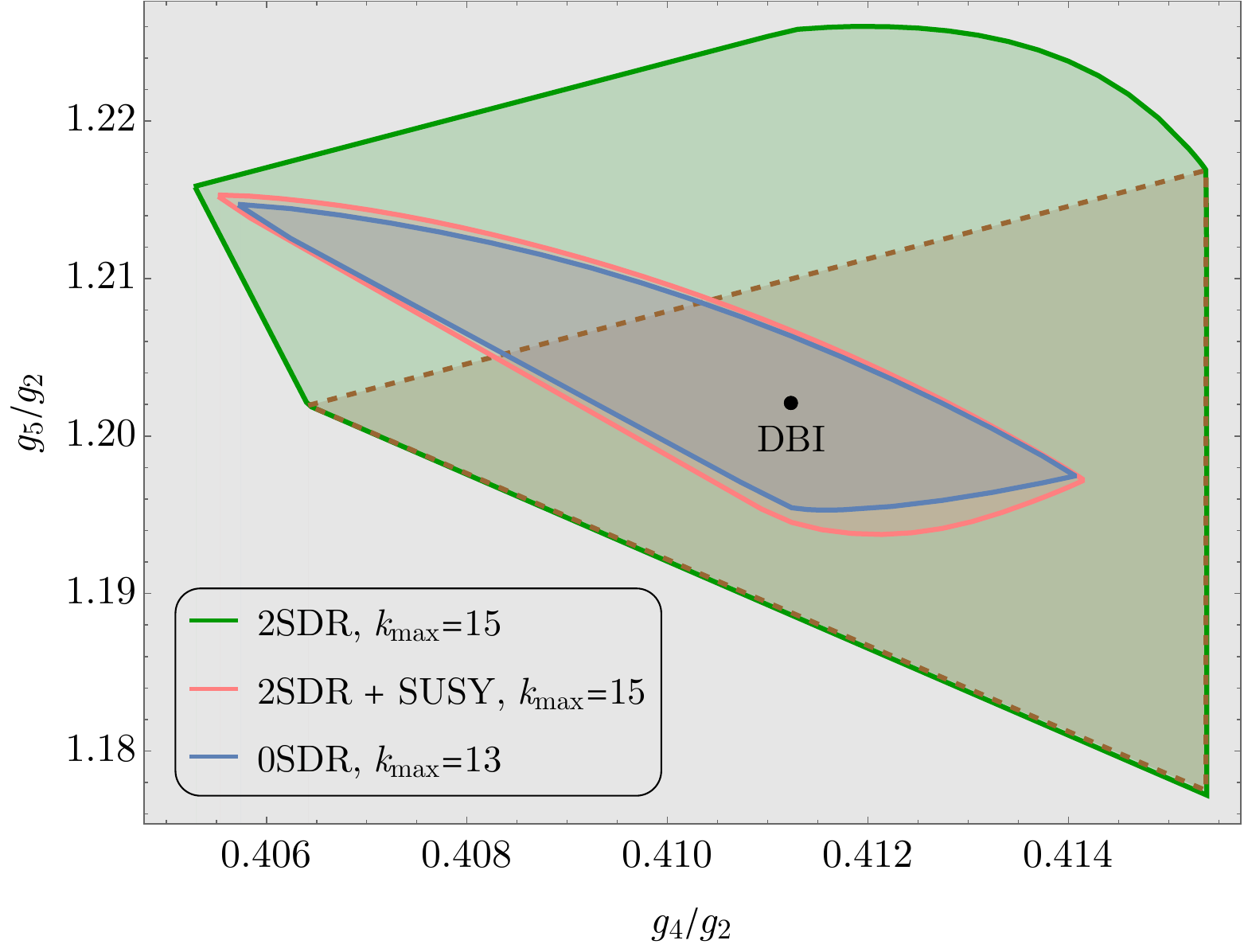}
    \end{subfigure}
    \hfill 
    \begin{subfigure}{0.48\textwidth}
        \includegraphics[width=70mm]{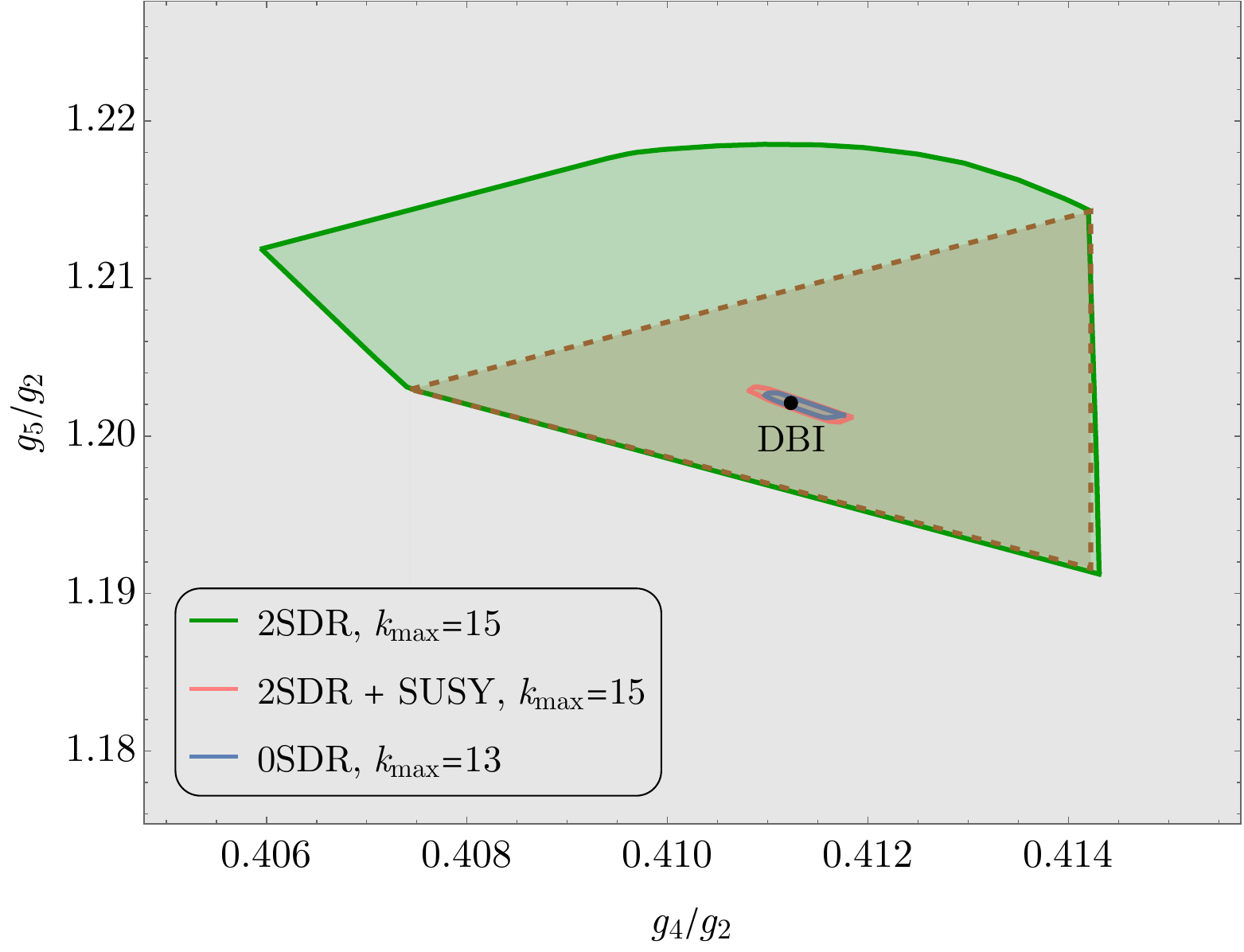}
    \end{subfigure}
    \caption{DBI islands in the $(g_4/g_2,g_5/g_2)$-plane  obtained by fixing the couplings of states at $M_\text{gap}^2$ to their DBI values in three different bootstrap approaches in $D=4$ (left) and $D=10$ (right). The brown subregion can be constructed in parallel with that in Figure \ref{fig:bothfixed} and that explains why the non-SUSY island cannot shrink.} 
\label{fig:islands_susyvsnonsusy}
\end{figure}

Finally, let us compare the scale of three islands around DBI obtained by fixing the lowest massive state coupling(s) in three different approaches: 
first, the island obtained from the non-SUSY 2SDR bootstrap (Section \ref{sec:DBInosusy}); second, the island obtained from the 2SDR bootstrap with SUSY imposed via null constraints (Section \ref{sec:2sdrsusy_nulls}); and third, the island obtained from the 0SDR supersymmetric bootstrap (Section~\ref{sec:0sdrsusy}). Note that for the 0SDR bootstrap, the bounds are placed on $g_2/g_0$ and $g_3/g_0$ of the stripped amplitude, but restoring the SUSY factor  $(s^2+t^2+u^2)$ changes these coefficients to 
$g_4/g_2$ and $g_5/g_2$ in the 2SDR boostrap. Hence, the 2SDR coefficients are used as the axis labels in Figure \ref{fig:islands_susyvsnonsusy}. (For this same reason we also shift $k_\text{max}$ by 2 to appropriately compare the 2SDR and 0SDR {\em supersymmetric} bootstrap results.) The two supersymmetric islands are clearly smaller than the non-SUSY islands, and we expect the two distinct approaches to imposing supersymmetry to yield the same bounds in the $k_\text{max} \to \infty$ limit.

\section{Bootstrapping the Closed Superstring}\label{sec:gravamps}

In Section \ref{sec:exampleamps} we showed that a general real-scalar 4-point amplitude in a theory with $\mcN=8$ supersymmetry can be written as
\begin{equation}\label{eq:sugra}
    A^{\mathcal{N}=8}(s,u)=(s^2+t^2+u^2)^2 \, f(s,t,u)\;,
\end{equation}
where $f(s,t,u)$ is fully permutation-symmetric. A key example in this class of theories is the closed superstring four-dilaton amplitude. In that case,  the function $f$ is the Virasoro-Shapiro amplitude, 
\be
\label{eq:VSsec7}
f^\text{VS}(s,t,u)=
A^\text{VS}(s,t,u)=-\alpha'^2\frac{\Gamma(-\alpha's)\Gamma(-\alpha't)\Gamma(-\alpha'u)}{\Gamma(1+\alpha's)\Gamma(1+\alpha't)\Gamma(1+\alpha'u)}
    \, .
\ee

Let us assume that $A^{\mathcal{N}=8}$ satisfies the standard Froissart bound  with $n_0=2$. 
In the derivation of the dispersion relations, we learned in \reef{stpolecontrib} that the graviton pole generates a $(1/u)$-term which obstructs the derivation of the dispersive representation of the Wilson coefficients $a_{k,q}$ with  $k-q=2$. 

Let us instead consider the 4-point amplitude without the $(s^2+t^2+u^2)^2$-factor. This stripped amplitude satisfies the $n_0 = -2$ Froissart bound, 
\begin{equation}\label{eq:VSfroissartsec7}
    \lim_{\substack{ |s| \to \infty \\ \text{fixed } u<0 }}s^2 f(s,t,u)=0\;.
\end{equation}
This suggests that the stripped amplitude can be bootstrapped with ``$(-2)$-times subtracted'' dispersion relations. However, in this approach one again picks up a $(1/u)$-term from the graviton pole in the dispersion relations for certain Wilson coefficients. 
In \cite{Caron-Huot:2021rmr,Albert:2024yap}, a ``smearing'' procedure was used to get around the $1/u$ issue and thereby bootstrap SUGRA amplitudes in the $(-2)$SDR bootstrap. Instead of performing this smearing procedure, we simply avoid the $(1/u)$-problem by resorting to the familiar 0SDR bootstrap, where the graviton pole is not a problem.

\begin{figure}
    \centering
    \includegraphics[width=0.49\linewidth]{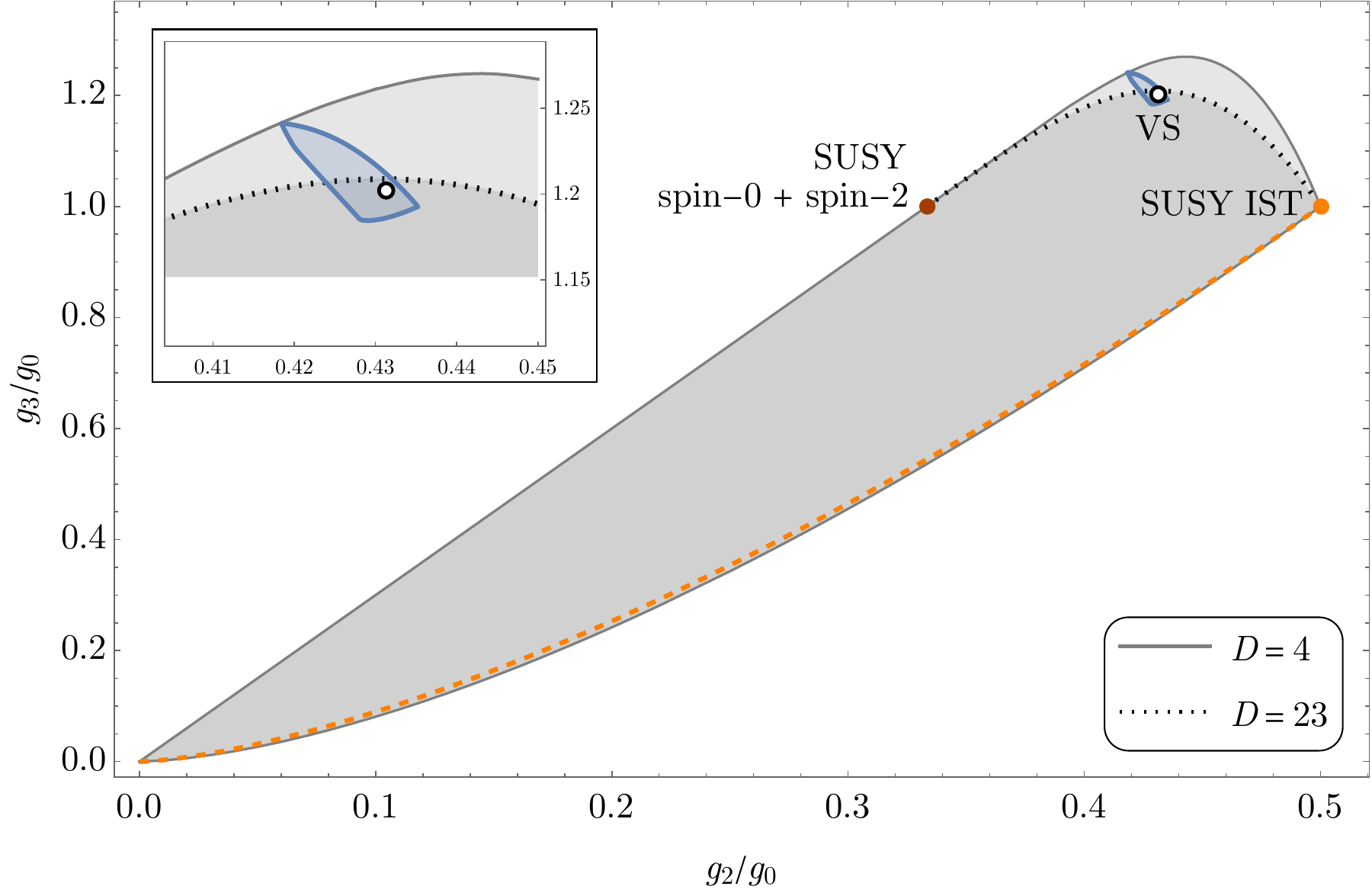}
    \includegraphics[width=.49\linewidth]{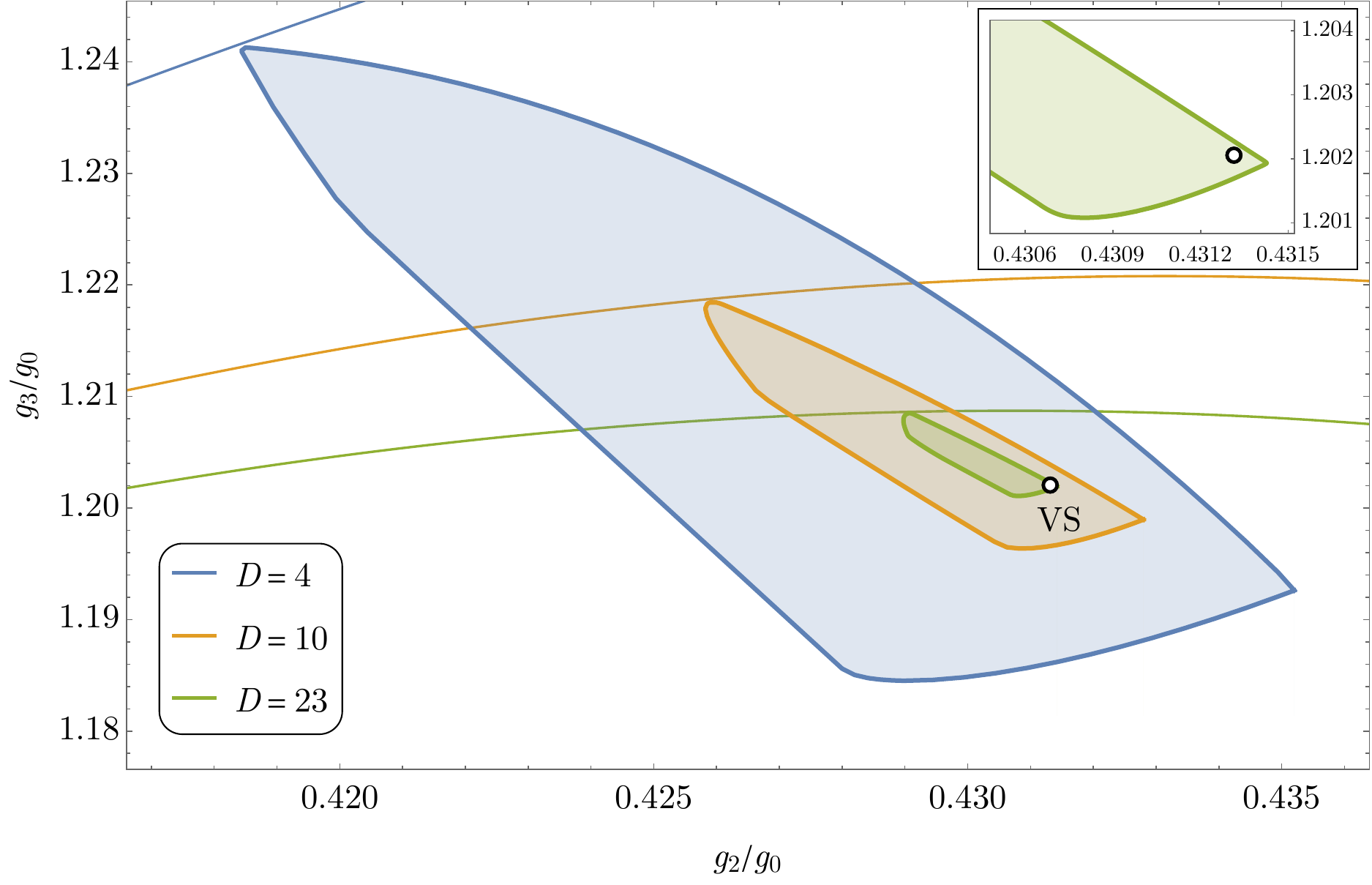}
    \caption{$D=4,10,23$ islands around Virasoro-Shapiro (VS) obtained by fixing the coupling of the scalar at the gap and setting the cutoff to be $\mu_c=2$ in the 0SDR bootstrap with $k_\text{max}=16$. In $D=23$, Virasoro-Shapiro sits close to a corner in the island. The island does not shrink more than $\mcO(10^{-3})$ from $k_\text{max}=10$ to 16. 
    We also show the upper bounds on the general 0SDR space in each~$D$.}
    \label{fig:VSisland}
\end{figure}

We proceed in parallel to the   0SDR bootstrap of DBI. Virasoro-Shapiro has Wilson coefficients
\be
  g_0 = 2 \zeta_3\,,~~~~
  g_2 = \zeta_5\,,~~~~
  g_3 = 2\zeta_3^2\,,~~~~\dots
\ee
which gives 
\be 
  \label{VSin0SDR}
  \frac{g_2}{g_0} = \frac{\zeta_5}{2\zeta_3} \approx 0.43131\,,~~~~
  \frac{g_3}{g_0} = \zeta_3 \approx 1.20206\,.
\ee
Note that the value of $g_3/g_0$ is exactly the same as for the stripped DBI amplitude \reef{DBIin0SDR}, but the values of $g_2/g_0$ differ slightly. 

The VS amplitude has a spin-0 state at the mass gap $M_\text{gap}^2 = 1/\alpha'$, the (normalized) coupling-squared of this state is
\begin{equation}
\label{VScoup}
    \frac{|\lambda_0|^2}{g_0}=\frac{1}{2\zeta_3}\;.
\end{equation}
There are no other states in the spectrum until $2M_\text{gap}^2= 2/\alpha'$. As noted in Section \ref{sec:exclosed}, the Virasoro-Shapiro amplitude has a peculiar  critical dimension, $D=23$, in which the spin-4 coupling at mass-level 4 vanishes. This fact motivates us to compute the $|\lambda_0|^2$-fixed island in $D=23$ as well as in the physically and string-theoretically motivated dimensions $D=4$ and $D=10$. Figure \ref{fig:VSisland} shows the islands for $D=4,10,23$ obtained by fixing $|\lambda_0|^2/g_0$ to the VS value \reef{VScoup} and setting the cutoff to $\mu_c=2$. The relevance of the critical dimension is clear, as the $D=23$ island is significantly smaller than the islands in sub-critical dimensions. Moreover, in  $D=23$, VS sits very close to the lower-right corner of the island, but the bounds do not shrink much in $k_\text{max}$.

\section{Discussion}\label{sec:disc}

In this paper we have examined various aspects of the S-matrix bootstrap of scattering amplitudes with four identitical scalars. This is essentially the simplest possible class of EFTs, yet it still contains non-trivial UV completions, such as the DBI amplitude and the scalar  Virasoro-Shapiro amplitude. We now summarize and discuss the results of the paper.

\vspace{2mm}
\noindent {\bf Convex Hull Conjecture.} 
We presented evidence that the allowed region of Wilson coefficients for unitary 4-point amplitudes obeying the standard Froissart bound is
spanned by the simple massive scalar amplitude and a one-parameter family of ``extremal'' amplitudes. Further, we proposed that the extremal amplitudes may be directly related to the extremal amplitudes with maximal supersymmetry.

If true, the Convex Hull Conjecture implies that the computationally intensive problem of bounding scalar EFTs reduces to understanding the convex hull of a single parameter family of theories. If it were possible to analytically construct the extremal amplitudes,  we would have a simple, closed-form way to map the entire space of UV complete theories. 
In other words, we would have a basis of UV complete amplitudes of which any amplitude satisfying the bootstrap assumptions could be written as a positive sum, and from that basis we could generate the entire space ``primally''. This highlights the practical importance of the secondary conjecture that the extremal curve is reconstructible from maximally supersymmetric amplitudes: these are the amplitudes which appear to be analytically tractable due to their possible relationship to string theory amplitudes \cite{Cheung:2023uwn,Cheung:2024uhn,Cheung:2024obl}. From a physical perspective, this conjecture is also highly relevant as it suggests that \textit{at 4 point, all amplitudes consistent with our physically-motivated bootstrap assumptions are constructible from maximally supersymmetric amplitudes.}

\vspace{2mm}
\noindent {\bf Extremal Theories: Spectrum, Single Regges, and Missing Daughters.} 
Let us summarize what we have learned about the extremal theories, focusing on the 0SDR case. We can think of these as maximally supersymmetric extremal theories with the $(s^2+t^2+u^2)$-factor stripped off. They have as the lowest massive state a scalar whose coupling to the massless external scalar $\phi$ is maximized for given gap to the next possible massive state at $\mu_2 M_\text{gap}^2$. We found in Section \ref{sec:extr0sdr} that the corresponding unique values of the Wilson coefficients were unaffected by whether or not scalars were allowed in the spectrum at and above $\mu_2 M_\text{gap}^2$. 

Next, in Section \ref{sec:bifrost}, we studied having  a scalar at $M_\text{gap}^2$, a spin 2 state at $\mu_2 M_\text{gap}^2$, and then a gap to the next possible state at or above $\mu_c M_\text{gap}^2$. With no couplings fixed, we found that max($g_2/g_0$) had a plateau for gaps in the range $\mu_2 \le \mu_c \lesssim \mu_c^\text{corner}(\mu_2)$. The value of the plateau and the location of the corner turns out to be {\em independent} of whether a spin 0 state is included at $\mu_2 M_\text{gap}^2$ and whether   spin 0 or spin 2 are included  above 
$\mu_c M_\text{gap}^2$. All that matters is that we have the spin 0 state at $M_\text{gap}^2$, the spin 2 at $\mu_2 M_\text{gap}^2$ and that we allow spin 4 and higher starting at $\mu_c M_\text{gap}^2$. This means that if the extremal amplitudes have daughter trajectories, they cannot include spin 0 or 2 states.

\begin{figure}
    \centering
\includegraphics[width=0.5\linewidth]{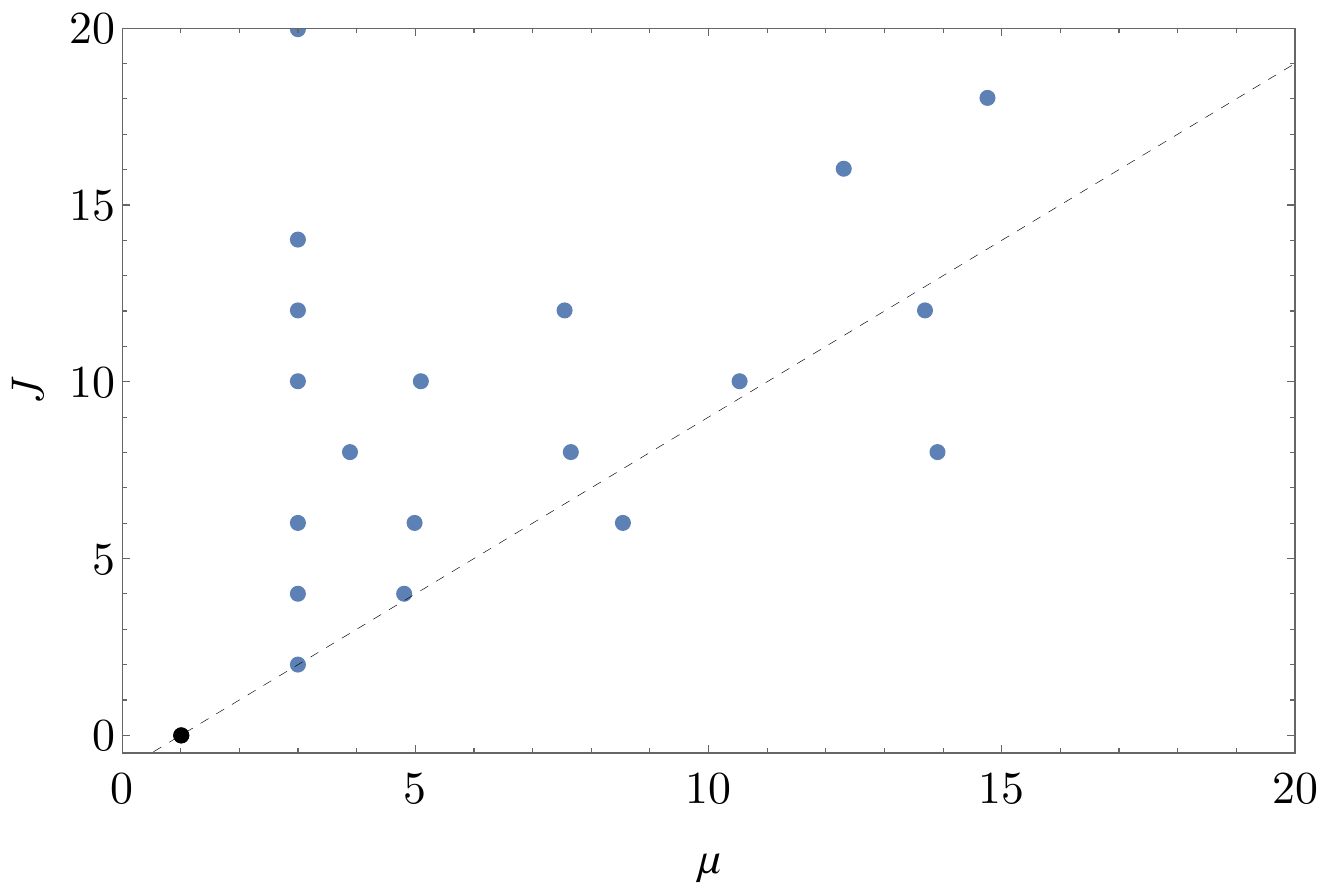}
    
    \caption{The $k_\text{max}=14$  spectrum of the 0SDR extremal theory with $\mu_c = 3$ where $\mu = M^2/M_{\gap}^2$ for states at mass $M^2$. The black dot in the lower left corner is the scalar that is inserted at $\mu = 1$ and the black dashed line shows the stringy ``linear'' spectrum.}
    \label{fig:spectrum}
\end{figure}

To examine the extremal theories further, we use SDPB to extract the  numerical spectrum of the amplitude that saturates the extremal bound \cite{Simmons-Duffin:2016wlq}. 
In Figure \ref{fig:spectrum}, we give an example of such a spectrum for the 0SDR extremal theory with $\mu_c = 3$ in $D=4$. In agreement with our previous discussion, it shows that there is indeed only one scalar in the spectrum, namely at $M_\text{gap}^2$ and only one spin 2 state, namely with mass-squared $3M_\text{gap}^2$. Considering $J=4$, the spectrum contains two states, one near $3M_\text{gap}^2$ and a second near $5M_\text{gap}^2$. The former, along with the spin tower also at $3M_\text{gap}^2$,  may be presumed to spurious: we showed in Section \ref{sec:maxspin} that the maximal spin at the second mass-state is $J=2$ in the 0SDR bootstrap; the fact that higher-spin states show up in the spectrum is a finite-$k_\text{max}$ effect and their couplings should be exponentially suppressed with increasing $k_\text{max}$.
(Similar spurious states have also been observed in previous work, e.g.~\cite{Albert:2023seb,Chiang:2023quf,Berman:2024wyt,Albert:2024yap}.)
However, the important observation for the discussion here is that the spectrum in Figure \ref{fig:spectrum} has a \textit{maximum} mass for states of a particular spin. This is in direct contrast to the infinite set of daughter trajectories in string  amplitudes; they have only a minimal mass at which states with particular spin can appear. 
For example, they have an infinite tower of $J=0$ states, $J=2$ states, etc.  

Based on the above observations and the discussion of the numerical spectrum, one may suspect that the numerical extremal theories do not have daughter trajectories, but that,  at finite $k_\text{max}$, SDPB is able to optimize the bounds by adding states along a single Regge trajectory as well as including spurious states with very small couplings. 
Single Regge trajectory theories have, however, been argued against in other works. 
The authors of  \cite{Eckner:2024pqt} showed that a subclass of $su$-symmetric amplitudes were required to have infinitely many Regge trajectories. This argument is expected to extend to the fully $stu$-symmetric case as well  \cite{Albert:2024yap}. An amplitude would be required to have states with finite spin at arbitrarily large masses, which directly contradicts the observation that there is a maximal mass for each spin in the SDPB spectrum. Therefore, it is possible that an infinite set of Regge trajectories would appear in these extremal spectra in the  $k_{\max} \to \infty$ limit, when states of large spin at low mass are no longer allowed.

The authors of \cite{Eckner:2024pqt} pointed out that subleading ``daughter'' Regge trajectories can be infinitely heavy, meaning they would have arbitrarily small contributions to the Wilson coefficients. If such an amplitude were the extremal one, then we would expect the SDPB spectrum to have only a single leading Regge trajectory, as was observed in \cite{Albert:2024yap}. A single Regge trajectory amplitude might then be allowed in some ``marginal'' sense similar to the spin-2 amplitude in the 2SDR bootstrap or scalar amplitude in the 0SDR bootstrap. Searching for a permutation symmetric single Regge trajectory amplitudes is, therefore, an intriguing avenue for analytic studies of these extremal amplitudes.\footnote{We thank Sasha Zhiboedov for discussions related to this point.} Such an amplitude would take the form
\be
    A(s,t,u) = \sum_{j \text{ even}}^{\infty} A^{(j)}(s,t,u)
\ee
for $A^{(j)}(s,t,u)$ defined as \reef{eq:AL} with $G_2\to G_j$: 
\begin{align}
\label{eq:AJ}
    A^{(j)}(s,t,u)
    =
    - \frac{1}{2} |\lambda_j|^2 \,
    \bigg(
    &
    \frac{M_j^2}{\makebox[\widthof{$u$}][c]{$s$}-M_j^2}
    \Big[
      G^{(D)}_j \big( 1+\tfrac{2t}{M_J^2} \big)
    + G^{(D)}_j \big( 1+\tfrac{2u}{M_J^2} \big)
    \Big]
    + \text{perms}
    \bigg)
    \,.
\end{align}
The masses and couplings would then need to be chosen such that the amplitude resums to obey the Froissart bound with $n_0 = 0$. Alternatively, we can assume resummation and use \reef{eq:akqsdr} to write the Wilson coefficients $a_{k,q}$ of this single Regge trajectory amplitude. These coefficients are parameterized by the couplings $|\lambda_j|^2$ and the masses $\mu_j$. With expressions for each coefficient, we can then by hand impose crossing symmetry and, in principle, solve for all the masses and couplings to see if there is a viable amplitude with only a single Regge trajectory. 

We do not know a way to solve this problem in general. In the specific case $\mu_J = 1$ for all $J$, there is a unique solution (up to rescaling), corresponding to the IST amplitude. Outside of this case, we have additionally tested amplitudes with one linear Regge trajectory,
\be  
  M^2_J = \gamma J+1 \,.
\ee
With this spectrum, we truncate the problem in $J$ and $k_{\max}$ and explicitly solve for all couplings $|\lambda_{J}|^2$ in terms of $\g$ and $|\lambda_0|^2$, which represents the overall scale in the problem. Upon doing so, however, we find that there is always some $|\lambda_{J}|^2$ which is required to be negative for crossing to be satisfied. In this sense, amplitudes with single \textit{linear} Regge trajectories appear to be incompatible with either crossing symmetry or unitarity, so they could not appear marginally in our bootstrap. If such amplitudes are truly on the extremal boundary of the $k_{\max}\to \infty$ region, then, they would likely be required to have nonlinear Regge trajectories.

Without any candidate other than the IST for a single Regge trajectory amplitude, though, it is still not clear whether we should expect them to be viable in the large $k_{\max}$ limit. 
Rather, one may suspect that the `true' $k_\text{max} \to \infty$ extremal theories have daughter trajectories. Given how close the 0SDR DBI amplitude is to the $D=10$ extremal bounds (e.g.~Figures \ref{fig:gcamps} (left) and \ref{fig:g4g5}), it is  tempting to think of them as part of the one-parameter family of extremal theories; specifically, it would be the case with $\mu_c=3$.
This motivated the study of deformed string amplitudes as analytic candidates for the extremal theories in Section \ref{sec:deformedamps}.  Specifically, we examined the hypergeometric deformations of the (Abelianized) Veneziano and Virasoro-Shapiro amplitudes of 
\cite{Cheung:2024obl}.  
We found that to match the extremal theories, a larger family of deformed amplitudes would be needed with at least one additional parameter that could allow one to fix the critical dimension while varying the slope of the leading Regge trajectory. One would presumably also need to generalize the deformations to include non-linear Regge trajectories.

\vspace{2mm}
\noindent {\bf Islands.} 
Supersymmetric DBI is a known analytic amplitude that satisfies all of our assumptions. We found that it lies very close to the boundary of the allowed regions in its critical dimension. Specifically, the DBI amplitude with the $(s^2 + t^2 + u^2)$ factor stripped off is very close to the  boundary of the general 0SDR allowed region in the critical dimension $D=10$ whereas in lower dimensions it is further from the bound. This makes sense since in the critical dimension, the amplitude is borderline from violating the unitary assumption, so it is natural that it becomes extreme with respect to the bounds. The property of being extremal is closely related to how well we can bootstrap the amplitude by specifying its lowest massive state and its coupling, as illustrated in Section \ref{sec:susyboot}.

When we allow the gravitational coupling to be non-zero, the closed superstring Virasoro-Shapiro (VS) amplitude becomes a target of the bootstrap. Similarly, we find that the VS amplitude is close the general 0SDR bound in the formal critical dimension ($D=23$), while for lower dimensions it is further in the interior of the allowed region. Again, extremality is key for obtaining a small island when the  spins and couplings of the lowest massive states are given as input. However, unlike the case of DBI, we do not have optimal bounds on the Virasoro-Shapiro amplitude because we are only using the zero-times-subtracted dispersion relations even though the amplitude satisfies the Froissart bound with $n_0=-2$, not just $n_0=0$. The graviton pole obstructs the simple version of $(-2)$SDR boostrap via divergent $1/u$ terms similar to those discussed around \reef{stpolecontrib}. This could in principle be dealt with using smearing as done in \cite{Caron-Huot:2021rmr,Albert:2024yap}.

\vspace{2mm}
\noindent {\bf Graviton Pole.} 
Recent work has pointed out that the appearance of the graviton pole in the twice-subtracted dispersion relations can actually tell us physical information about gravitational theories \cite{Hillman:2024ouy,Haring:2024wyz}. Such poles have also been explicitly included in the numerical bootstrap by studying smeared dispersion relations \cite{Caron-Huot:2021rmr,Albert:2024yap}. The authors of \cite{Albert:2024yap} in particular studied a similar kind of spectral input to that discussed here in the $(-2)$SDR bootstrap of $\mcN = 8$ SUGRA theories with the kinematic factor stripped off. It is unclear, though, if there is a more ``fundamental'' set of extremal theories in the $(-2)$SDR bootstrap that would be related to the 0SDR family we studied in Section \ref{sec:bifrost}. That bootstrap with smearing is qualitatively different from the bootstraps we have studied in this paper due to the fact that the $(-2)$SDR dispersion relations access the gravitational coupling $\kappa$. In the $(-2)$SDR bootstrap, the IST amplitude is marginal in the same way the scalar amplitude is marginal in the 0SDR bootstrap or the spin-2 amplitude is marginal in the 2SDR bootstrap. However, there is no marginal single-exchange amplitude, so there is only one theory that would naively be a part of the extremal set. If a family of $(-2)$SDR-extremal amplitudes existed, it could  relate the gravitational amplitudes like Virasoro-Shapiro to the non-gravitational 0SDR extremal theories including possibly DBI. 

\vspace{2mm}
\noindent {\bf Abelian Versus Non-Abelian.} 
Non-Abelian analogues of the corner theories described in Section \ref{sec:bifrost} were found in \cite{Berman:2024wyt}. The spectrum of the non-abelian corner theories was:  
 spin 0 at $M_\text{gap}^2$, spin 1 at $\mu_2' M_\text{gap}^2$, spin 2 at $\mu_3' M_\text{gap}^2$ etc. When these amplitudes are Abelianized, the odd-spin states are lost and we end up with a spectrum with spin 0 at  $M_\text{gap}^2$,  spin 2 at $\mu_3' M_\text{gap}^2$ etc. Thus the 3rd mass level in the non-Abelian extremal amplitudes matches the 2nd mass-level of the Abelian ones. It is 
 natural to expect the 0SDR extremal theories to be simply the Abelianization of the 
 non-Abelian corner theories in \cite{Berman:2024wyt}. 
Indeed, initial tests at low $\mu_c$ show strong agreement between the Abelianized corner theories and the lowest 0SDR extremal theory coefficient $g_2/g_0$, but we have not done a systematic analysis. 

It is an interesting question if analogous corner theories exist in a bootstrap of photon or graviton scattering. In these cases, there are multiple spectral densities and low-energy expansions associated with different helicities of the external particles, so it would be quite surprising if these different spectral densities and Wilson coefficients were all extremized by a one-parameter family of corner theories at the same time.

The procedure of reconstructing extremal theories in bootstraps with higher levels of subtraction could also be an important feature of other systems. For example, the 0SDR $su$-symmetric bootstrap of \cite{Berman:2023jys,Berman:2024wyt} describes color-ordered scalar amplitudes in maximally supersymmetric Yang-Mills theory. It was pointed out in \cite{Berman:2024wyt} that the crossing-symmetry null constraints that apply to this problem are the same as those used in the bootstrap of large-$N$ pion amplitudes \cite{Albert:2022oes}, except the pion bootstrap uses singly subtracted dispersion relations (1SDR). Colored scalars in SYM and large $N$ pions are two very different EFTs, so that makes it interesting to ask whether the same type of scalar-subtraction construction relates the extremal theories of the two bootstraps. More precisely, one could check whether the 0SDR $su$-symmetric bootstrap could be used to reconstruct the 1SDR $su$-symmetric bootstrap bounds (analogous to how we use the 0SDR $stu$-symmetric bootstrap to reconstruct the 2SDR $stu$-symmetric bootstrap bounds via scalar-subtraction and rescaling). A relationship between the two bootstraps would shed light on what the corners in the space of pion EFTs are and how they relate to non-Abelian scalar EFTs with SUSY.

\vspace{2mm}
\noindent {\bf The 6d $a$-Theorem.}
The results of the 0SDR bootstrap have an interesting connection to the $a$-theorem, because the dilaton effective action is a special case of the Abelian scalar EFT. 
In the $D=6$ dilaton effective action \cite{Schwimmer:2010za,Komargodski:2011vj,Elvang:2012st}, the Wilson coefficient $g_3$ is proportional to $\Delta a = a_\text{UV}-a_\text{IR}$. It may appear from Figure \ref{fig:0sdr_extr} that the numerical 0SDR bootstrap implies
$g_3 \ge 0$. However, the lower bound on $g_3/g_0$ is for finite $k_\text{max}$ actually slightly negative; for $k_\text{max}=8$ it is $-4 \cdot 10^{-6}$ while for $k_\text{max}=14$ it is $-4 \cdot 10^{-7}$. Further, if the Convex Hull Conjecture holds for the 0SDR bootstrap, then since all extremal theories have $g_{3}> 0$, then no 0SDR amplitude can have $g_3 < 0$. Thus, there is a numerical indication that the the lower bound on $g_3/g_0$ goes to zero as a power law as $k_\text{max} \to \infty$. However, this would only give a rather weak numerical version of the 6d $a$-theorem for the case where the dilaton 4-point amplitude satisfies the stronger Froissart constraint with $n_0 = 0$. This includes the  supersymmetric case for which the 6d $a$-theorem has already been established \cite{Cordova:2015fha}. 

The near-positivity of the Wilson coefficients can be understood in the 0SDR bootstrap from the fact that the lightest massive state must be a scalar (unless the theory is the IST), so the Wilson coefficients are dominated by scalar contributions, which are always positive. However, this is only true in the infinite $k_\text{max}$ limit, and thus it is compatible with the finding that at any finite $k_\text{max}$ there is still a tiny region of allowed space with $g_3<0$. This is due to higher spins at the gap still being allowed with tiny couplings at finite $k_\text{max}$, even though they are exponentially suppressed as $k_\text{max}$ increases.

\section*{Acknowledgements}
We would like to thank Jan Albert, Emil Bjerrum-Bohr, Poul Henrik Damgaard, Paolo Di Vecchia, and Sasha Zhiboedov 
for useful comments and discussions.  
All four authors would like to thank the Niels Bohr International Academy at the Niels Bohr Institute in Denmark for hospitality during the final stages of this work. 
This research was supported in part through computational resources and services provided by Advanced Research Computing at the University of Michigan, Ann Arbor. 
JB, HE, and NG are supported in part by Department of Energy grant DE-SC0007859.  JB was supported in part by the Cottrell SEED Award number 
CS-SEED-2023-004 from the Research Corporation for Science Advancement. NG was supported in part by the Van Loo and the Leinweber Postdoctoral Fellowships from the University of Michigan.  
LLL is supported by a Graduate Research Fellowship from the National Science Foundation under grant DGE-2241144, a Rackham Merit Fellowship from the University of Michigan, and a Career Development Fellowship from Out to Innovate.



\appendix

\section{Supersymmetry Constraints}
\label{app:SUSY}
In this appendix, we present details of how to obtain the DBI amplitude from the Veneziano amplitude and we discuss the general constraints of maximal supersymmetry.

\subsection*{Supersymmetry, 
Abelianization, and DBI}
Consider $\mathcal{N}=4$ supersymmetry in $D=4$. The 4-gluon SYM EFT amplitude can be written \cite{Berman:2023jys} as
\be
\label{4ptglueplus}
A[1^- 2^-  3^+ 4^+ ] 
  = \<12\>^2 [34]^2 g(s,u)\,,
\ee 
where the function $g$ is required by supersymmetry to be symmetric in $s$ and $u$. As examples, we have
\be
\label{gfctex}
  \begin{array}{lll}
    \text{Pure YM:}
    &&
    \displaystyle
    g(s,u) = -\frac{1}{su}\,,
    \\[3mm]
    \text{Veneziano:}
    &&
    \displaystyle
    g(s,u) =
    -\a'^2 \frac{\Gamma(-\a' s)\Gamma(-\a' u)}{\Gamma(1+\a' t)}\,,
  \end{array}
\ee
 and in a general low-energy EFT expansion, $g(s,u) = -1/(su) + $higher-derivative terms symmetric in $s$ and $u$.

Obtaining the Abelian amplitude from a color-ordered one with the same external states is done by summing over the inequivalent color-orderings:
\be
  \label{u1decoup}
  A^\text{Abelian}(1234)=A[1234]+ A[1342] + A[1423]  \,.
\ee
Applying this to the general MHV gluon amplitude \reef{4ptglueplus}, we get a photon amplitude
\be
  \label{Abelianproj}
  \begin{split}
  A^\text{Abelian}(1^-2^-3^+4^+)
  &=A[1^-2^-3^+4^+]+ A[1^-3^+4^+2^-] + A[1^-4^+2^-3^+]
  \\
  &= 
\<12\>^2 [34]^2\Big(
g(s,u) + g(t,s) + g(u,t) \Big)\,.
  \end{split}
\ee 
In this sum, the leading gluon self-interaction adds up to zero, as it should since photons do not self-interact in a 2-derivative theory; indeed the RHS of \reef{u1decoup} being zero is simply the $U(1)$ decoupling identity.

The $\mathcal{N}=4$ supersymmetry connect all 16 massless states of the CTP self-conjugate supermultiplet, and the resulting Ward identities imply that all 4-point amplitudes are proportional. In particular, picking two conjugate complex scalars $z$ and $\bar{z}$ among the 3 pairs of complex scalars, the amplitude
$A(zz\bar{z}\bar{z})$ is  the same as $A(1^-2^-3^+4^+)$ with 
the polarization spinor-helicity factor $\<12\>^2[34]^2$ replaced by $s^2$. (This is true for both the non-Abelian and Abelian amplitudes.) Hence, 
\be
   A^\text{Abelian}(zz\bar{z}\bar{z})
   =s^2 \Big(
g(s,u) + g(t,s) + g(u,t) \Big)\,.
\ee
Note that because $g(u,s) = g(s,u)$, it follows that any  maximally supersymmetric Abelian complex scalar 4-point amplitude 
can be written as
\be
  \label{complxA}
   A^\text{Abelian}(zz\bar{z}\bar{z})
   =s^2 f(s,t,u) \,,
\ee
where $f(s,t,u)=g(s,u) + g(t,s) + g(u,t)$ is fully symmetric in $s,t,u$. 

If $\phi$ is the real part of $z$, we obtain the real scalar amplitude from
\be 
  \label{zbarztophi}
  A^\text{Abelian}(\phi\phi\phi\phi)
  = 
 A^\text{Abelian}(zz\bar{z}\bar{z})
  + A^\text{Abelian}(z\bar{z}z\bar{z})
   +A^\text{Abelian}(\bar{z}zz\bar{z})
\ee
Thus, we find that the real scalar amplitude in any $\mathcal{N}=4$ SYM EFT can be written 
\be
 A^\text{Abelian}(\phi\phi\phi\phi)
 = (s^2 +t^2 + u^2) \,f(s,t,u)
\ee
for $f$ is fully symmetric in $s,t,u$. This will be important in Section \ref{sec:susyboot}.

Applying the Abelianization \reef{u1decoup} to the open superstring tree amplitude (Veneziano) in \reef{gfctex} we obtain the complex scalar DBI amplitude \reef{Astripped}. Subsequently,  transformation from complex to real scalars as in \reef{zbarztophi}, we find the  real scalar DBI amplitude \reef{eq:dbi}.

\subsection*{Supergravity}
The MHV closed string amplitude  
for scattering of four gravitons can in $D=4$ be written
\begin{equation}
 A(++--) = -\a'^4 [12]^4
 \<34\>^4 \frac{\Gamma(-\a's)\Gamma(-\a't)\Gamma(-\a'u)}{\Gamma(1+\a's)\Gamma(1+\a't)\Gamma(1+\a'u)}\,.
\end{equation}
In $\mathcal{N} \ge 4$ supergravity, this amplitude is related to a scalar amplitude with two complex conjugate scalars $z$ and $\bar{z}$
via the SUSY Ward identity
\begin{equation}
  A(zz\bar{z}\bar{z})= \frac{[34]^4}{\<12\>^4} A(++--)
  \,.
\end{equation}
Using $s = \<12\>[12]=\<34\>[34]$, this gives
\begin{equation}
  A(zz\bar{z}\bar{z}) = - s^4 \frac{\Gamma(-s)\Gamma(-t)\Gamma(-u)}{\Gamma(1+s)\Gamma(1+t)\Gamma(1+u)} \,.
\end{equation}
This has $t \leftrightarrow u$ symmetry but not full $s,t,u$ symmetry because the external states are complex scalar (so it is symmetric only in the identical states, just as in \reef{complxA}).

To get the amplitude for four real scalars, so we use \reef{zbarztophi} to get
\begin{equation}
  A(\phi\phi\phi\phi)
  =
  -2(s^4+t^4+u^4)\frac{\Gamma(-s)\Gamma(-t)\Gamma(-u)}{\Gamma(1+s)\Gamma(1+t)\Gamma(1+u)}
\end{equation}
Since $s+t+u = 0$, we can write
\begin{equation}
  \label{s4tos22}
  (s^4+t^4+u^4) =\frac{1}{2}\big(s^2+t^2+u^2\big)^2
\end{equation}
so that we have as the final result 
\begin{equation}
  A(\phi\phi\phi\phi)
  =
  -\big(s^2+t^2+u^2\big)^2\frac{\Gamma(-s)\Gamma(-t)\Gamma(-u)}{\Gamma(1+s)\Gamma(1+t)\Gamma(1+u)}\,.
\end{equation}
This is the supergravity amplitude \reef{susygravitonAmp}. The real scalar can be thought of as the dilaton.

\section{Gegenbauer Polynomials}\label{app:gegenbauer}

In this appendix, we review some useful properties of the Gegenbauer polynomials. We use the conventions of~\cite{Correia:2020xtr}. Many more identities can be found in~\cite{Gradshteyn:1702455}.

In $D \geq 3$ spacetime dimensions, the Gegenbauer polynomials $ G^{(D)}_j(\cos\theta)$ diagonalize the Lorentz group Casimir operator. The index $j$ is often called the spin. In $D=4$, they reduce to the familiar Legendre polynomials. The Gegenbauer polynomials $ G^{(D)}_j(x)$ with real ${D > 3}$ and integer ${j \geq 0}$ are defined by the following generating function,
\begin{align}
    \frac{1}{ (1-2xt+t^2)^{\frac{d-3}{2}} }
    =
    \sum_{j=0}^\infty
    \frac{ \Gamma(j+d-3) }
         { \Gamma(j+1) \Gamma(d-3) }
     G^{(D)}_j(x)
    \,
    t^j
    \, .
\end{align}
The generating function definition can be analytically continued in $D$ and $j$ using the ${}_2 F_1$ hypergeometric function,
\begin{align}
\label{eq:Gegdef}
     G_j^{(D)}(x)
    &=
    {}_2 F_1
    \Big(
    {-j}, j+D-3; \frac{D-2}{2}; \frac{1-x}{2}
    \Big)
    \, .
\end{align}
This definition is convenient since it extends down to~${D=3}$ and manifestly normalizes the Gegenbauer polynomials by $ G_j^{(D)}(1) = 1$ for all $D$ and $j$ for which they are defined. Hence, we take~\eqref{eq:Gegdef} as the definition of the Gegenbauer polynomials and restrict to real ${D > 2}$ and integer ${j \geq 0}$.

Several useful properties and identities follow directly from the definition~\eqref{eq:Gegdef} and the series representation of the hypergeometric function.  Each $ G_j^{(D)}(x)$ is a polynomial in~$x$ of degree~$j$ with parity~$j$,
\begin{align}
     G_j^{(D)}(-x)
    =
    (-)^j \,
     G_j^{(D)}(x)
    \, .
\end{align}
The first few Gegenbauer polynomials are
\begin{align}
     G_0^{(D)}(x)
    &=
    1
    \vphantom{\tfrac{1}{D-2}}
    \, ,
\no \\
     G_1^{(D)}(x)
    &=
    x
    \vphantom{\tfrac{1}{D-2}}
    \, ,
\no \\
     G_2^{(D)}(x)
    &=
    \tfrac{1}{D-2}
    \big( (D-1) x^2 - 1 \big)
    \, ,
\no \\
     G_3^{(D)}(x)
    &=
    \tfrac{1}{D-2}
    \big( (D+1) x^3 - 3x \big)
    \, ,
\no \\
     G_4^{(D)}(x)
    &=
    \tfrac{1}{D(D-2)}
    \big( (D+3)(D+1) x^4 + 3(D+2) x^2 + 3 \big)
    \, .
\end{align}
The product of two Gegenbauer polynomials with spins $j_1$ and $j_2$ may be expanded into a positive sum of spin-$j$ Gegenbauer polynomials with $|j_1-j_2| \leq j \leq j_1+j_2$ as follows,
\begin{align}
\label{eq:GegProd}
     G_{j_1}^{(D)}(x) \,
     G_{j_2}^{(D)}(x)
    &=
    \sum_{j = |j_1-j_2|}^{j_1+j_2}
    c_{j_1,j_2;j}^{(D)} \,
     G_{j}^{(D)}(x)
    \, ,
\end{align}
where
\begin{align}
    c_{j_1,j_2;j}^{(D)}
    &=
    \begin{cases}
    \frac{ ( j+\frac{D-3}{2} ) \,
           \Gamma(D-3) \,
           \Gamma(J+D-3)
           \vphantom{\frac{1}{2}} }
         { \Gamma(\frac{D-3}{2})^2 \,
           \Gamma(J+\frac{D-1}{2}) \,
           \vphantom{\frac{1}{2}} }
    \frac{ \Gamma(J-j+\frac{D-3}{2})
           \vphantom{\frac{1}{2}} }
         { \Gamma(J-j+1)
           \vphantom{\frac{1}{2}} }
    &
    j_1+j_2+j \text{ even}
    \\
    \times
    \frac{ \Gamma(J-j_1+\frac{D-3}{2})
           \vphantom{\frac{1}{2}} }
         { \Gamma(J-j_1+1)
           \vphantom{\frac{1}{2}} }
    \frac{ \Gamma(J-j_2+\frac{D-3}{2})
           \vphantom{\frac{1}{2}} }
         { \Gamma(J-j_2+1)
           \vphantom{\frac{1}{2}} }
    \frac{ \Gamma(j_1+1)
           \vphantom{\frac{1}{2}} }
         { \Gamma(j_1+D-3) 
           \vphantom{\frac{1}{2}} }
    \frac{ \Gamma(j_2+1) 
           \vphantom{\frac{1}{2}} }
         { \Gamma(j_2+D-3) 
           \vphantom{\frac{1}{2}} }
    \, ,
    \\[2ex]
    \hfill
    0 \, ,
    \hfill \hfill
    &
    j_1+j_2+j \text{ odd} 
    \end{cases}
\end{align}
and $J=\half(j_1+j_2+j)$.

For arguments of the form $x = 1+2\delta$, we can expand the Gegenbauer polynomials in powers of $\delta$ as follows:
\begin{align}
\label{eq:Gegv}
     G_j^{(D)}(1+2\delta)
    &=
    \sum_{n=0}^j
    v_{j,n}^{(D)} \, 
    \delta^n
    \, ,
    \quad \text{where} \quad
    &
    v_{j,n}^{(D)}
    &=
    \binom{j}{n}
    \frac{ \Gamma(j+D-3+n) \, \Gamma(\frac{D-2}{2}) }
         { \Gamma(j+D-3) \, \Gamma(\frac{D-2}{2}+n) }
    \, .
\end{align}
For real $D > 2$ and integer $n \geq 0$, each coefficient $v_{j,n}^{(D)}$ is a polynomial in $j$ of degree $2n$. The leading coefficient is given by
\begin{align}
    \lim_{j \to \infty}
    \frac{ v_{j,n}^{(D)} }
         { j^{2n} }
    =
    \frac{1}{n!}
    \frac{ \Gamma(\frac{D-2}{2}) \phantom{{}+n} }
         { \Gamma(\frac{D-2}{2}+n) }
    \, .
\end{align}
For numerical efficiency at large $j$, it is convenient to represent these coefficients without using the gamma function. We can write
\begin{align}
    v_{j,n}^{(D)}
    &=
    \binom{j}{n}
    \frac{ (j+D-3)_n }
         { ( \frac{D-2}{2} )_n }
    \, ,
\end{align}
where $(z)_n = \Gamma(z+n)/\Gamma(z)$ is the rising Pochhammer symbol.

The Gegenbauer polynomials also satisfy a number of useful integral identities. First, we have the orthogonality relation,
\begin{align}
\label{eq:GegOrthInt}
    \int_{-1}^{1}
    \mathrm{d}x \,
    (1-x^2)^{\frac{D-4}{2}} \,
     G_{j}^{(D)}(x) \,
     G_{j'}^{(D)}(x)
    =
    \frac{ 2 \, \delta_{j,j'} }
         { \mcN^{(D)}_{\phantom{j}} n_j^{(D)} }    
    \, ,
\end{align}
where the two normalization factors are given by
\begin{align}
\label{eq:Gegn}
    \mcN^{(D)}
    =
    \frac{ 1 }
         { (16\pi)^{\frac{D-2}{2}} \, \Gamma(\frac{D-2}{2}) }
    \quad \text{and} \quad
    &
    n_j^{(D)}
    =
    \frac{ (4\pi)^{\frac{D}{2}} (2j+D-3) \, \Gamma(j+D-3) }
         { \pi \, \Gamma(\frac{D-2}{2}) \, \Gamma(j+1) }
    \, .
\end{align}
Another useful integral identity is
\begin{align}
\label{eq:GegxInt}
    \int_{-1}^{1}
    \mathrm{d}x \,
    (1-x^2)^{\frac{D-4}{2}} \,
     G_{j}^{(D)}(x) \,
    x^ J
    &=
    \begin{cases}
    \dfrac{ \Gamma(\frac{D-2}{2}) \,
            \Gamma( J+1) \,
            \Gamma(\frac{ J-j+1}{2}) }
          { 2^j \, 
            \Gamma( J-j+1) \,
            \Gamma(\frac{ J+j+D-1}{2}) }
    \, ,
    &
    \text{even }  J-j \geq 0 
    \\[2ex]
    \hfill
    0 \, ,
    \hfill \hfill
    &
    \text{otherwise}
    \end{cases}
    \, .
\end{align}
These two integrals,~\eqref{eq:GegOrthInt} and~\eqref{eq:GegxInt}, may be used to write the residues of any tree-level four-point amplitude in terms of Gegenbauer polynomials.

\section{Independent Null Constraints}\label{app:indnulls}
In this appendix, we derive the null constraints for both the 0SDR and 2SDR set ups. Our approach differs in spirit from (but is mathematically equivalent to) the contour integral approach present in much of literature~\cite{Caron-Huot:2020cmc}.

Null relations are a direct consequence of crossing symmetry. These relations may be simply derived by relating the two bases of Wilson coefficients, $g_{k}$ and $a_{k,q}$, which occur in the two equivalent forms of the low-energy expansion of $A(s,u)$. The $g_{k}$ Wilson coefficients appear in the manifestly permutation symmetric form of the low-energy expansion:
\begin{align}
    A(s,u)
    &=
    A_\text{massless}(s,u)
    +
    \sum_{m,n=0}^\infty
    g_{2m+3n; \, m,n} \, 
    (s^2+t^2+u^2)^m (stu)^n
    \, ,
\end{align}
with $t=-s-u$. In the 0SDR set up, we have dispersion relations for all the $g_k$. In the 2SDR set up, we have dispersion relations for the $g_k$ with $k \geq 2$. The $a_{k,q}$ Wilson coefficients appear as follows:
\begin{align}
    A(s,u)
    &=
    A_\text{massless}(s,u)
    +
    \sum_{k=0}^\infty
    \sum_{q=0}^k 
    a_{k,q} \, s^{k-q} u^q
    \, .
\end{align}
In the 0SDR set up, we have dispersion relations for all the $a_{k,q}$. In the 2SDR set up, we have dispersion relations for the $a_{k,q}$ with ${k-q \geq 2}$.

At each $k \geq 1$, the number of $a_{k,q}$ is larger than the number of $g_k$. For instance, in the $g_k$ basis, it is clear that there are no coefficients at Mandelstam order $k=1$, but in the $a_{k,q}$ basis, there are two coefficients with $k=1$. By comparing the two low-energy expansions, we can relate the two bases as follows:
\begin{equation}
\label{eq:akqtogmn}
    a_{k,q}
    =
    \frac{1}{q!}
    \frac{\p^q}{\p u^q}
    \bigg[
    \sum_{\substack{ m,n \geq 0 \\ 2m+3n = k }}
    2^m (-1)^n \,
    g_{k; \, m,n} \, 
    ( u^2 + u + 1 )^m
    ( u^2 + u )^n
    \bigg]_{u=0}
    \, .
\end{equation}
For each $k$, this expression defines a linear transformation from the $g_k$ to the $a_{k,q}$. We denote this transformation by $\bbT_k$ so that
\begin{align}
\label{eq:aTg}
    \vec{a}_k
    &=\
    \bbT_k \cdot \vec{g}_k
    \, ,
\end{align}
where $\vec{a}_k$ and $\vec{g}_k$ are column vectors of the respective Wilson coefficients. Each $\vec{a}_k$ has $k+1$ entries, and each $\vec{g}_k$ has $N(k)$ entries, where
\begin{align}
    N(k)
    &=
    \begin{cases}
        \floor{ \frac{1}{3} \floor{ \frac{k}{2} }} + 1
        & k \text{ even}
        \\
        \floor{ \frac{1}{3} \floor{ \frac{k-2}{2} }} + 1
        & k \text{ odd}
    \end{cases}
    \, .
\end{align}
This expression for $N(k)$ is simply the number of unordered partitions of $k$ into $2$ and $3$.

The $(k+1) \times N(k)$-dimensional matrix $\bbT_k$ is not square and therefore not invertible.  Moreover, the $a_{k,q}$ at fixed $k$ are themselves not all linearly independent! This fact is a consequence of the dispersion relation for $a_{k,q}$ and properties of the $w_{j;k,q}^{(D)}$ coefficients. For even $k$, the $a_{k,q}$ with even $q$ form a linearly independent basis (before taking into account crossing symmetry). For odd $k$, the $a_{k,q}$ with odd $q$ similarly form a linearly independent basis. To enforce this reduction to even or odd $q$, we define $(k+1) \times N(k)$-dimensional projectors $\bbP^{(\text{ind})}_{k}$ which project each $\vec{a}_k$ onto its first $N(k)$ entries with $q$ even/odd for $k$ even/odd, respectively. Explicitly, these projectors are given by
\begin{equation}
    \bbP^{(\text{ind})}_{k\;\text{even}}=
    \begin{pmatrix}
    1 & 0 & 0 & 0 & \cdots & 0 \\
    0 & 0 & 1 & 0 & \cdots & 0 \\
    \vdots & \vdots & \vdots & \vdots & \ddots & \vdots \\
    0 & 0 & 0 & 0 & \cdots & 0 \\
    \end{pmatrix}
    \, ,
    \qquad
    \bbP^{(\text{ind})}_{k\;\text{odd}}=
    \begin{pmatrix}
    0 & 1 & 0 & 0 & \cdots & 0 \\
    0 & 0 & 0 & 1 & \cdots & 0 \\
    \vdots & \vdots & \vdots & \vdots & \ddots & \vdots \\
    0 & 0 & 0 & 0 & \cdots & 0 \\
    \end{pmatrix}
    \, .
\end{equation}
Each $\bbP^{(\text{ind})}_{k}$ has exactly $N(k)$ entries equal to $1$. Moreover, the product $\bbP^{(\text{ind})}_{k} \cdot \bbT_k$ is an invertible $N(k) \times N(k)$-dimensional matrix. By acting with $\bbP^{(\text{ind})}_k$ on~\eqref{eq:aTg} and then inverting, we find
\begin{align}
\label{eq:gTa}
    \vec{g}_k
    &=
    (
    \bbP^{(\text{ind})}_k \cdot
    \bbT_k
    )^{-1}
    \cdot
    \bbP^{(\text{ind})}_k
    \cdot
    \vec{a}_k
    \, .
\end{align}
Hence, these projectors reduce $\vec{a}_k$ to an $N(k)$-dimensional sub-vector which can be uniquely transformed to $\vec{g}_k$. The remaining components of $\vec{a}_k$ can be turned into null relations. 

To compute the null relations we define $(k+1) \times N(k)$-dimensional projectors $\mathbb{P}^\mathrm{(null)}_{k}$ which project each $\vec{a}_k$ onto its entries which were not included in $\mathbb{P}^\mathrm{(ind)}_{k} \cdot \vec{a}_k$ and which have both $k-q \geq n_0$ (in the $n_0$SDR set up) and $q$ even/odd for $k$ even/odd, respectively. This definition uniquely defines $\mathbb{P}^\mathrm{(null)}_{k}$, so we omit its explicit expression. Next, we use~\eqref{eq:aTg} and~\eqref{eq:gTa} to write each component of $\vec{a}_k$ in terms of the $N(k)$ sub-components which we have mapped to $\vec{g}_k$:
\begin{align}
    \vec{a}_k 
    &=
    \bbT_k \cdot
    (
    \bbP^{(\text{ind})}_k \cdot
    \bbT_k
    )^{-1}
    \cdot
    \bbP^{(\text{ind})}_k
    \cdot
    \vec{a}_k
    \, .
\end{align}
This equation enforces crossing symmetry among the $a_{k,q}$ Wilson coefficients. To derive the null equations at Mandelstam order $k$, we simply act on both sides with $\bbP^{(\text{null})}_k$ and rearrange:
\begin{align}
\label{eq:nulleqs}
    \vec{0}
    &=
    \bbP^{(\text{null})}_k \cdot
    \big(
    \mathds{1}
    -
    \bbT_k \cdot
    (
    \bbP^{(\text{ind})}_k \cdot
    \bbT_k
    )^{-1}
    \cdot
    \bbP^{(\text{ind})}_k
    \big)
    \cdot
    \vec{a}_k
    \, .
\end{align}
Here $\vec{0}$ is a column vector of zeroes with the appropriate length, and $\mathds{1}$ is the $(k+1) \times (k+1)$-dimensional identity matrix. This final expression provides the null relations for any $k$ and for any number of subtractions. Because it is a simple linear-algebraic expression, we can compute the null relations to arbitrarily high Mandelstam order near-instantaneously using Mathematica.

\section{Fixed Coupling Analysis for DBI}
\label{app:fixcoupling}

\begin{figure}
     \centering
     \includegraphics[width=.49\linewidth]{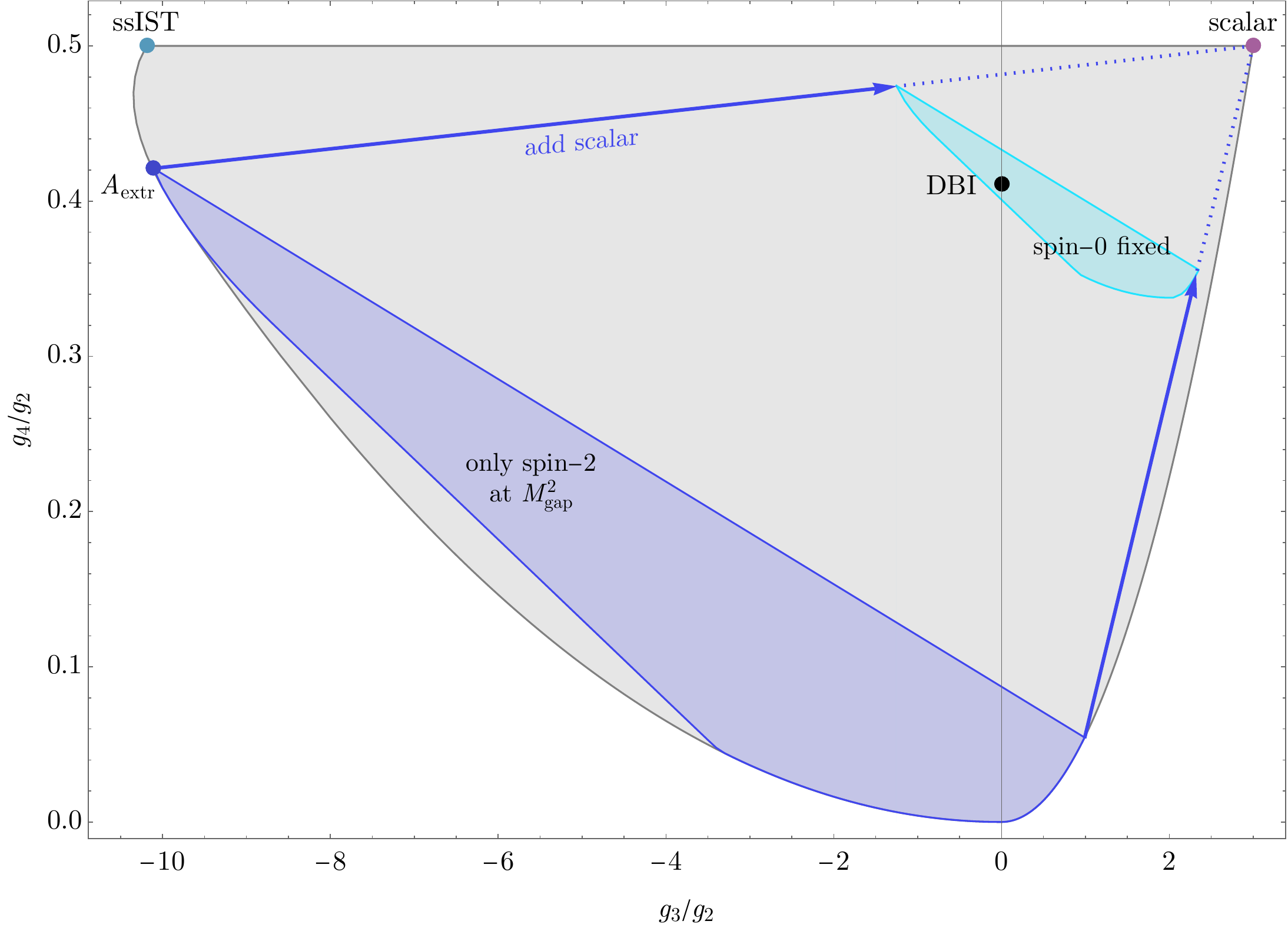}~
     \includegraphics[width=.49\linewidth]{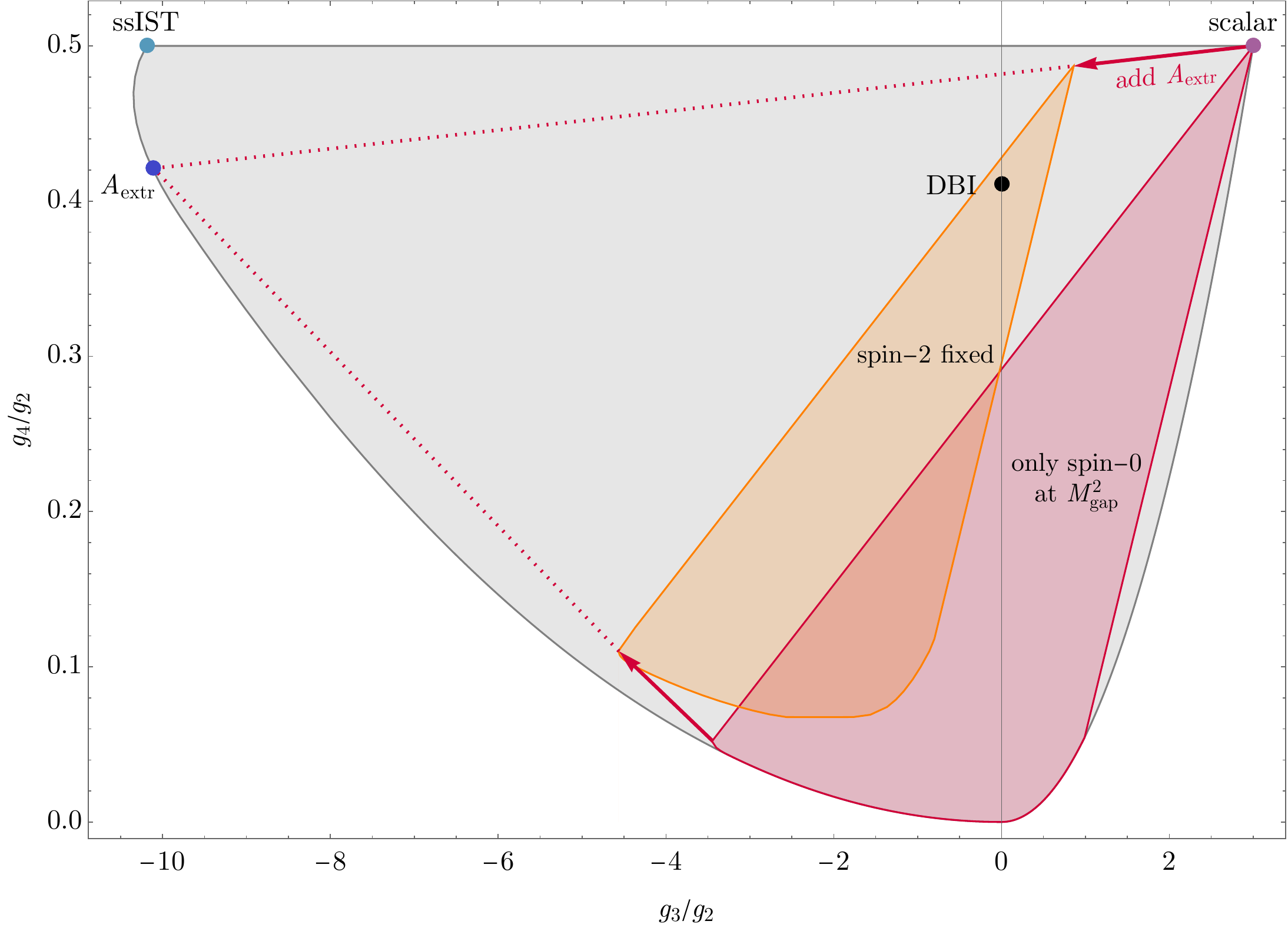}
    \caption{In $D=4$, allowed regions for a mass spectrum with spin 0 and 2 at $M_\text{gap}^2$ and no other states until 
    $3 M_\text{gap}^2$ in the non-supersymmetric 2SDR bootstrap. Orange region: spin 2 coupling fixed to the DBI value while the spin 0 coupling is unfixed. Cyan region: spin 0 coupling fixed to the DBI value and the spin 0 coupling is unfixed. Each of these region can be obtained by adding the spin 0 theory or the $\mu_c=3$ extremal $A_\text{extr}$ theory to a region that has, respectively, only spin 2 (blue) or only spin 0 (red) at $M_\text{gap}^2$, hence they have the same shape, only scaled in size. 
    }
    \label{fig:2sdr_islands}
\end{figure}

In this appendix, we provide additional details for the DBI non-SUSY fixed coupling analysis of Section \ref{sec:DBInosusy}.

To begin, we allow a spin 0 and spin 2 at $M_\text{gap}^2$ while assuming that there are no other states until $3M_\text{gap}^2$.  Suppose we now fix the spin-0 coupling $|\lambda_0|^2/g_2$ to its DBI value \reef{DBIcoup02} while letting the spin-2 coupling be unfixed. The result is the cyan region shown in Figure \ref{fig:2sdr_islands}.
The shape of this region is exactly a scaled-down version of the dark-blue region for which we have input only a spin-2 state with unfixed coupling at the mass gap and not allowed any spin-0 states. The reason is simply that the full cyan region with fixed spin-0 coupling can be obtained by simply adding the scalar to the amplitudes in the allowed only-spin-2 purple region. 

Similarly, if we instead fix the spin 2 coupling $|\lambda_2|^2/g_2$ to its DBI value \reef{DBIcoup02}, while leaving the spin 0 coupling to vary freely, we find the orange region in Figure \ref{fig:2sdr_islands}. Any theory with spin 2 fixed to the DBI value can be obtained by adding the extremal amplitude  $A_\text{extr}$ with maximal spin 2 coupling at the gap (and $\mu_c=3$) to the amplitudes in the region with only spin 0 at the mass gap (red). Therefore the orange region looks like a scaled-down version of the red region.

\bibliography{bootstrap}

\end{document}